\documentclass[aps,prd,onecolumn,nofootinbib,superscriptaddress]{revtex4}
\usepackage{graphicx}    
\usepackage{xcolor}      
\usepackage{caption}     
\usepackage{subcaption}  
\usepackage{amsmath, amssymb} 
\usepackage{float}       
\usepackage{ragged2e}    
\usepackage{tabularray}   
\usepackage{etoolbox}    
\usepackage{microtype} 
\usepackage[normalem]{ulem}
\pdfpageheight=\paperheight
\pdfpagewidth=\paperwidth
\makeatletter
\renewcommand{\@makecaption}[2]{
  \parbox{\linewidth}{
    \bfseries #1\quad\normalfont
    \justifying
    \fussy
    \tolerance=2000
    \emergencystretch=1em
    \hyphenpenalty=200
    \setlength{\parfillskip}{0pt plus 1fil}
    #2
  }
  \vspace{\belowcaptionskip}
  \vspace{-\baselineskip}
}
\makeatother
\usepackage[colorlinks=true,
            linkcolor=blue,
            urlcolor=blue,
            filecolor=black,
            citecolor=red]{hyperref}

\usepackage{nameref}
\makeatletter
\let\orglabel\label
\renewcommand{\label}[1]{\orglabel{#1}}
\makeatother

\begin{document}
\baselineskip=0.4 cm

\title{Images from disk and spherical accretions of Bardeen black hole surrounded by perfect fluid dark matter}
\author{Hui Zeng}
\affiliation{School of Electronic Information Engineering, China West Normal University, Nanchong, 637002, China}


\author{Yuan Meng}
\email{mengyuanphy@163.com}
\affiliation{School of Physics and Astronomy, China West Normal University, Nanchong, 637002, China}

\date{\today}

\begin{abstract}
\baselineskip=0.5 cm
In this paper, we investigate the shadow and optical appearance of the Bardeen black hole surrounded by perfect fluid dark matter (PFDM) illuminated by various static accretions. First, we find that as the dark matter parameter $\left|\alpha\right|$ increases, the fundamental characteristic quantities of the black hole, the event horizon $r_h$, the photon sphere radius $r_{ph}$, and the critical impact parameter $b_{ph}$ all increase, while the peak of the effective potential $V_{\text{eff}}$ decreases and shifts toward the direction of increasing $r$. In contrast, the magnetic charge parameter $g$ suppresses $r_h$, $r_{ph}$, and $b_{ph}$, while increases the peak of $V_{\text{eff}}$ and shifts it toward the direction of decreasing $r$. This indicates a competing effect between the dark matter parameter $\left|\alpha\right|$ and the magnetic charge parameter $g$ on the fundamental properties of the black hole. Furthermore, we use the EHT observational data to constrain the dark matter parameters $\alpha$ and find that the constraint range given by the supermassive black hole SgrA* is stricter than that of the black hole M87*. Finally, we investigate black hole images under different accretion models. The results reveal that both dark matter parameters $\alpha$ and accretion models significantly influence the black hole images. For larger dark matter parameters $\left|\alpha\right|$, the inner shadow or central faint illuminating region of the Bardeen black hole surrounded by PFDM is larger, but the bright ring of the image is fainter. This provides a potential method for us to distinguish between classical Bardeen black holes and Bardeen black holes surrounded by PFDM. These preliminary results may provide some clues for future investigations of dark matter using black hole shadows and images.

\end{abstract}


\maketitle
\tableofcontents

\section{Introduction}

Recent breakthroughs in Very Long Baseline Interferometry by the Event Horizon Telescope (EHT) collaboration have opened new avenues for probing the fundamental nature of gravity in the strong field regime. The most significant achievement is that the EHT successfully captured and published the first images of two supermassive black holes: one is the supermassive black hole at the center of the M87* galaxy \cite{EventHorizonTelescope:2019dse,EventHorizonTelescope:2019uob,EventHorizonTelescope:2019jan,EventHorizonTelescope:2019ths,EventHorizonTelescope:2019pgp,EventHorizonTelescope:2019ggy}, and the other is the supermassive black hole SgrA* at the center of the Milky Way system \cite{EventHorizonTelescope:2022wkp,EventHorizonTelescope:2022apq,EventHorizonTelescope:2022wok,EventHorizonTelescope:2022exc,EventHorizonTelescope:2022urf,EventHorizonTelescope:2022xqj}. Due to the strong gravitational lensing effect of the black hole, the surrounding light rays are captured and form the central shadow in the image \cite{Virbhadra:1999nm,Virbhadra:2002ju,Virbhadra:2007kw,Synge:1966okc,Bardeen:1972fi,Bozza:2010xqn}. The boundary of the black hole shadow is determined by the critical impact parameter corresponding to the unstable circular orbit of the photon \cite{Perlick:2021aok}. Moreover, the bright ring-like structure results from the superposition of multiple light rays. This is closely related to the distribution and structure of accretion matter, and strong gravitational lensing effects around the black hole.

In fact, originally, the black hole shadow was termed the escape cone of the photon \cite{Synge:1966okc}. Subsequently, the angular radius of the photon capture region of the Schwarzschild black hole was derived \cite{Luminet:1979nyg}. Bardeen's research further showed that a rotating Kerr black hole would produce a $D$-shaped shadow structure, which is attributed to the dragging effect of the black hole, resulting in an asymmetry in the orbital radii of prograde and retrograde photons \cite{Bardeen:1972fi}. This has given rise to extensive discussion on black hole shadows, and various numerical simulations of black hole shadows have been discussed \cite{Shen:2005cw,Yumoto:2012kz,Atamurotov:2013sca,Papnoi:2014aaa,Abdujabbarov:2015xqa,Kumar:2018ple}, which have shown that black hole shadows depend on spacetime geometry. In addition, black hole shadows in various modified theories of gravity and different spacetime dimensions have been extensively investigated, such as the Gauss-Bonnet theory \cite{Ma:2019ybz,Guo:2020zmf}, conformal gravity \cite{Meng:2022kjs}, the Chern-Simons type theory \cite{Meng:2023wgi,Ayzenberg:2018jip,Amarilla:2010zq}, $f(R)$ gravity \cite{Addazi:2021pty,Dastan:2016vhb}, and so on \cite{Amarilla:2011fx,Amarilla:2013sj,Amir:2017slq,Mizuno:2018lxz,Eiroa:2017uuq,Vagnozzi:2019apd,Banerjee:2019nnj,Chowdhuri:2020ipb,Konoplya:2019goy,Younsi:2016azx,Olmo:2023lil}. Furthermore, the shadows of naked singularities \cite{Shaikh:2018lcc,Joshi:2020tlq,Dey:2020bgo} and wormholes \cite{Rahaman:2021web,Kasuya:2021cpk,Shaikh:2018oul,Wielgus:2020uqz,Peng:2021osd,Neto:2022pmu,Tsukamoto:2012xs} have been discussed, respectively. It is worth noting that the EHT has provided observational constraints on the shadow characteristics of the supermassive black holes M87* and SgrA*. For example, EHT observations have constrained the angular diameter of the SgrA* black hole shadow to be $\theta_d=48.7\pm7 \mu as$ \cite{EventHorizonTelescope:2022wkp}, and the angular diameter of the supermassive M87* black hole shadow to be $29.32\mu as<\theta_d<51.06\mu as$ \cite{Kuang:2022ojj,EventHorizonTelescope:2021dqv}. Although the supermassive black holes M87* and SgrA* are generally considered to be Kerr black holes, the limited resolution of the EHT still leaves room for alternative black hole models, or those predicted by modified theories of gravity. In addition, the astronomical observations of the supermassive black holes M87* and SgrA* can be used to test and constrain both alternative general relativistic black hole models and modified theories of gravity. Consequently, using the black hole shadow observed by the EHT to test and constrain the modified gravity theories has been widely discussed in \cite{Vagnozzi:2022moj,Afrin:2021imp,Kuang:2022ojj,Tang:2022hsu,Tang:2022bcm,Kuang:2022xjp,Kumar:2019pjp,Shaikh:2021yux,Wu:2023yhp,Capozziello:2023tbo,Sui:2023rfh,Pantig:2022qak,Ghosh:2023kge,Tsukamoto:2014tja}.

The bright ring-like structure in EHT images is closely associated with accreting matter around the black hole, and its spatial distribution is significantly influenced by the surrounding luminous accretion flow. This accretion material is mainly composed of gas and dust, forming a rapidly rotating accretion disk around the black hole. The formation of black hole images arises from the combined effects of accretion disk radiation, photon rings, gravitational lensing, and relativistic effects, which generally require numerical simulations using general relativistic magnetohydrodynamics (GRMHD) \cite{EventHorizonTelescope:2019pcy}. In fact, some simple accretion disk models are sufficient to capture key features of black hole images and reveal the properties of strong gravitational field regions. Wald et al. first investigated the imaging features of Schwarzschild black hole using a simplified thin accretion disk model \cite{Gralla:2019xty}, and defined three types of emission from the accretion disk based on the number $m$ of intersections of light rays with the accretion disk: direct emission ($m=1$), lensed ring emission ($m=2$), and photon ring emission ($m\geq 3$). Their study revealed that the observed intensity of the accretion disk primarily originates from direct emission. The contribution of lensed ring emission is relatively small, and the contribution of photon ring emission is negligible. Subsequently, the thin accretion disk models and spherically symmetric accretion models for various types of black holes have been widely studied \cite{Dokuchaev:2019pcx,Peng:2020wun,He:2021htq,Eichhorn:2021iwq,Li:2021riw,Wang:2023vcv,Wang:2024lte,Meng:2024puu,Zeng:2020dco,Capozziello:2025wwl,Alloqulov:2025pex,Alloqulov:2024zln,Saurabh:2020zqg,Zeng:2020vsj,Qin:2020xzu,Narayan:2019imo}. These studies demonstrate that the black hole images are primarily determined by the properties of spacetime geometry rather than the details of accretion. It is noteworthy that black holes featuring a double photon sphere structure exhibit additional ring features in their images, which differs from the typical case of a single photon sphere \cite{Gan:2021xdl,Gan:2021pwu,Meng:2023htc,Wang:2025hzu}. Moreover, considering more realistic astrophysical conditions, the accretion images of various types of rotating black holes have also been extensively studied \cite{Hou:2022eev,Wang:2023fge,Liu:2021yev,Zhang:2023bzv,He:2024amh,Zheng:2024ftk,Guo:2024mij,Heydari-Fard:2023kgf,Donmez:2024lfi,Yang:2024nin,Li:2024ctu,Wang:2024uda,Meng:2025ivb}. In general, using black hole image features to distinguish between general relativistic black holes \cite{Guo:2022iiy,Guo:2022ghl,Archer-Smith:2020hqq,Okyay:2021nnh,Uniyal:2022vdu,Hou:2022eev,Uniyal:2023inx,Akbarieh:2023kjv,Gao:2023mjb,Theodosopoulos:2023ice,Meng:2023uws}, other compact objects \cite{Boshkayev:2022vlv,Xavier:2023exm,Sakai:2014pga,Bacchini:2021fig,Destounis:2023khj}, and black holes in modified gravity theories is a powerful detection method, but this does not always work, as
even in GR, the images of black holes may exhibit degeneracies. For example, sufficiently dense boson stars \cite{Liebling:2012fv,Rosa:2023qcv} may have shadows similar to those of classical black holes \cite{Rosa:2022tfv} and images of very relativistic rotating boson stars may resemble Kerr black holes \cite{Vincent:2015xta}.

General relativity (GR) has achieved remarkable success since its establishment, but it still faces challenges in explaining cosmological observations. For example, the observed cosmic acceleration expansion cannot be directly explained by the original theory, which may require either modifications to GR or the introduction of new physical components \cite{SupernovaSearchTeam:1998fmf,SupernovaCosmologyProject:1998vns,Hao:2025utc}. Currently, the standard cosmological model ($\Lambda$CDM), which rests on dark energy and dark matter, is capable of fitting a wide range of observational data with extremely high accuracy. This framework usually requires retaining the cosmological constant term in the Einstein field equation or introducing other dynamical models \cite{Planck:2018vyg,Peebles:2002gy}. In addition, the shadow and gravitational lensing effects produced by black holes surrounded by dark matter halos have been extensively studied. For example, the rotating charged black hole surrounded by perfect dark matter \cite{Das:2021otl}, the Schwarzschild black hole surrounded by perfect dark matter \cite{Atamurotov:2021hoq}, the rotating nonlinear charged black hole in perfect fluid dark matter (PFDM) \cite{Ma:2020dhv}, the Euler-Heisenberg black hole surrounded by PFDM \cite{Su:2024lvs}, etc \cite{Liu:2023xtb,Dike:2022heo}. Moreover, various accretion images of black holes surrounded by dark matter have also been widely discussed. Such as the rotating Bardeen black hole surrounded by PFDM \cite{He:2024amh}, Brane-World black hole by PFDM \cite{Zeng:2021mok}, rotating black holes in PFDM \cite{Heydari-Fard:2022xhr}, Euler-Heisenberg black hole in the cold dark matter halo \cite{You:2024uql}, etc \cite{Claros:2024atw,Pathrikar:2025sin,Deligianni:2020tyz,Mizuno:2018lxz,Sun:2023woa}. These studies show that dark matter alters the spacetime geometry around the black hole, which has a significant impact on the black hole shadow, gravitational lensing effects, and accretion images. Therefore, we aim to investigate regular Bardeen black holes surrounded by PFDM and study the imprints of dark matter on black hole accretion images.

This paper is organized as follows. In Sec.\ref{sec:null geodesic}, we study the effective potential of photons around the black hole to analyze the photon sphere radius and the impact parameter, and also constrain the range of the coupling parameter by means of EHT observations. In Sec.\ref{sec:photon deflection}, utilizing the Euler-Lagrange equation and the ray-tracing method, we calculated the deflection trajectory of photons around the Bardeen black hole surrounded by PFDM. Furthermore, we employed two simplified emission functions to investigate the total observed intensity of black hole illuminated by the thin disk accretion. In Sec.\ref{sec:thin spherical accretion}, we presented the shadow and image of a black hole illuminated by a static spherical accretion. Sec.\ref{sec:conclusion} presents the conclusion.

\section{Photons sphere and observational constraints of Bardeen black hole surrounded by PFDM}\label{sec:null geodesic}

\begin{figure}[htbp]
    \centering
    \begin{minipage}{\textwidth}
    \begin{subfigure}[t]{0.4\textwidth}
        \centering
        \includegraphics[width=\textwidth]{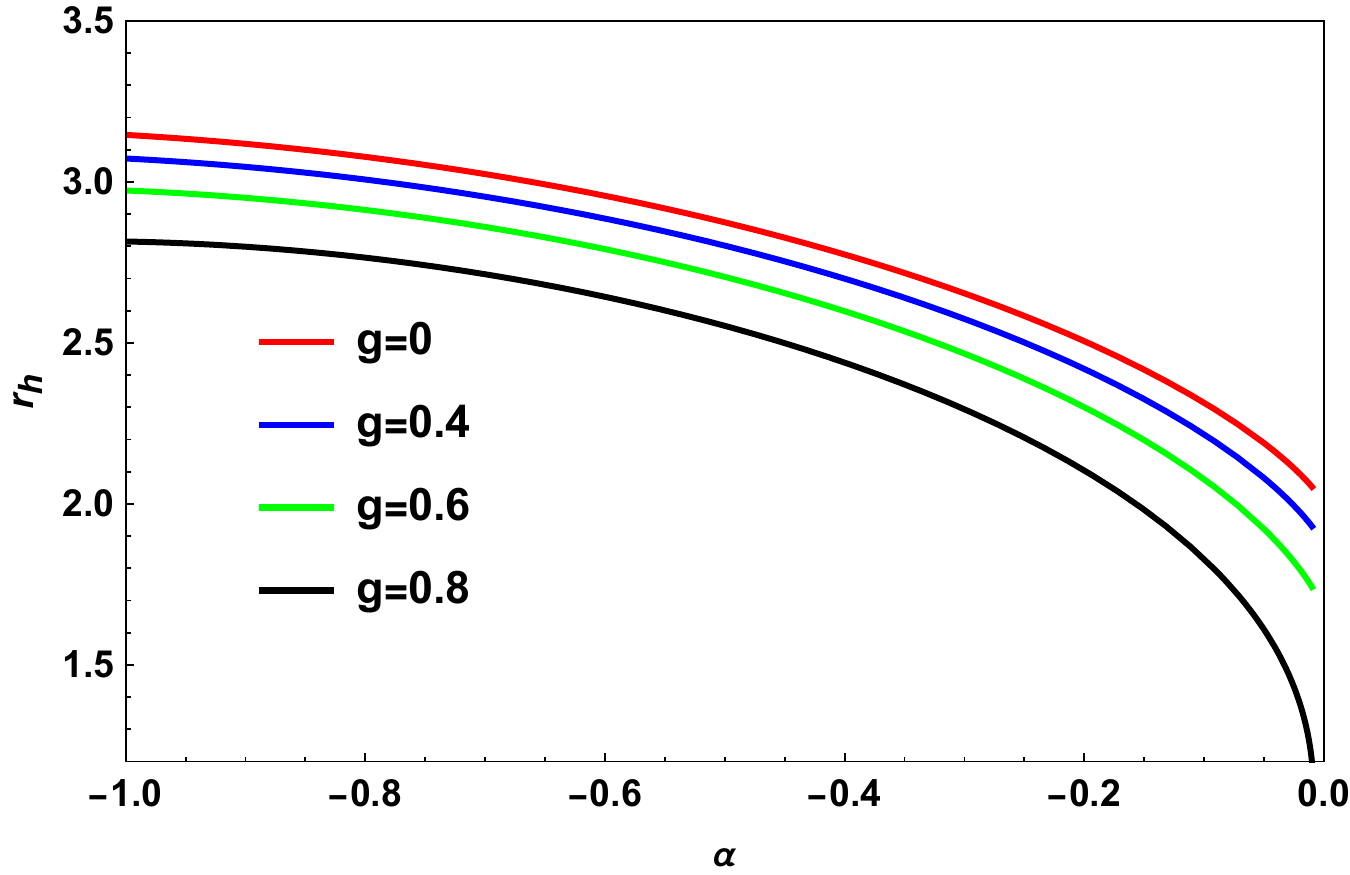}
        \caption{}
        \label{subfig:rh_a}
    \end{subfigure}
    \hspace{1mm}
\begin{subfigure}[t]{0.4\textwidth}
        \centering
        \includegraphics[width=\textwidth]{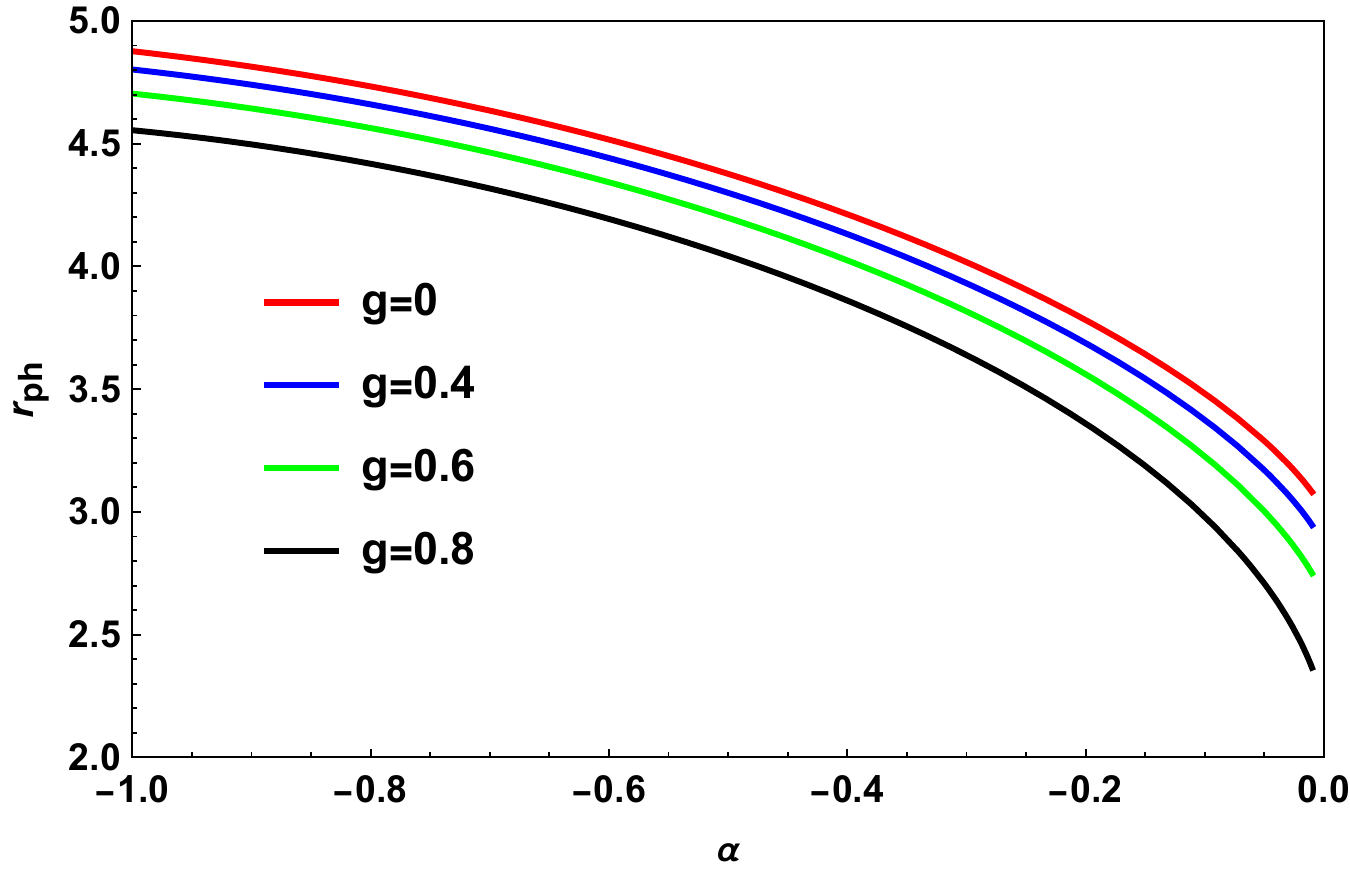}
        \caption{}
        \label{subfig:rh_g}
    \end{subfigure}
    \end{minipage}

    \vspace{0.1em}
    
\begin{minipage}{\textwidth}
\begin{subfigure}[t]{0.4\textwidth}
        \centering
        \includegraphics[width=\textwidth]{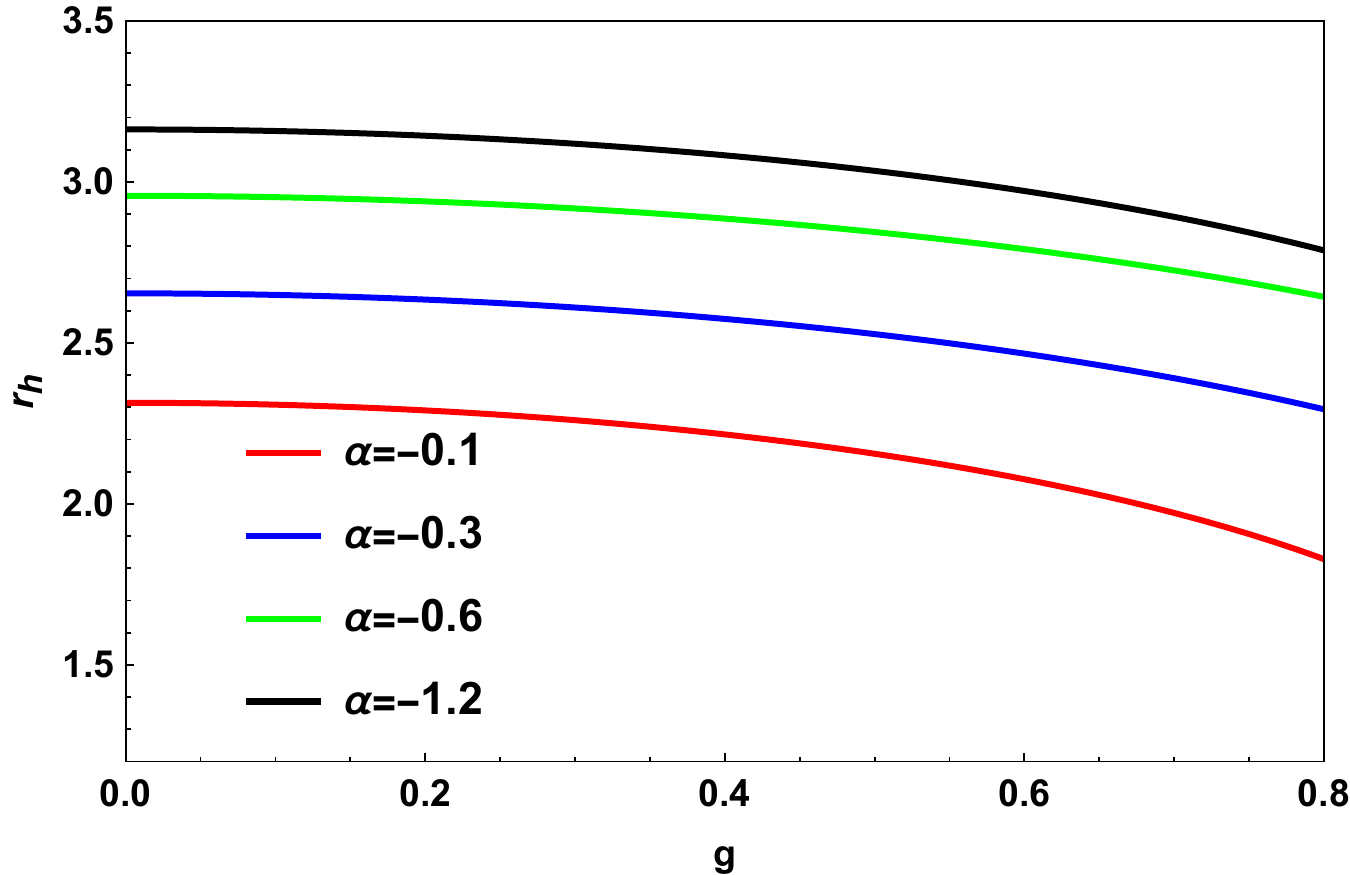}
        \caption{}
        \label{subfig:rph_a}
    \end{subfigure}
    \hspace{1mm}
\begin{subfigure}[t]{0.4\textwidth}
        \centering
        \includegraphics[width=\textwidth]{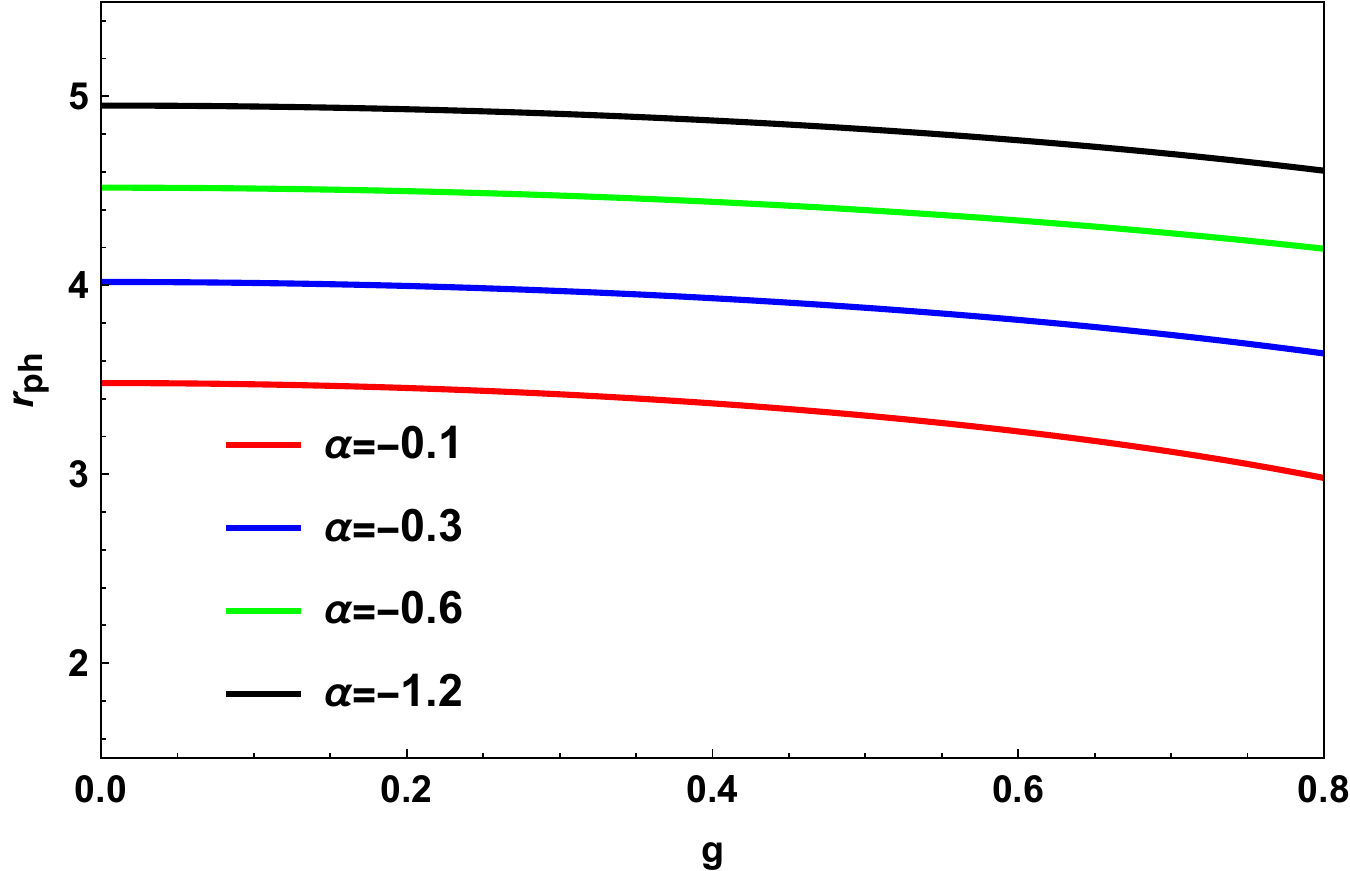}
        \caption{}
        \label{subfig:rph_g}
    \end{subfigure}
  \end{minipage}
    \caption{%
     \justifying 
     The event horizon $r_h$ and the photon sphere radius $r_{ph}$ as functions of the dark matter parameter $\alpha$ and the magnetic charge parameter $g$ of the Bardeen black hole surrounded by PFDM. Here we set $M=1$.  
   } 
    \label{fig:rh-rph}
\end{figure}

For the static spherically symmetric spacetime, considering the coupling between the gravitational field and the non-linear electromagnetic field, a Bardeen black hole surrounded by PFDM can be obtained \cite{Zhang:2020mxi} 
\begin{eqnarray}
ds^2=-f(r)dt^2+\frac{1}{f(r)}dr^2+r^2(d\theta^2+\sin^2\theta d\phi^2)\quad \text{with} \quad 	f\left(r\right)=1-\frac{2Mr^{2}}{\left(r^{2}+g^{2}\right)^{\frac{3}{2}}}+\frac{\alpha}{r}\ln\frac{r}{\left|\alpha\right|}, \label{eq:metric}
\end{eqnarray}
where $g$, $M$, and $\alpha$ are the magnetic charge, mass, and dark matter parameter of the black hole, respectively. In addition, to ensure the validity of the weak energy condition, the dark matter parameter $\alpha$ must satisfy $\alpha< 0$ \cite{Sun:2023woa}. When the dark matter parameter $\alpha=0$, the black hole reduces to the regular Bardeen black hole. For magnetic charge $g=0$, the black hole reduces to the Schwarzschild black hole surrounded by PFDM \cite{Li:2012zx,Ayon-Beato:2000mjt}, and when $\alpha=0$ and $g=0$, the black hole reduces to the Schwarzschild black hole. In the following study, we will rescale all the physical quantities by the mass $M$ and set $M=1$ for simplicity. By solving equation $f(r)=0$, we can determine the event horizon $r_h$ of the black hole. However, since the explicit expressions for the solutions of the event horizon $r_h$ is too complicated, we show the relationship between the event horizon $r_h$ and the magnetic charge parameter $g$ and the dark matter parameter $\alpha$ in Fig.\ref{fig:rh-rph}. Obviously, the dark matter parameter $\alpha$ and the magnetic charge parameter $g$ exhibit distinctly different influences on the event horizon. As the coupling parameter $\left|\alpha\right|$ decreases or $g$ increases, the event horizon $r_h$ shows a monotonically decreasing trend. 

\begin{figure}[htbp]
    \centering
    \begin{minipage}{\textwidth}

     \begin{subfigure}[t]{0.4\textwidth}
        \centering
        \includegraphics[width=\textwidth]{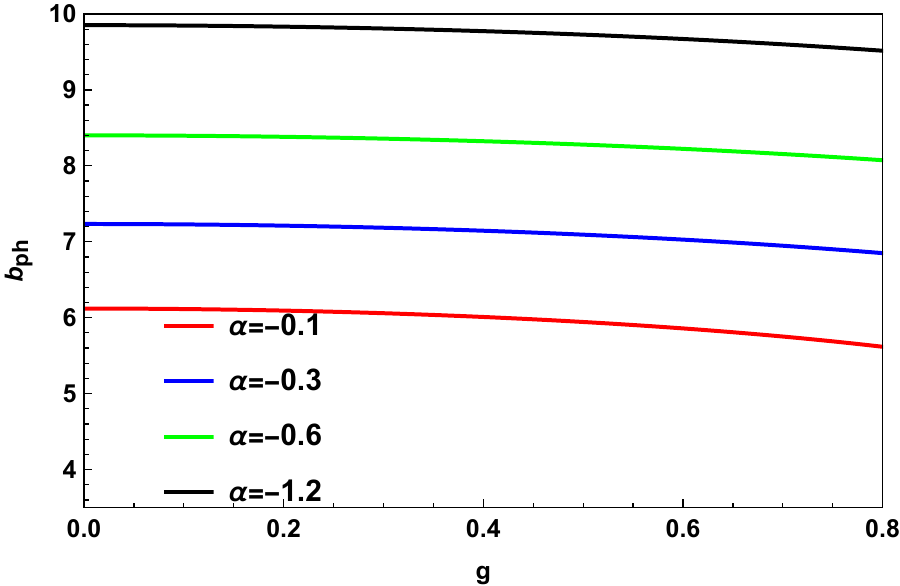}
        \label{subfig:bph-a}
    \end{subfigure}
    \hspace{1mm}
\begin{subfigure}[t]{0.4\textwidth}
        \centering
        \includegraphics[width=\textwidth]{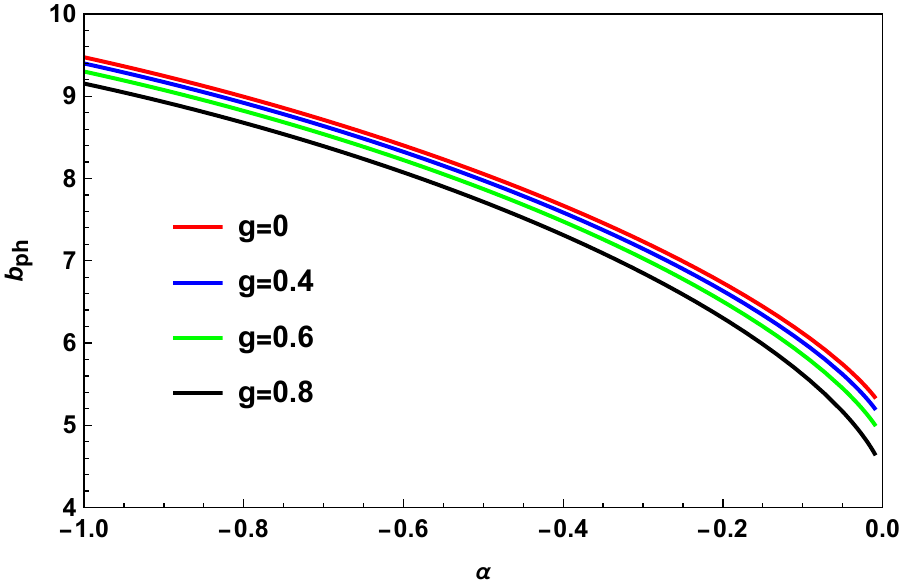}
        \label{subfig:bph-g}
    \end{subfigure}
    \begin{subfigure}[t]{0.4\textwidth}
        \centering
        \includegraphics[width=\textwidth]{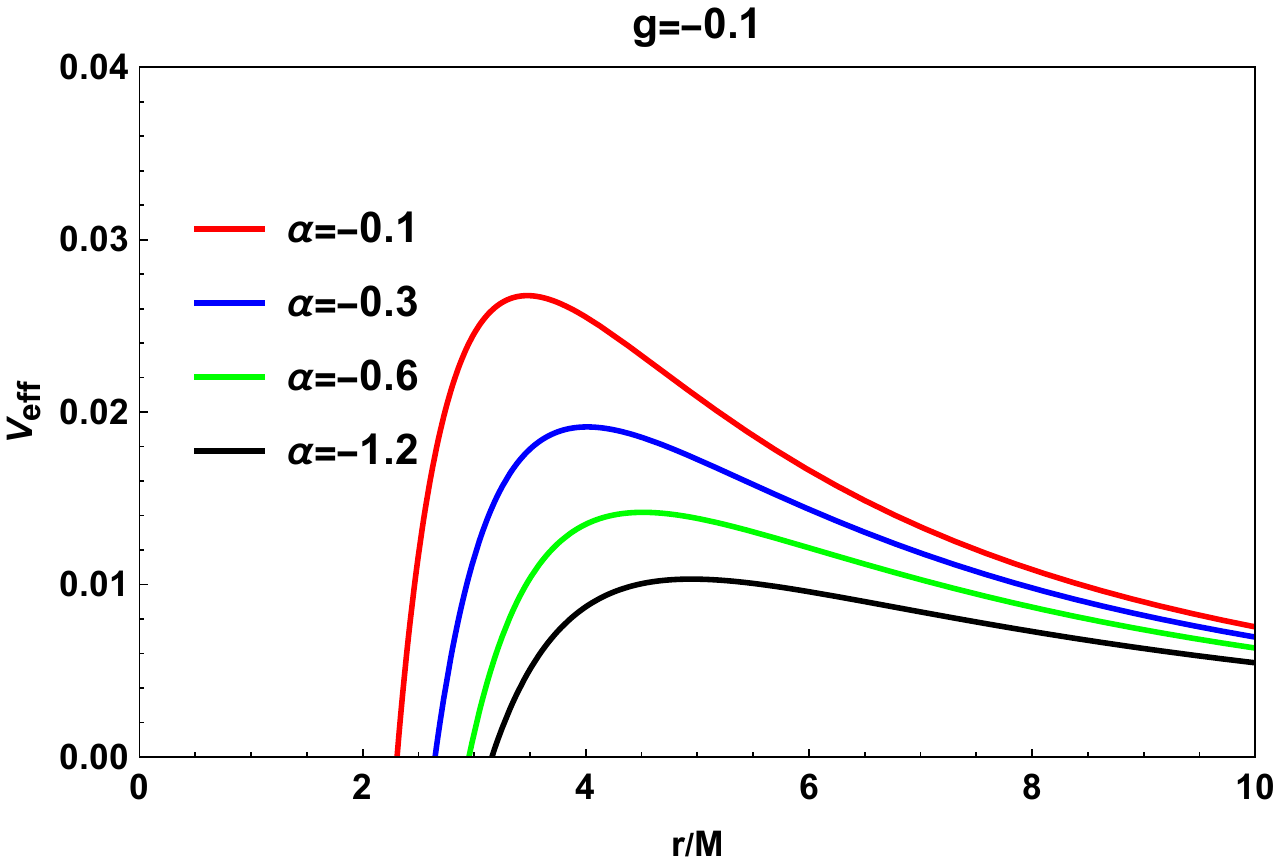}
        \label{subfig:Veff-a}
    \end{subfigure}
    \hspace{1mm}
\begin{subfigure}[t]{0.4\textwidth}
        \centering
        \includegraphics[width=\textwidth]{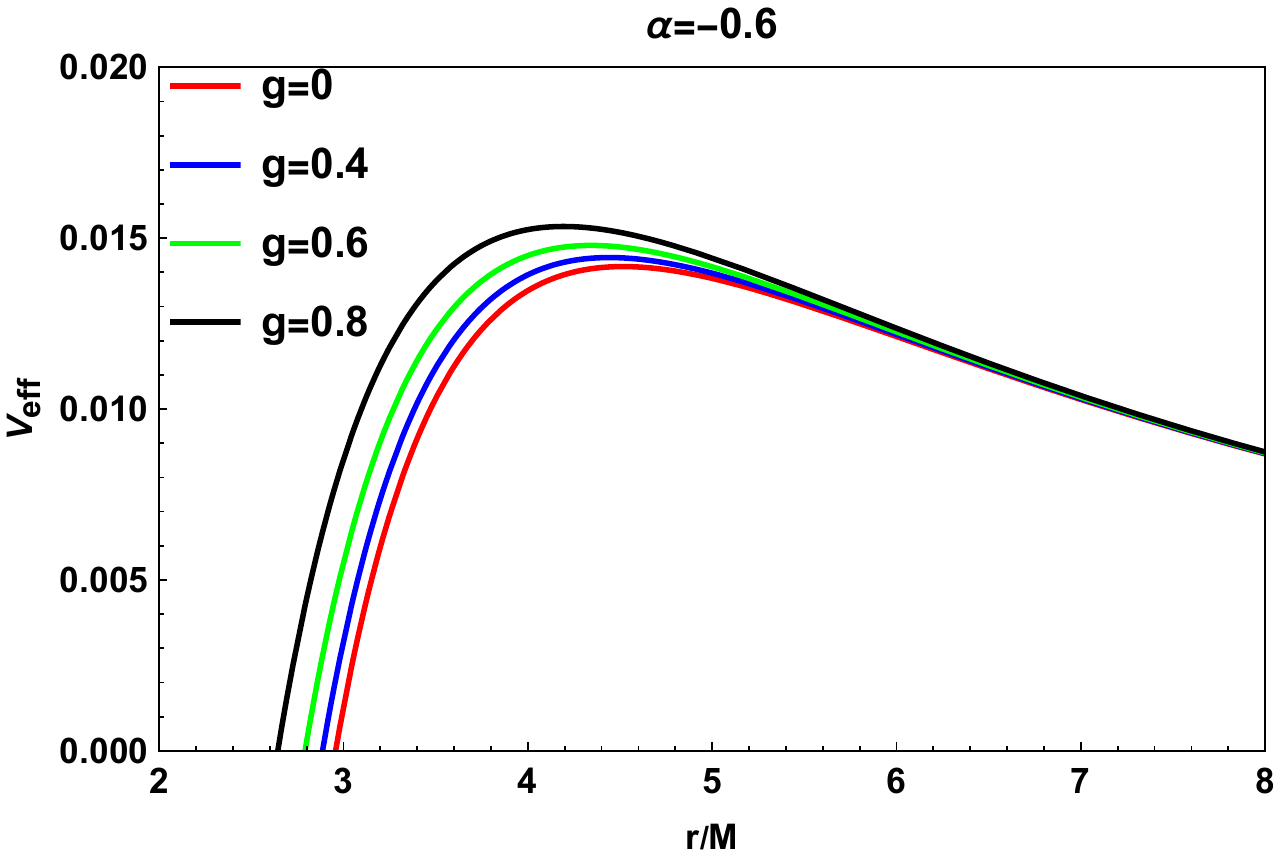}
        \label{subfig:Veff-g}
    \end{subfigure}
    \end{minipage}
    \caption{
        The critical impact parameter $b_{ph}$ and the effective potential $V_{\text{eff}}$ as a function of the dark matter parameter $\alpha$ and the magnetic charge parameter $g$ of the Bardeen black hole surrounded by PFDM. Here we set $M=1$.
        \hspace*{\fill}  
    }
    \label{fig:veff-bph}
\end{figure}

Next, we investigate the null geodesic motion and shadow of the Bardeen black hole surrounded by PFDM. The trajectory of photon in curved spacetime can be described by the Euler-Lagrange equation as
\begin{eqnarray}
\frac{d}{d\lambda}\Big(\frac{\partial \mathcal{L}}{\partial \dot{x}^\mu}\Big)-\frac{\partial \mathcal{L}}{\partial x^\mu}=0,
\end{eqnarray}
where $\dot{x}^\mu={dx^\mu/d\lambda}$ represents the four-velocity of the photon and $\lambda$ is the affine parameter. Furthermore, in the static spherically symmetric Bardeen black hole surrounded by PFDM, the Lagrangian of the photon takes the form
\begin{eqnarray}
\mathcal{L}=\frac{1}{2}g_{\mu\nu}\dot{x}^\mu\dot{x}^\nu=\frac{1}{2}\big[-f(r)\dot{t}^2+\frac{1}{f(r)}\dot{r}^2+r^2(\dot{\theta}^2+\sin^2\theta\dot{\phi}^2)\big].
\end{eqnarray}
Due to the spherical symmetry of spacetime, there are two conserved quantities: the energy $E$ and the $z$ component of the angular momentum $L_z$ 
\begin{eqnarray}
E\equiv-\frac{\partial \mathcal{L}}{\partial \dot{t}}=f(r)\dot{t},~~~~~~~~~~L_z=\frac{\partial \mathcal{L}}{\partial \mathcal{\dot{\phi}}}=r^2 \sin^2 \theta\dot{\phi}.
\end{eqnarray}
In addition, we consider the motions of photons moving on the equatorial plane $\theta=\pi/2$ without losing generality. Furthermore, considering $\mathcal{L}=0$ for the photon and defining the impact parameter $b=L_z/E$, we can obtain three first-order differential equations of motion for the photons
\begin{eqnarray}
\dot{t}=\frac{1}{bf(r)}, ~~~\dot{\phi}=\pm\frac{1}{r^2},~~~\dot{r}^2=\frac{1}{b^2}-V_{\mathrm{eff}}(r),
\label{eq:radial equation}
\end{eqnarray}
where the $\pm$ sign represents the clockwise and counterclockwise motion of photons, and the effective potential is given as
\begin{eqnarray}
V_{\mathrm{eff}}(r)=\frac{f(r)}{r^2}.
\label{eq-veff}
\end{eqnarray}
Then, we will discuss the circular orbit and photon sphere, which requires $\dot{r}=0$ and $\ddot{r}=0$. Therefore, according to Eq. (\ref{eq:radial equation}), the radius $r_{ph}$ of the photon sphere can be obtained by 
\begin{eqnarray}
b_{ph}=\frac{1}{\sqrt{V_{\mathrm{eff}}(r_{ph})}},~~~~~~~~~V'_{\mathrm{eff}}(r_{ph})=0,
\label{eq:rph}
\end{eqnarray}
where $b_{ph}$ is the critical impact parameter corresponding to the radius $r_{ph}$ of photon sphere. Subsequently, the relationship between the photon sphere radius $r_{ph}$ and the parameters (the magnetic charge $g$ and the dark matter parameter $\alpha$) is shown below in Fig.\ref{fig:rh-rph}. Similar to the event horizon, the smaller the dark matter parameter $\left|\alpha\right|$ or the larger magnetic charge parameter $g$, the smaller the corresponding photon sphere radius $r_{ph}$. In addition, we give the curves of critical impact parameters and effective potential changing with parameters $\alpha$ and $g$ in Fig.\ref{fig:veff-bph}. Obviously, a smaller $\left|\alpha\right|$ and a larger $g$ correspond to a smaller critical impact parameter $b_{ph}$, a larger peak in the effective potential, and a leftward shift of that peak. Overall, for the event horizon $r_h$, the radius of the photon sphere $r_{ph}$, and the effective potential $V_{\text{eff}}$, there is a competitive effect between the dark matter parameter $\left|\alpha\right|$ and the magnetic charge parameter $g$.

\begin{figure}[htbp]
    \centering
    \begin{minipage}{\textwidth}

     \begin{subfigure}[t]{0.4\textwidth}
        \centering
        \includegraphics[width=\textwidth]{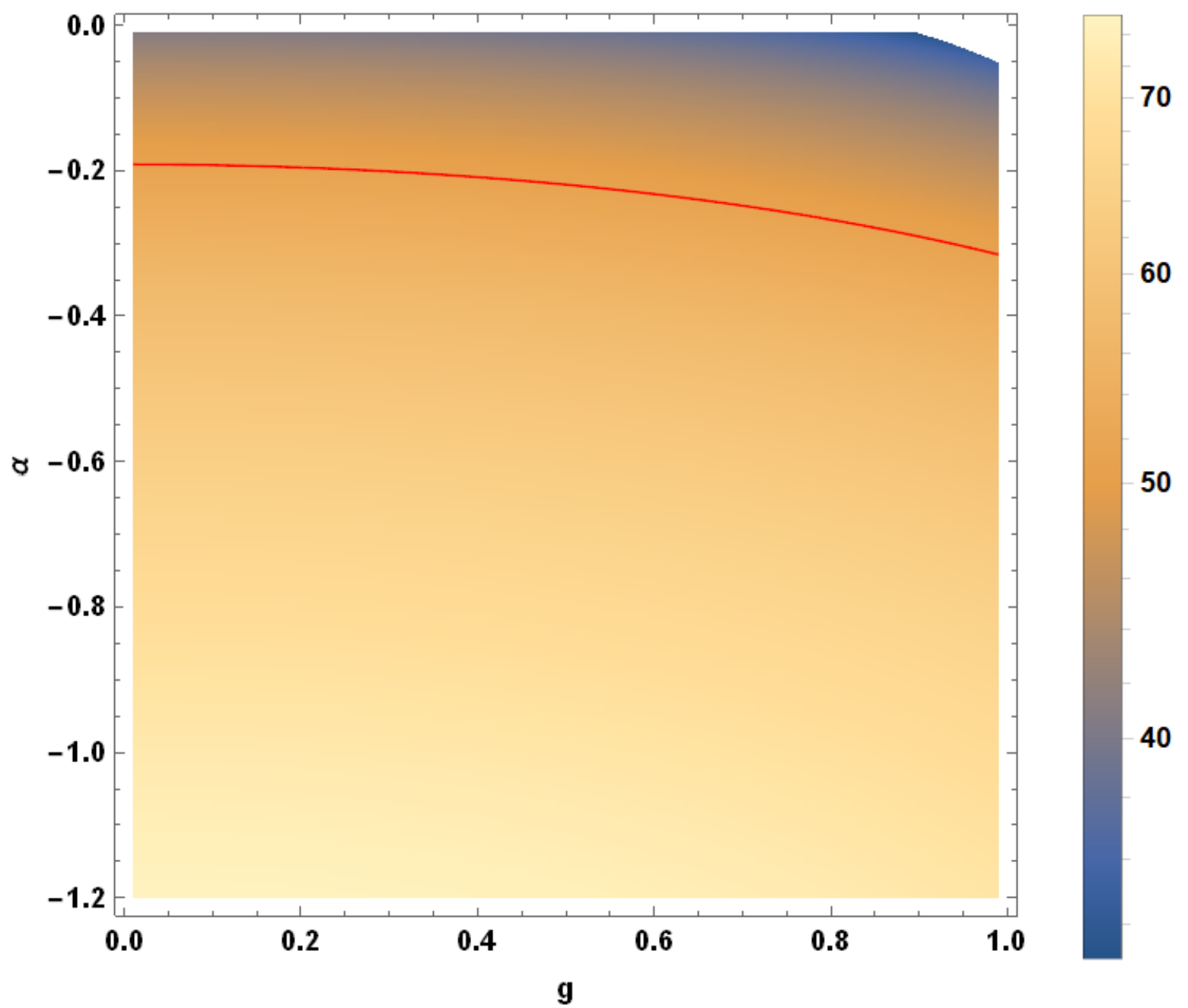}
        \label{subfig:M87}
    \end{subfigure}
    \hspace{1mm}
\begin{subfigure}[t]{0.4\textwidth}
        \centering
        \includegraphics[width=\textwidth]{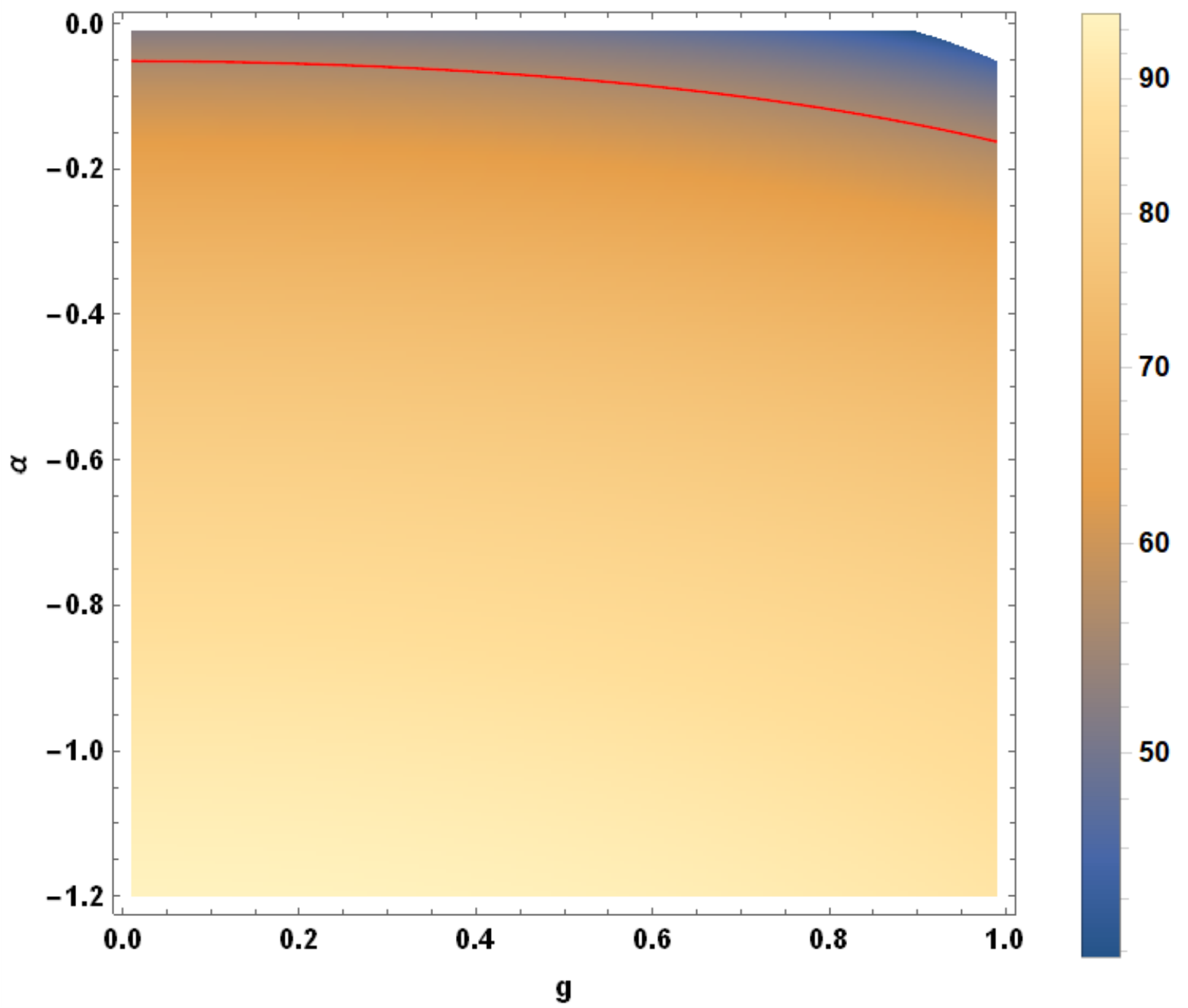}
        \label{subfig:SgrA}
    \end{subfigure}
    \end{minipage}
    \caption{Left: the constraints of the coupling parameters $\alpha$ and $g$ by the observational shadow of M87* black hole. The red curve corresponds to the angular diameter $\theta_d=51.06 ~\mu as$. Right: the constraints of the coupling parameters $\alpha$ and $g$ by the observational shadow of SgrA* black hole. The red curve corresponds to the angular diameter $\theta_d=55.7 ~\mu as$.}
    \label{fig:M87-SgrA}
\end{figure}

On the other hand, for the distant observer, the angular diameter $\theta_{d}$ of black hole shadow can be given by \cite{Kumar:2020owy}
\begin{equation}
\theta_d=2\frac{b_{ph}}{d_L},
\label{eq:thetad}
\end{equation}
here, $d_L$ is the distance between the black hole and the observer. According to EHT observation data, the mass of the supermassive black hole M87* is $M=6.5\times 10^{9}M_{\odot}$ and the distance is $d_L = 16.8\mathrm{Mpc}$; the mass of SgrA* is $M = 4.0 \times 10^{6}M_{\odot}$ and the distance is $d_L = 8.15\mathrm{kpc}$. Their angular diameters $\theta_d$ are 29.32-51.06 $\mu as$ and 48.7±7 $\mu as$ respectively. Based on these data, we can roughly constrain the range of the magnetic charge parameter $g$ and dark matter parameter $\alpha$ \cite{EventHorizonTelescope:2022wkp,Kuang:2022ojj,EventHorizonTelescope:2021dqv}. Considering that the Bardeen black holes surrounded by PFDM are supermassive black holes M87* and SgrA*, the density plots of the angular diameter of black holes are given in Fig. \ref{fig:M87-SgrA}. The red curves $\theta_d=51.06\mu as$ and $\theta_d=55.7\mu as$ are the upper limits of the observational data of M87* and SgrA*, respectively. Therefore, the parameter space ($g,\alpha$) above the curves satisfies the observational constraints of the EHT. It is worth noting that our results for the constraint of the dark matter parameter $\alpha$ are consistent with the horizon-scale image of SgrA* at 2$\delta$ constraints \cite{Vagnozzi:2022moj}. Clearly, the dark matter parameter $\alpha$ is constrained to a very narrow range. However, in order to better reflect the influence of dark matter parameter $\alpha$ on the black hole accretion image, we will not limit the dark matter parameter $\alpha$ to the above parameter space in the following discussion. 

\section{Images of Bardeen black hole surrounded by PFDM illuminated by thin accretions disk}\label{sec:photon deflection}

In this section, we use the ray-tracing method to first analyze the light ray propagation properties around the Bardeen black hole surrounded by PFDM, and then investigate the image of this black hole illuminated by a geometrically and optically thin accretion disk. We assume that the accretion disk is located in the equatorial plane of the black hole, and the observer is located at the North pole. Due to the strong gravitational lensing effect of the black hole, light rays emitted from the accretion disk may intersect the accretion disk multiple times before reaching the observer, significantly altering the optical appearance of the black hole. Therefore, we adopt the method proposed in \cite{Gralla:2019xty} to first classify the light trajectories around the Bardeen black hole surrounded by PFDM, and then investigate the influence of dark matter parameter $\alpha$ and magnetic charge parameter $g$ on the light ray propagation around the black hole.

\subsection{Classification of the light rays}\label{light rays}

According to Eq.(\ref{eq:radial equation}), the equation of motion for the photon can be rewritten as
\begin{eqnarray}
	\frac{dr}{d\phi}=\pm \frac{1}{r^2}\sqrt{\frac{1}{b^2}-V_{\mathrm{eff}}(r)}.
\label{de}
\end{eqnarray}
Obviously, the trajectory of motion for the light ray is determined by the impact parameter $b$. One can define the number of light ray orbits $n=\phi/ (2\pi)$, where $\phi$ is the total change in azimuthal angle. For $n=3/4$, the corresponding impact parameters are denoted as $b_{1}$ and $b_{4}$ (where $b_{1} < b_{4}$); for $n=5/4$, they are denoted as $b_{2}$ and $b_{3}$ (where $b_{2} < b_{3}$). Furthermore, based on the number of times the light ray intersects the accretion disk, light rays can be classified into three types \cite{Gralla:2019xty}.
(1)Direct emission: the light rays intersect the accretion disk at most once, $1/4<n<3/4\Rightarrow b\in(0,b_{1})\cup(b_{4},+\infty)$. (2)Lensed ring emission: the light rays intersect the accretion disk only twice, $3/4<n<5/4\Rightarrow b\in(b_{1},b_{2})\cup(b_{3},b_{4})$. (3) Photon ring emission: the light rays intersect the accretion disk at least three times, $n>5/4\Rightarrow b\in(b_{2},b_{3})$. 
 
\begin{figure}[h]
    \centering
    \begin{subfigure}[b]{0.23\textwidth}
        \centering
        \includegraphics[width=0.85\textwidth]{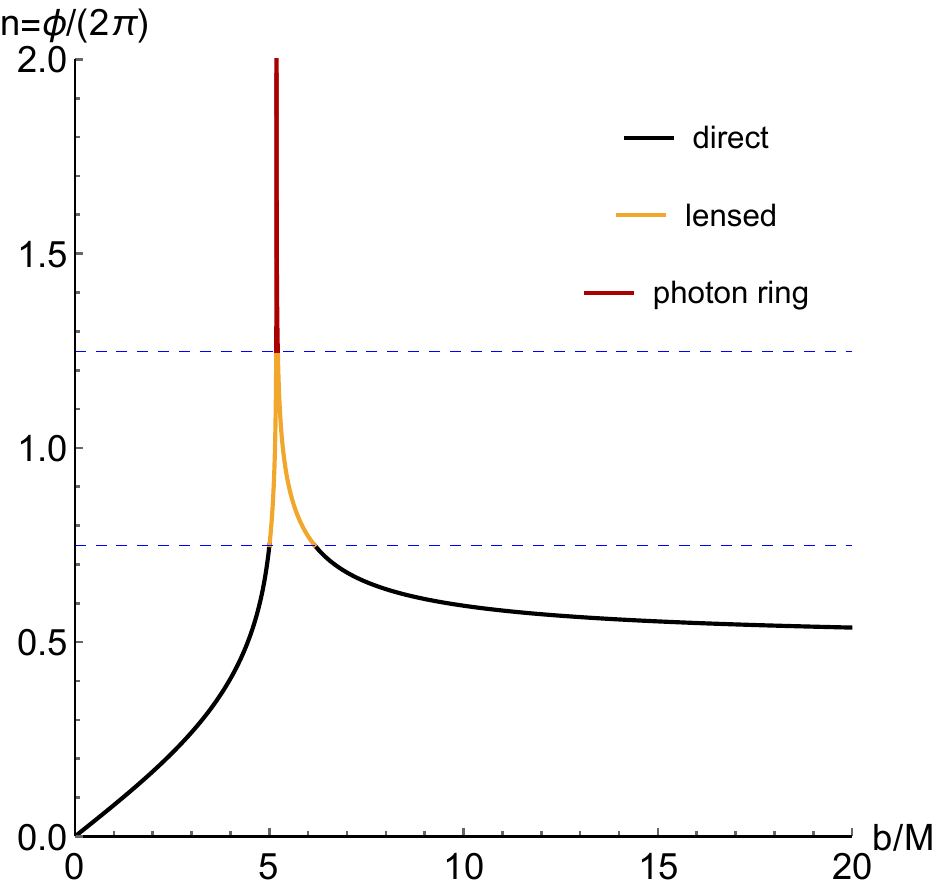}
        \includegraphics[width=0.85\textwidth]{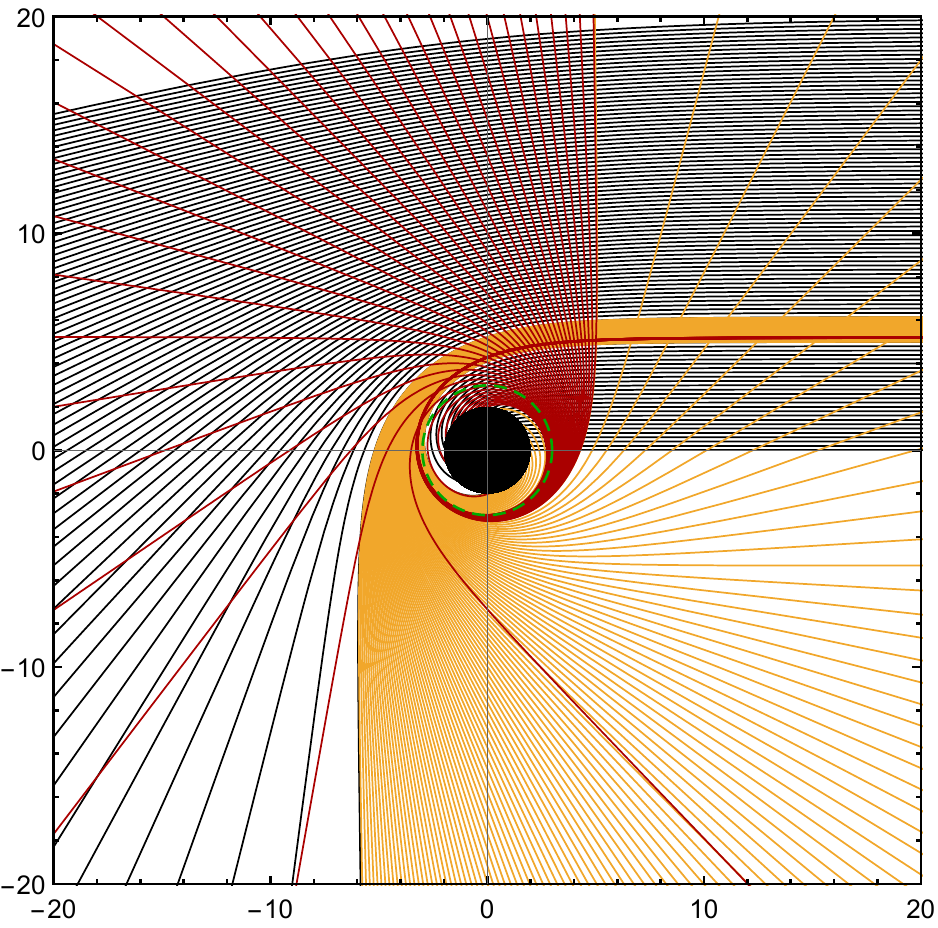}
        \caption{$g=0.1,\alpha=0$}
    \end{subfigure}
    \hfill
\begin{subfigure}[b]{0.23\textwidth}
        \centering
        \includegraphics[width=0.85\textwidth]{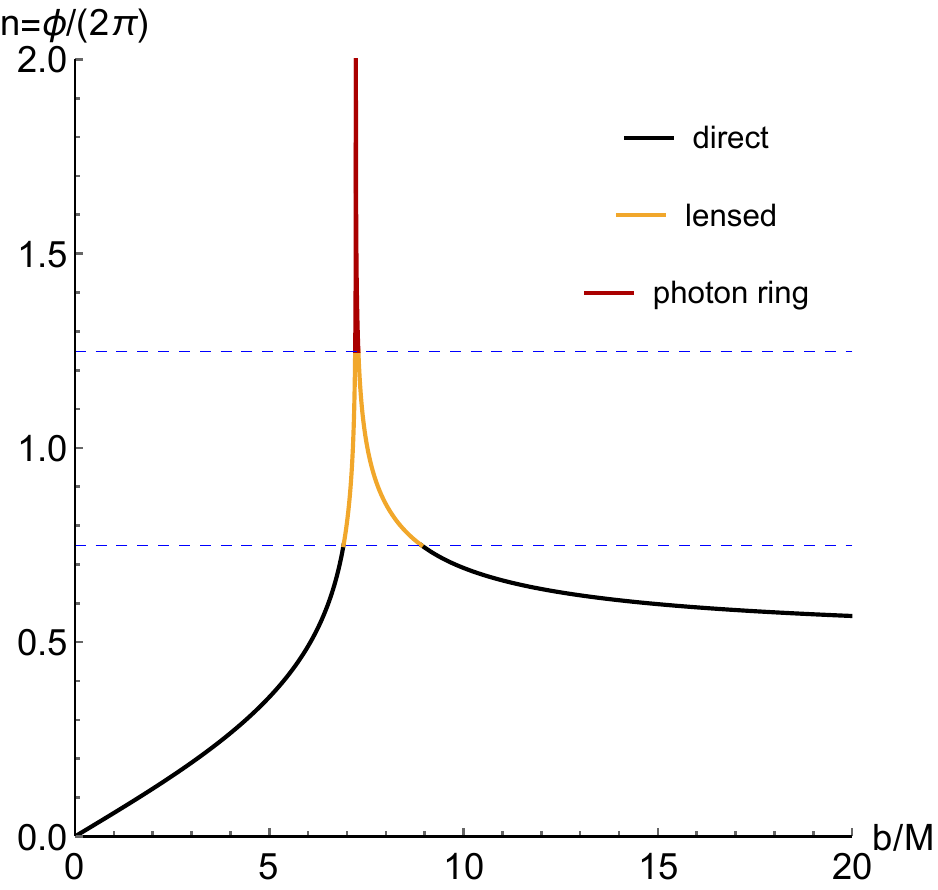}
        \includegraphics[width=0.85\textwidth]{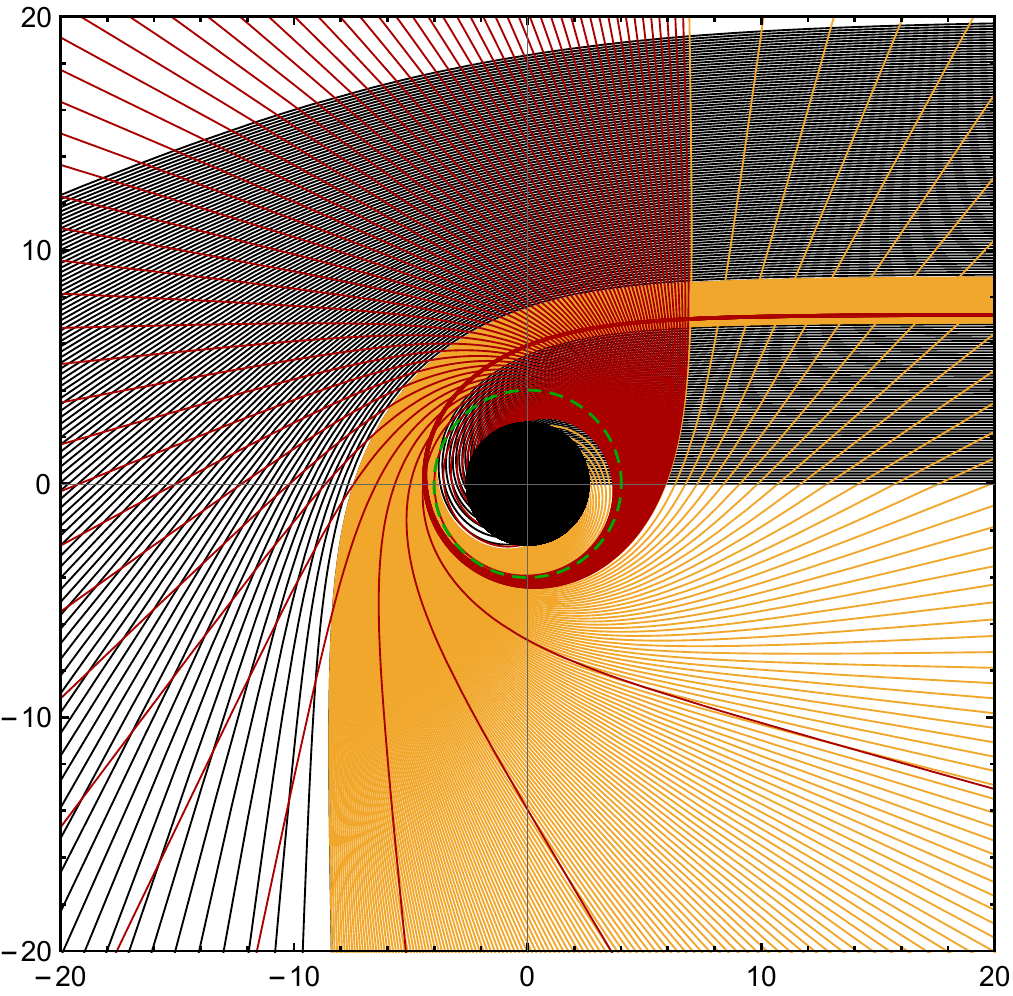}
        \caption{$g=0.1,\alpha=-0.3$}
    \end{subfigure}
    \hfill
\begin{subfigure}[b]{0.23\textwidth}
        \centering
        \includegraphics[width=0.85\textwidth]{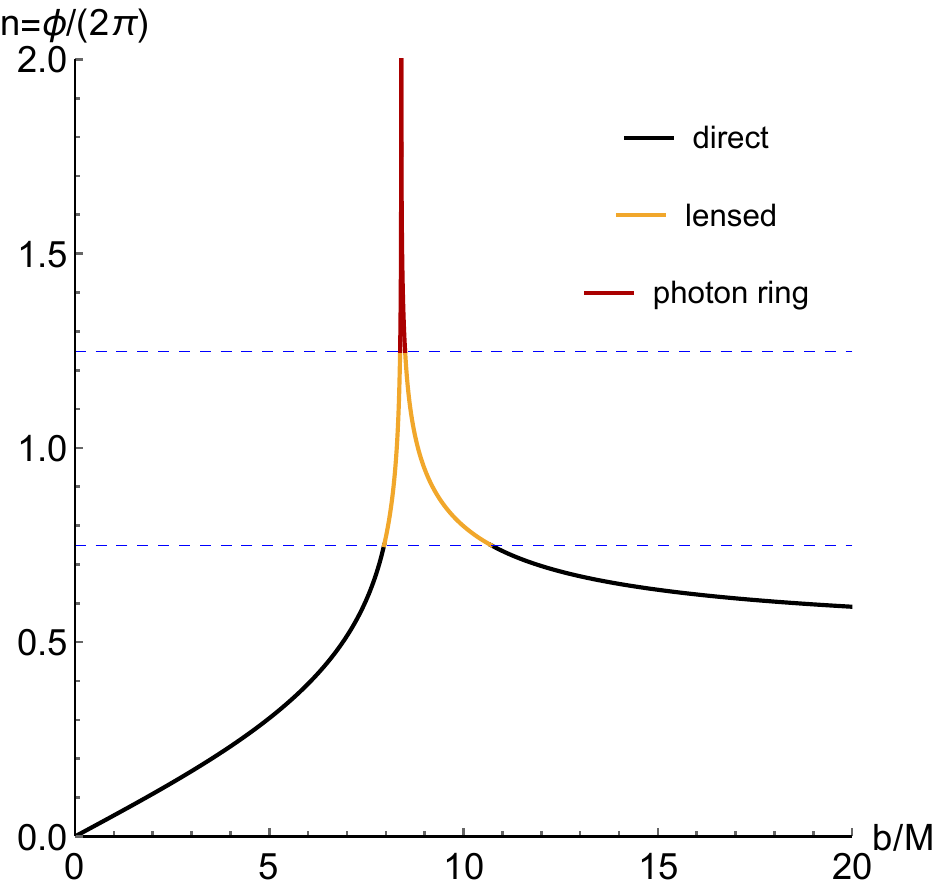}
        \includegraphics[width=0.85\textwidth]{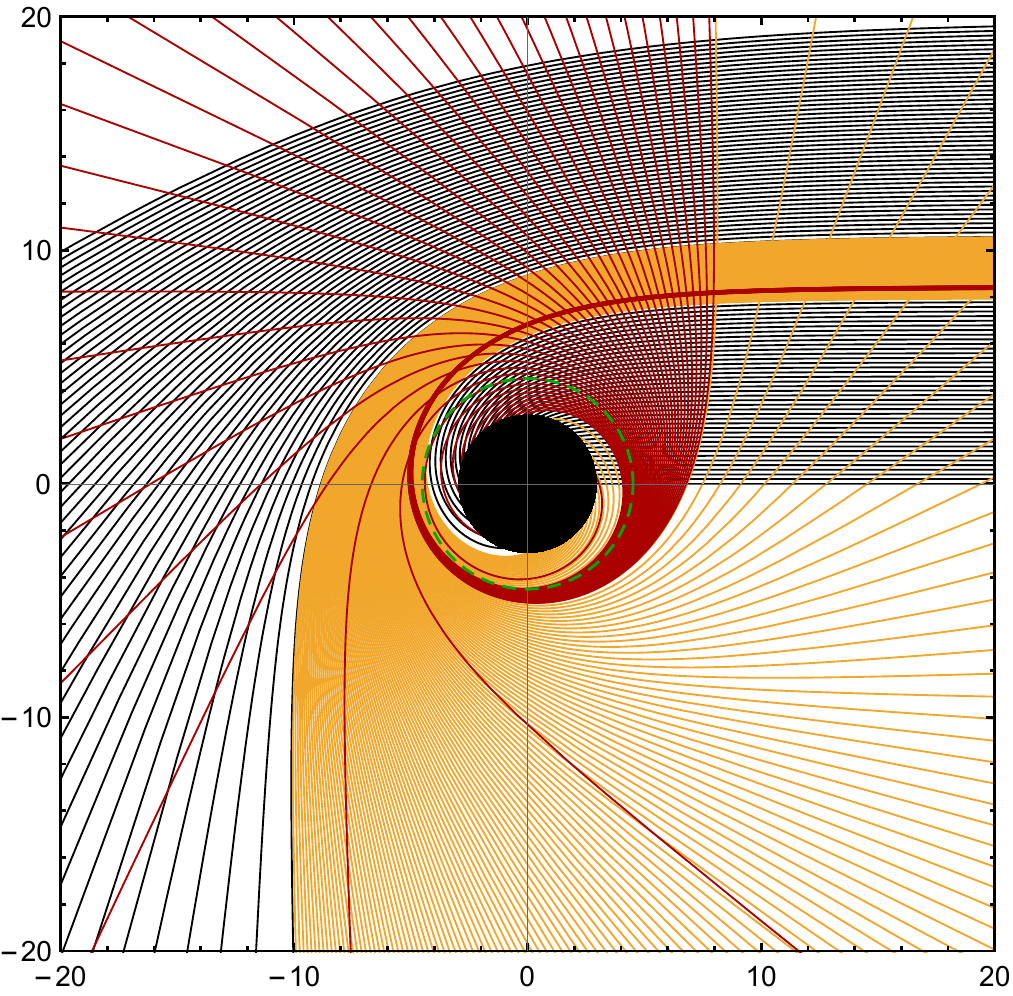}
        \caption{$g=0.1,\alpha=-0.6$}
    \end{subfigure}
    \hfill
\begin{subfigure}[b]{0.23\textwidth}
        \centering
        \includegraphics[width=0.85\textwidth]{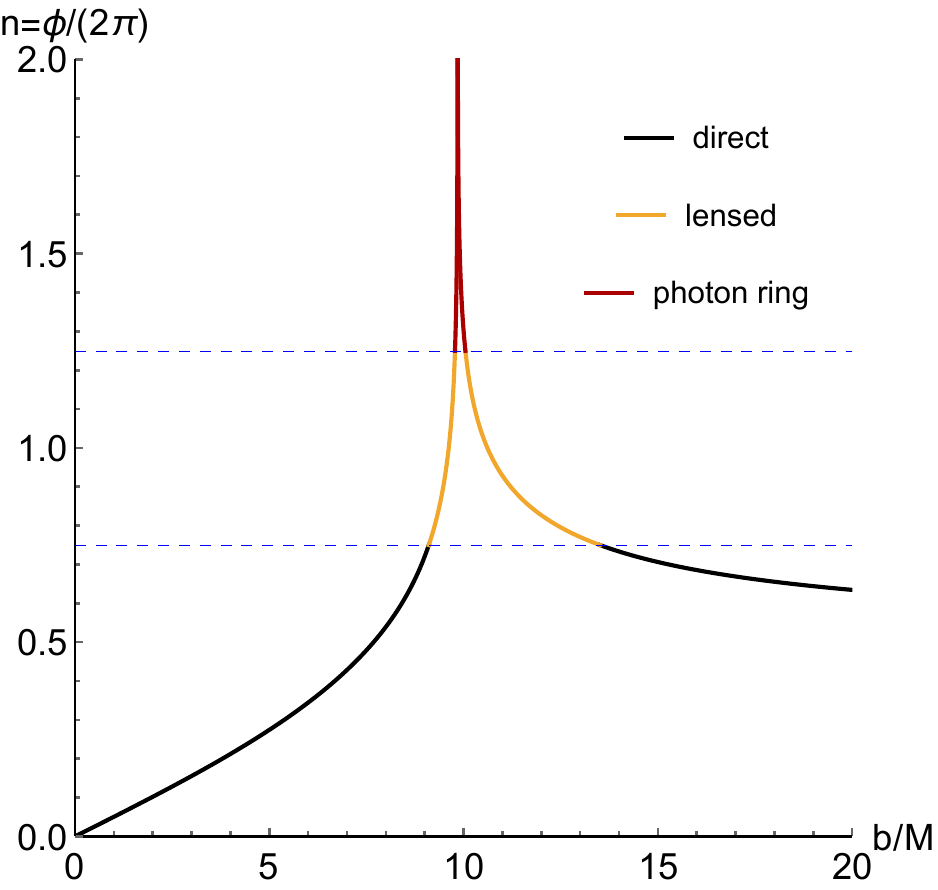}
        \includegraphics[width=0.85\textwidth]{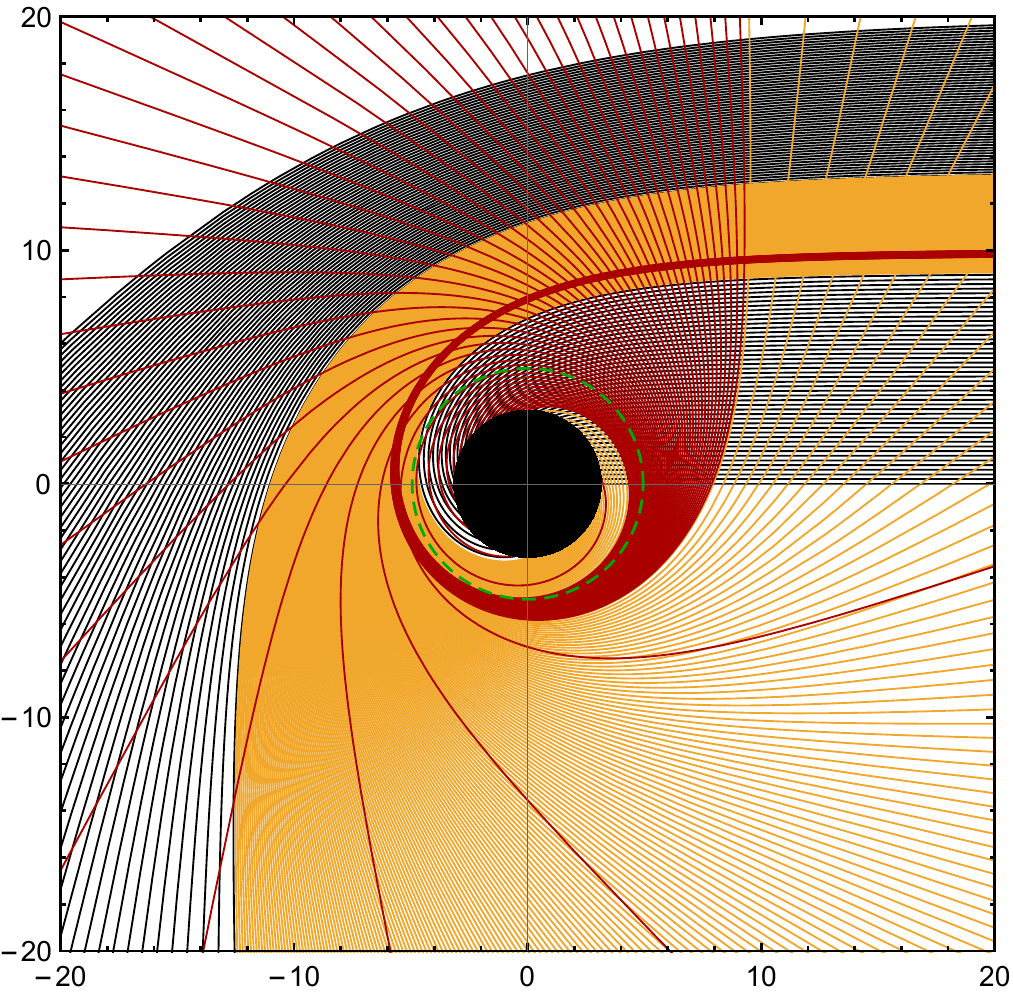}
        \caption{$g=0.1,\alpha=-1.2$}
    \end{subfigure}
    \vspace{0.5cm}  
\begin{subfigure}[b]{0.23\textwidth}
        \centering
        \includegraphics[width=0.85\textwidth]{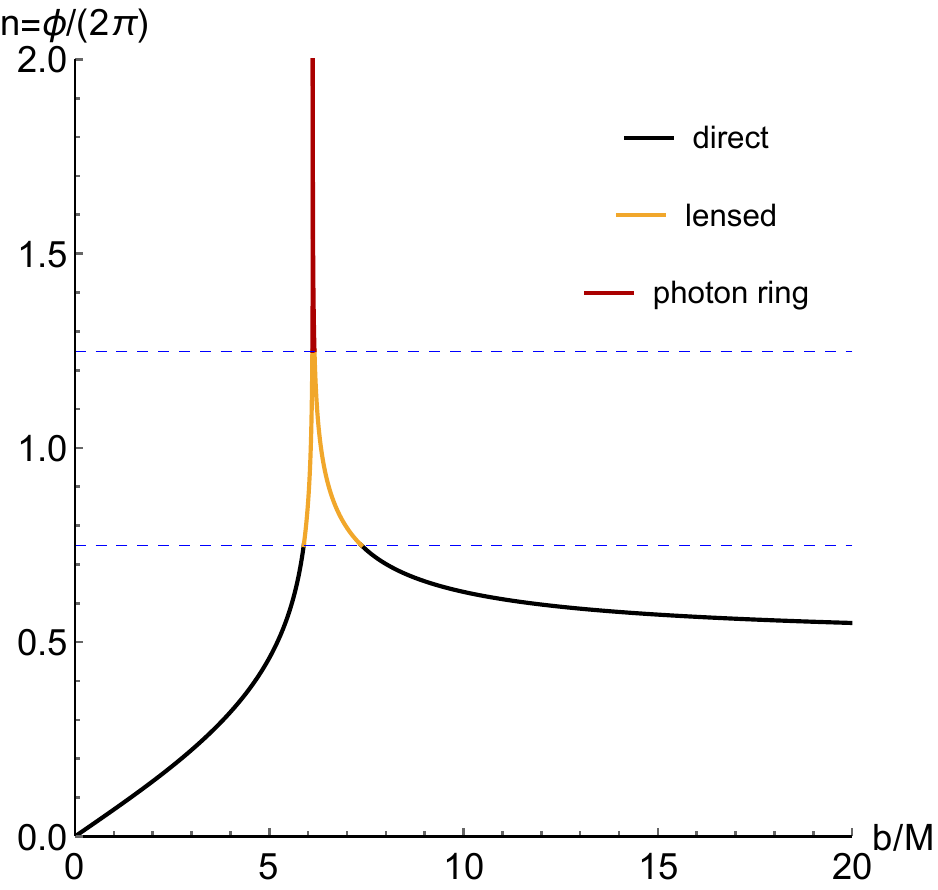}
        \includegraphics[width=0.85\textwidth]{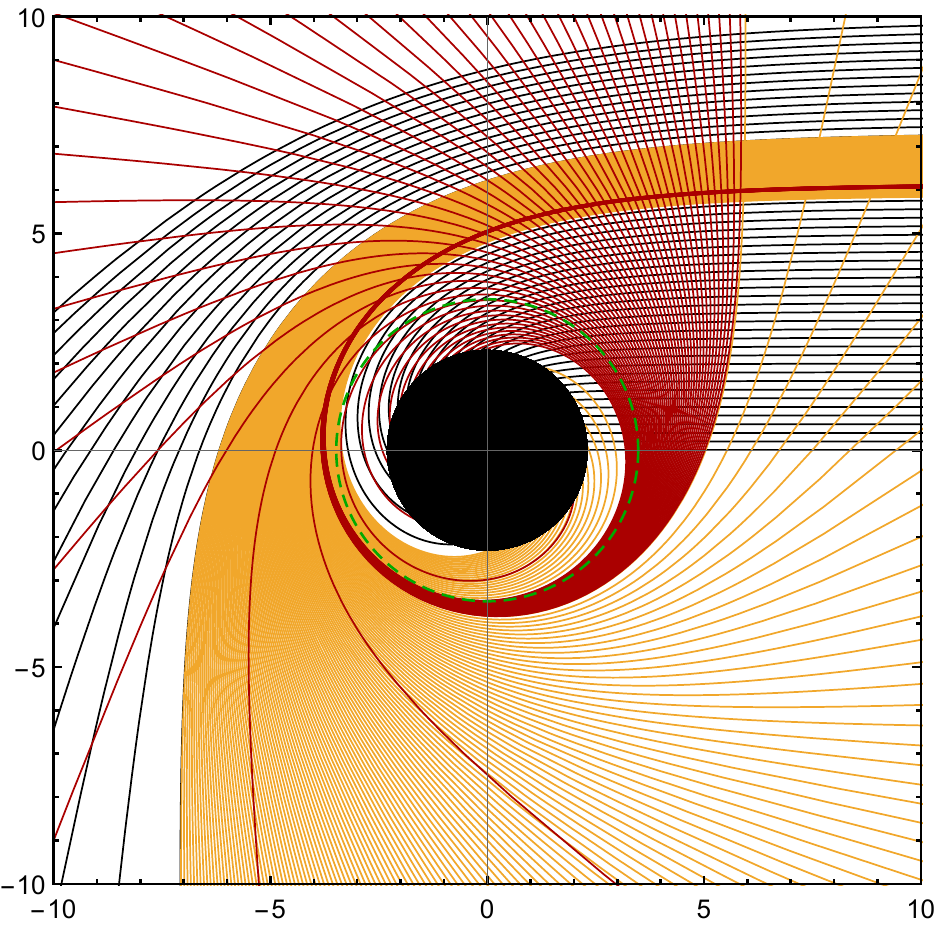}
        \caption{$g=0,\alpha=-0.1$}
    \end{subfigure}
    \hfill
\begin{subfigure}[b]{0.23\textwidth}
        \centering
        \includegraphics[width=0.85\textwidth]{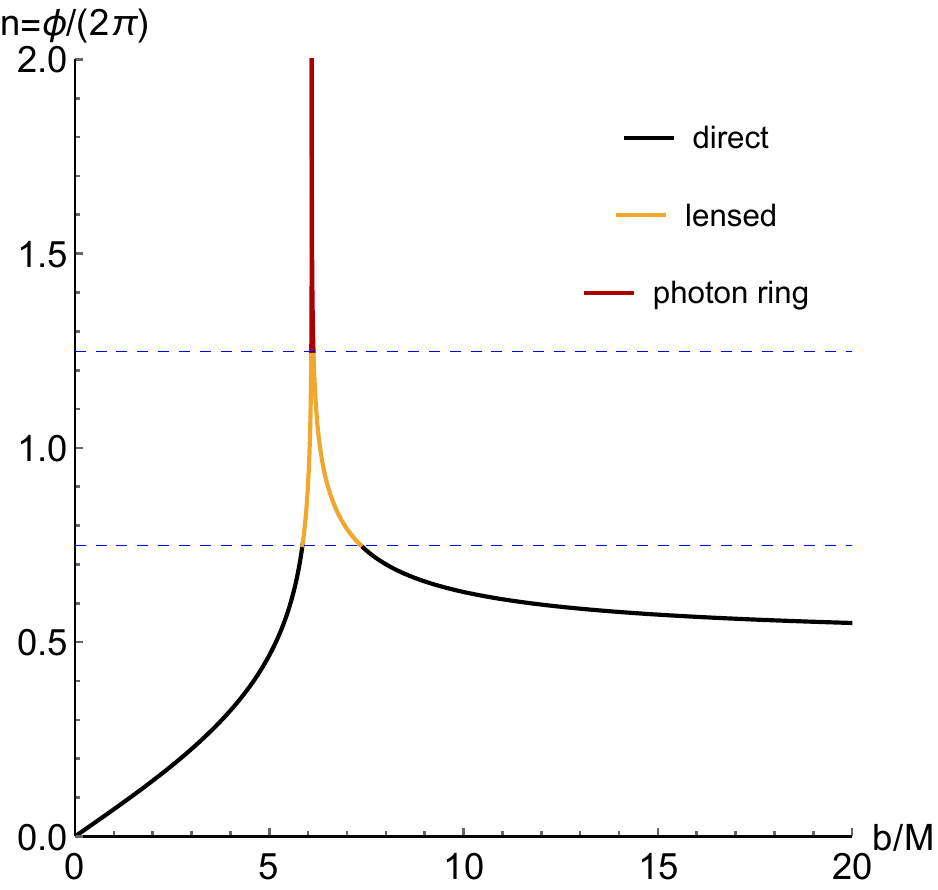}
        \includegraphics[width=0.85\textwidth]{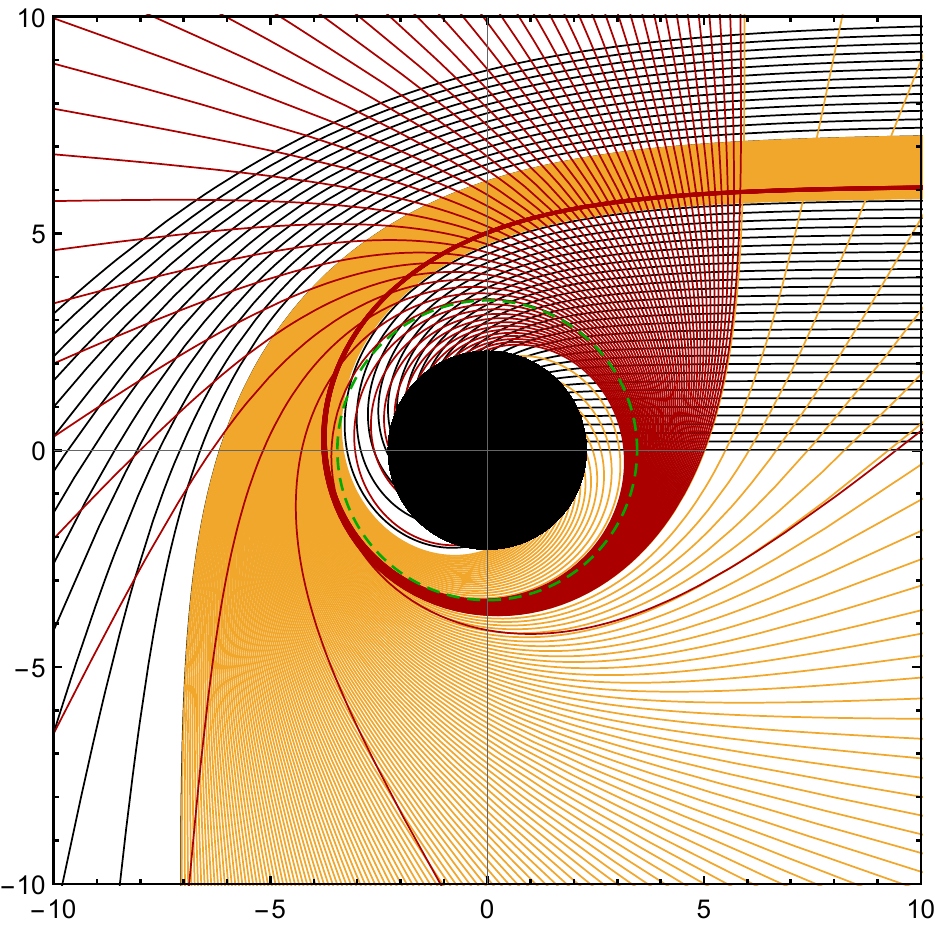}
        \caption{$g=0.2,\alpha=-0.1$}
    \end{subfigure}
    \hfill
\begin{subfigure}[b]{0.23\textwidth}
        \centering
        \includegraphics[width=0.85\textwidth]{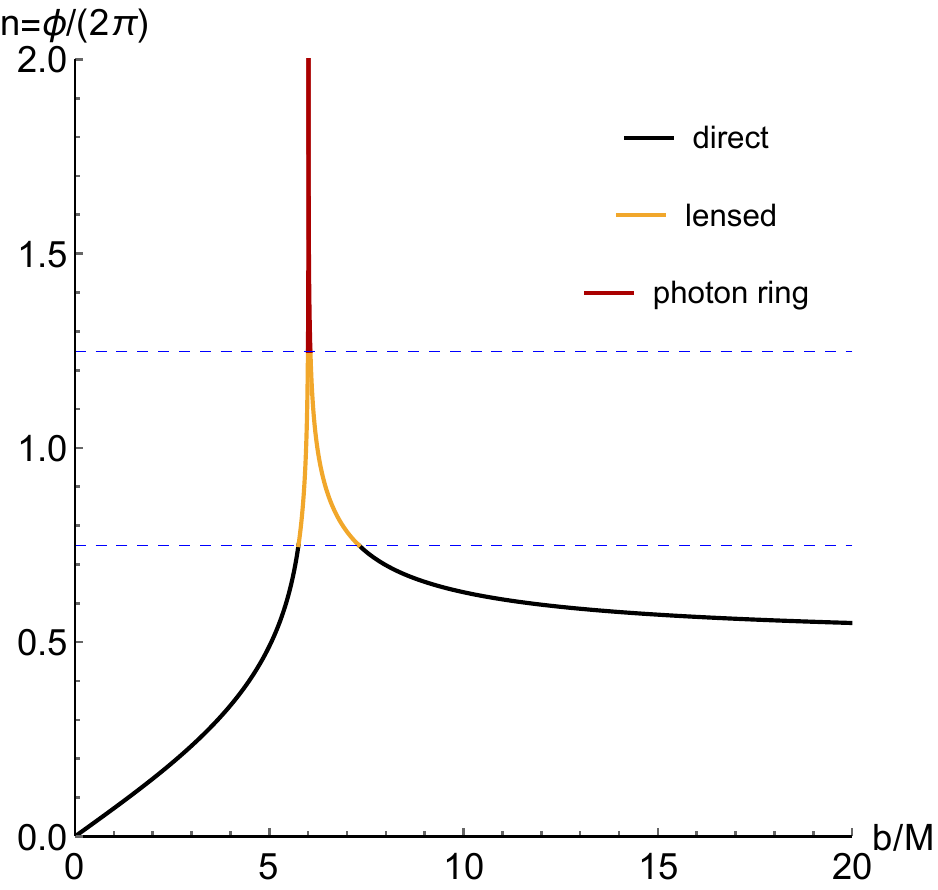}
        \includegraphics[width=0.85\textwidth]{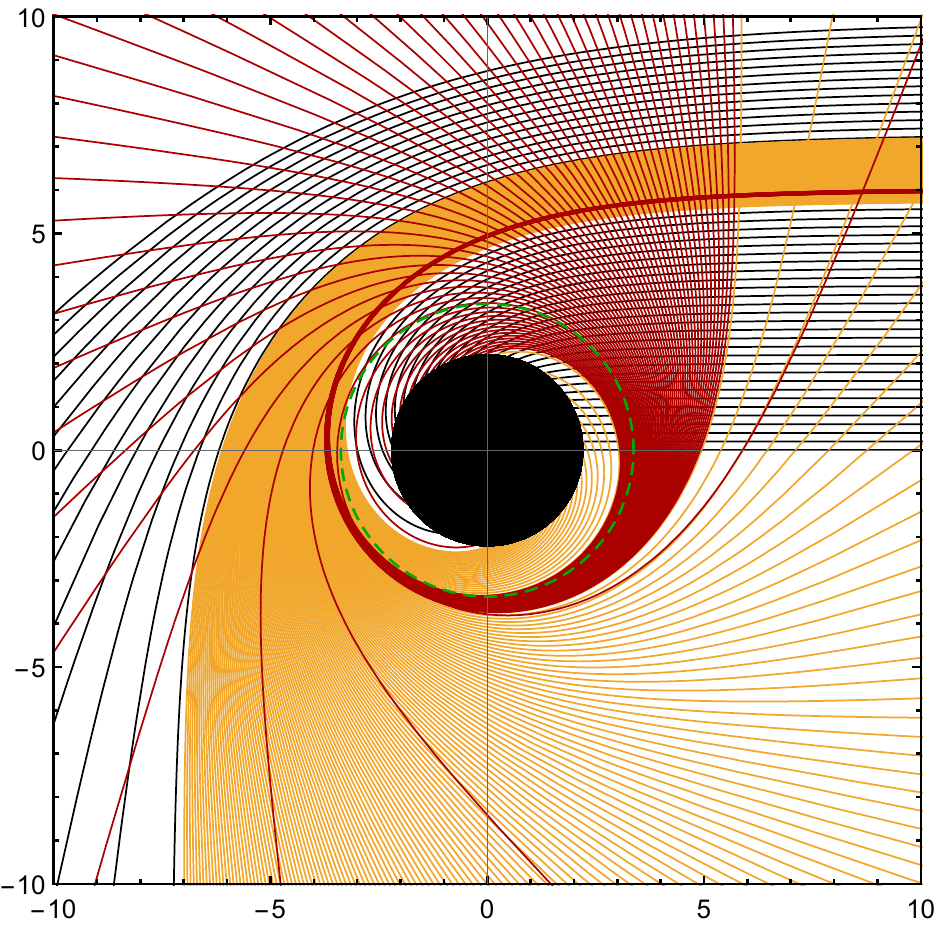}
        \caption{$g=0.4,\alpha=-0.1$}
    \end{subfigure}
    \hfill
\begin{subfigure}[b]{0.23\textwidth}
        \centering
        \includegraphics[width=0.85\textwidth]{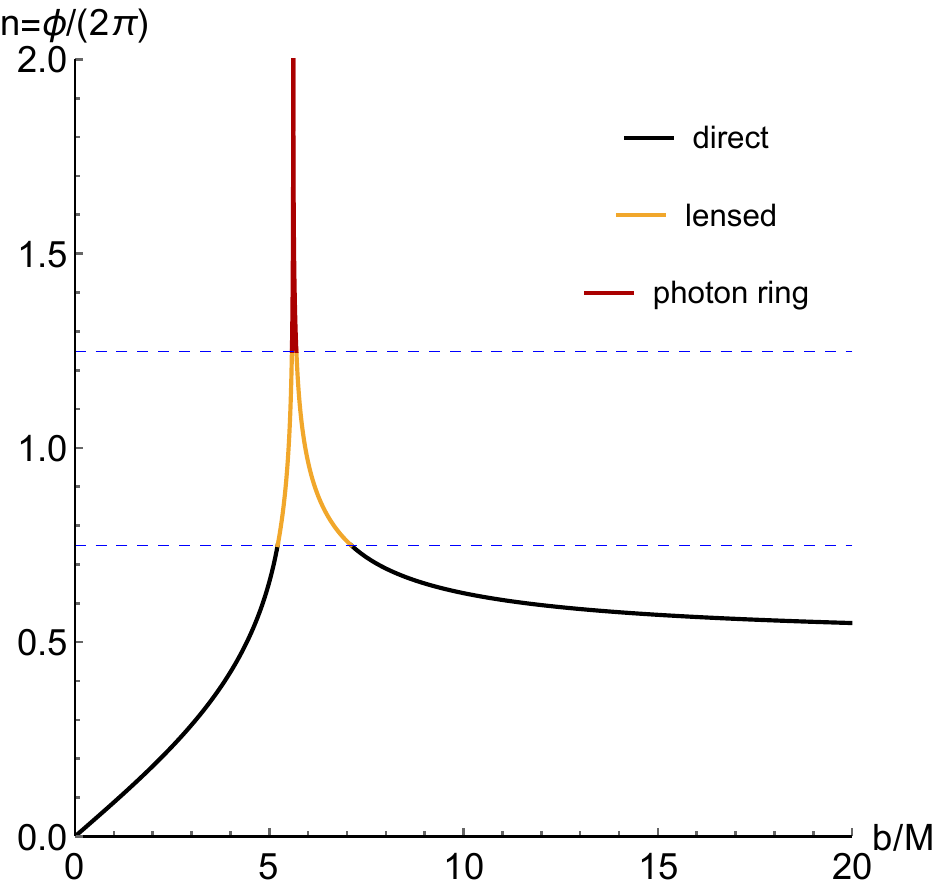}
        \includegraphics[width=0.85\textwidth]{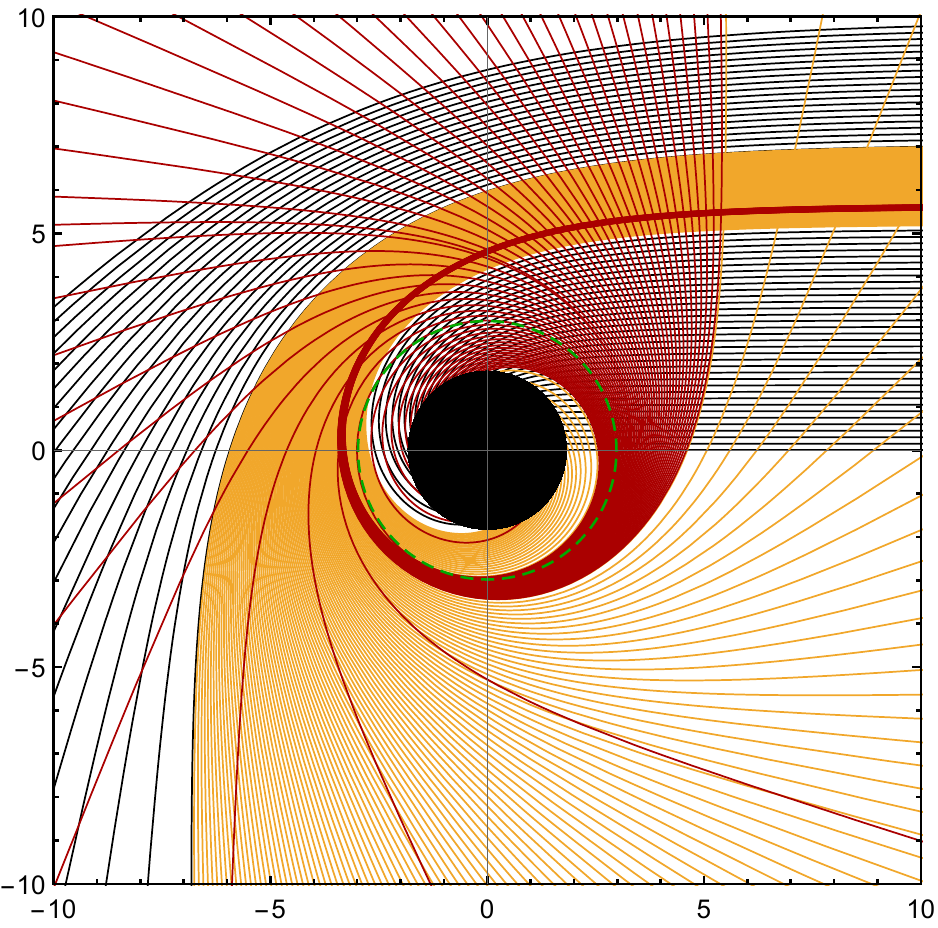}
        \caption{$g=0.8,\alpha=-0.1$}
    \end{subfigure}
    \caption{Top of (a-h): the number of photon orbits $n$ of the Bardeen black hole surrounded by PFDM as a function of the impact parameter $b$ for different $g$ and $\alpha$ with $M=1$.  Bottom of (a-h): a selection of associated photon trajectories in the Euclidean polar coordinates $(r,\phi)$. The black curves, gold curves, and red curves correspond to the direct emissions ($n<3/4$), lensed ring emissions ($3/4<n<5/4$), and photon ring emissions ($n>5/4$), respectively. The black disks and the green dashed curves denote the black holes and photon spheres, respectively. 
}
    \label{fig:orbits-traject}
\end{figure}

In order to more intuitively illustrate the relationship between the number of photon orbits $n$ and the impact parameters, we plot the number of photon orbits as a function of the impact parameter in Fig.\ref{fig:orbits-traject} and present the light ray trajectories around the black hole. As the impact parameter increases, the photon orbit number $n$ gradually increases and reaches a peak at the critical impact parameter $b_{ph}$, and then gradually decreases. In addition, different values of dark matter parameters and magnetic charge parameters significantly affect the distribution width of the three types of emission light rays around the black hole. In Table \ref{table01}, we present the key characteristic parameters of the Bardeen black hole surrounded by PFDM for various values of the dark matter and magnetic charge parameters. These include the event horizon, photon sphere radius, critical impact parameter, and the widths of the photon ring and lensed ring emissions. Obviously, as the dark matter parameter $\left|\alpha\right|$ and the magnetic charge parameter $g$ increase, the width of the lensed ring emission and the photon ring emission around the black hole will become wider. However, compared with the dark matter parameter $\alpha$, the effect of the magnetic charge parameter $g$ on the width of lensed ring emission and photon ring emission is relatively smaller. Moreover, the effect of magnetic charge on the optical appearance of black holes has also been extensively discussed \cite{He:2024amh,Guo:2023grt,He:2021htq,Akbarieh:2023kjv}. The previous studies show that for the accretion images of static Bardeen black holes, the magnetic charge parameter $g$ decreases the size of the inner shadow and increases the total observed intensity in \cite{He:2021htq}.
Therefore, in the following studies, we will focus on the influence of dark matter parameter $\alpha$ on the optical appearance of Bardeen black holes surrounded by PFDM.

\begin{table}[htbp]
	\centering
	\resizebox{\textwidth}{!}{%
		\begin{tblr}{
				cells = {c},
				hline{1-2,10} = {-}{},
			}
			g   & $\alpha$    & $r_h$       & $r_{ph}$     & $b_{ph}$     & $b_1$       & $b_2$       & $b_3$      & $b_4$      & photon ring width     & lensed ring width   \\
			0.1 & 0    & 1.99247  & 2.99164 & 5.18747 & 5.00485   & 5.17897 & 5.21962  & 6.16226  & 0.04065 & 1.11676   \\
			0.1 & -0.3 & 2.64920   & 4.01235 & 7.22857 & 6.91197   & 7.21083 & 7.29242   & 8.90938   & 0.08159 & 1.91582   \\
			0.1 & -0.6 & 2.95269  & 4.51216 & 8.39639 & 7.95604   & 8.36751 & 8.49510   & 10.70267 & 0.12759 & 2.61904   \\
			0.1 & -1.2 & 3.15822  & 4.94632 & 9.84744 & 9.10548   & 9.78017 & 10.04839 & 13.51178  & 0.26822 & 4.13808    \\ \hline\hline
			0   & -0.1 & 2.31416 & 3.48255 & 6.12047 & 5.88831 & 6.10897 & 6.16345  & 7.36368  & 0.05448 & 1.42089 \\
			0.2 & -0.1 & 2.29048  & 3.45646 & 6.09340  & 5.85566   & 6.08133 & 6.13770   & 7.34797    & 0.05637 & 1.43594   \\
			0.4 & -0.1 & 2.21591   & 3.37504 & 6.00945 & 5.75284   & 5.99534 & 6.05822  & 7.30029  & 0.06288 & 1.48457   \\
			0.8 & -0.1 & 1.82854  & 2.98019 & 5.61690  & 5.21760    & 5.58279 & 5.69680   & 7.10060   & 0.11401 & 1.76899   
		\end{tblr}
	}
\caption{The event horizon $r_h$, photon sphere radius $r_{ph}$, critical impact parameter $b_{ph}$, the position of three emission types $b_i$ ($i=1,2,3,4$), photon ring width, and lensed ring width with different magnetic charges $g$ and dark matter parameters $\alpha$. Here, we fix $M=1$.}
	\label{table01}
\end{table}

\subsection{Observed intensities and optical appearances}\label{sec:thin disk accretions}

Light rays emitted from the accretion disk may pass through the disk arbitrary times after being deflected by the black hole's strong gravitational field. Each time they pass through the accretion disk, the light rays extract energy from it, significantly contributing to the optical appearance of the black hole. Due to the varying number of intersections between the three types of light ray and the accretion disk, they exhibit distinctly different observational signatures. Therefore, this section aims to investigate the effects of the three types of light on the optical appearance of the black hole and the observed images of the total intensity.

\begin{figure}[h]
    \centering
    \begin{subfigure}[b]{0.23\textwidth}
        \centering
        \includegraphics[width=\linewidth]{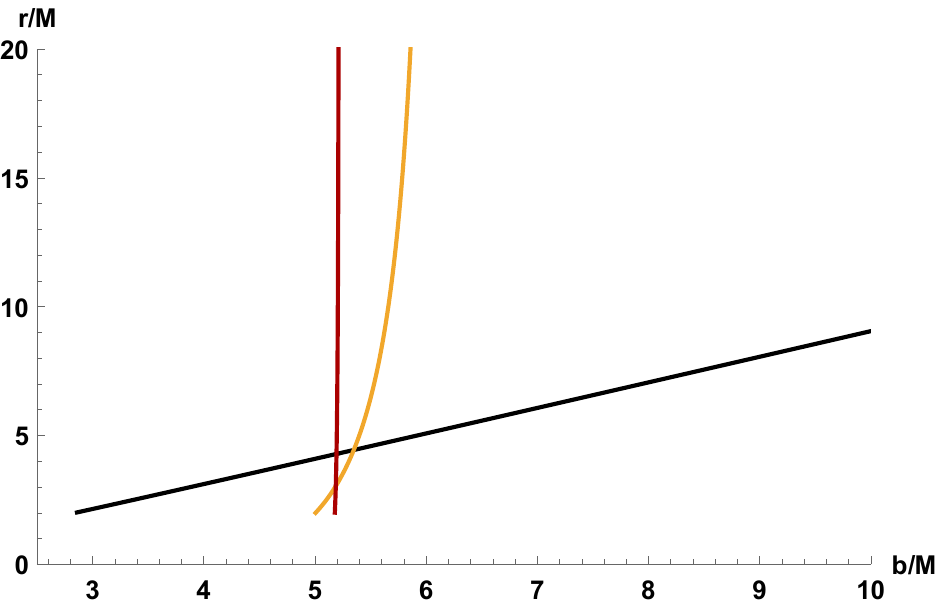}
        \caption{$g=0.1,\alpha=0$}
    \end{subfigure}
    \hfill   
\begin{subfigure}[b]{0.23\textwidth}
        \centering
        \includegraphics[width=\linewidth]{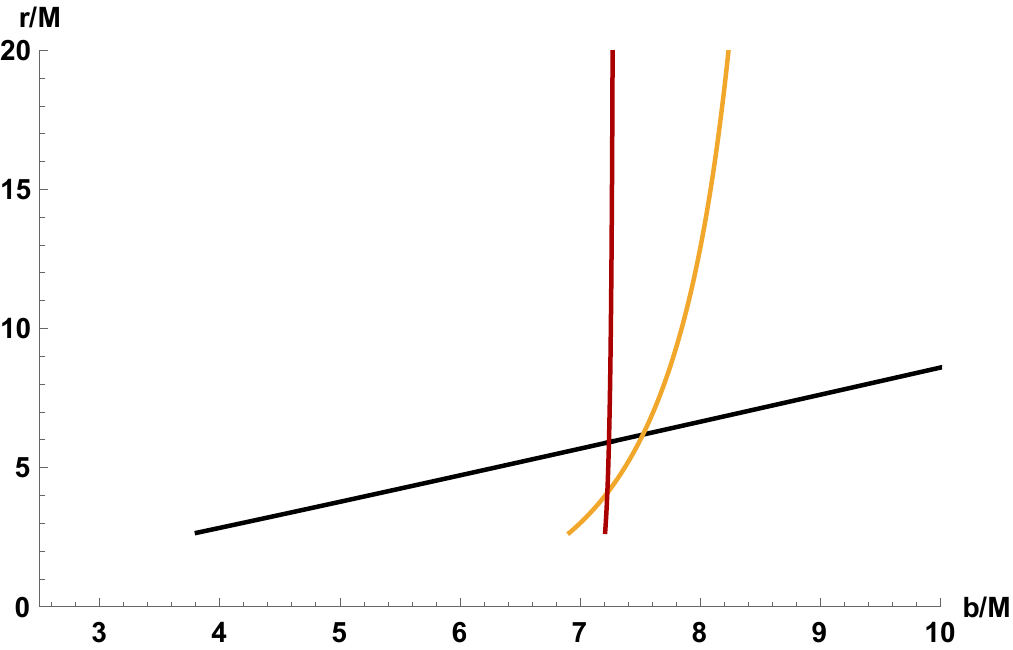}
        \caption{$g=0.1,\alpha=-0.3$}
    \end{subfigure}
    \hfill 
\begin{subfigure}[b]{0.23\textwidth}
        \centering
        \includegraphics[width=\linewidth]{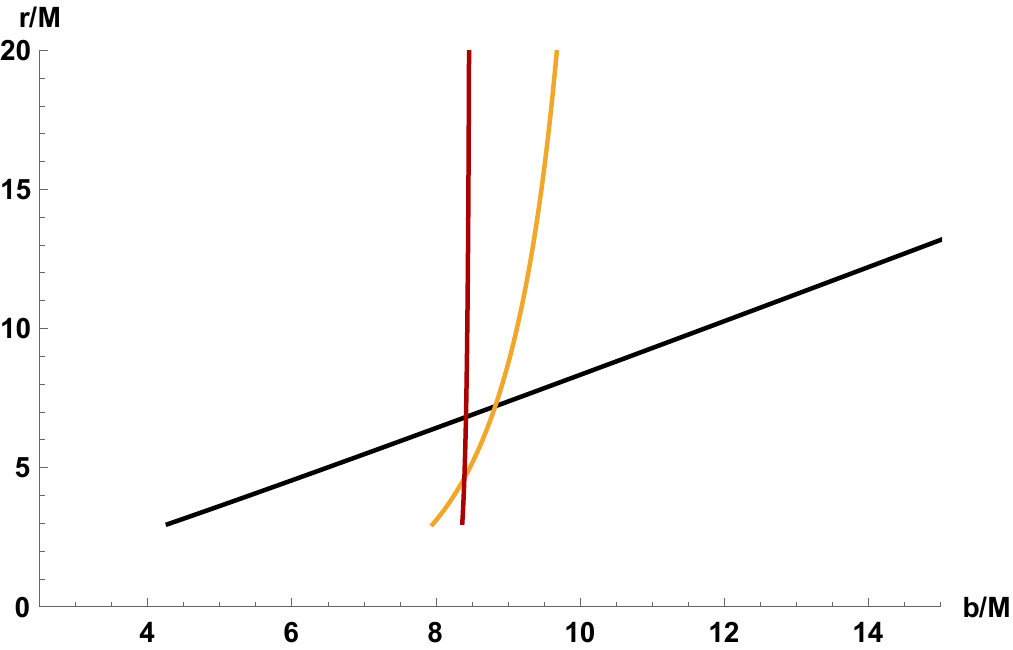}
        \caption{$g=0.1,\alpha=-0.6$}
    \end{subfigure}
    \hfill
\begin{subfigure}[b]{0.23\textwidth}
        \centering
        \includegraphics[width=\linewidth]{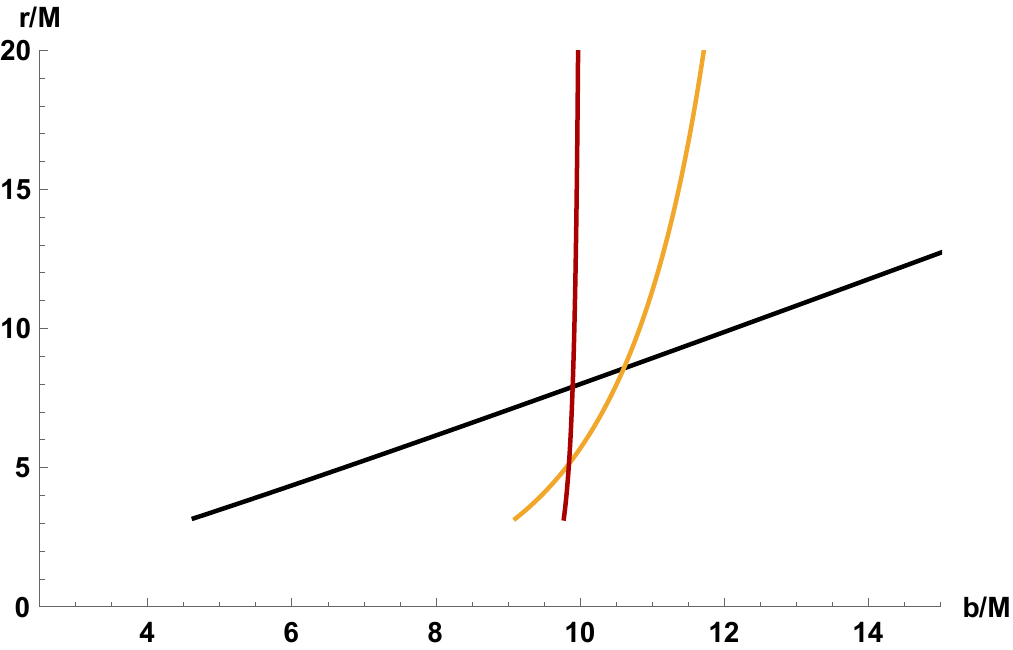}
        \caption{$g=0.1,\alpha=-1.2$}
    \end{subfigure}
    \caption{
        The first three transfer functions of a Bardeen black hole surrounded by PFDM with different magnetic charges $g$ and dark matter parameters $\alpha$.
        They represent the radial coordinates of the first (black), second (gold), and third (red) intersections with the emission. Here, we fix $M=1$.
    }
    \label{fig:transfer}
\end{figure}

Considering that the thin accretion disk emits isotropically, the specific intensity of the emission frequency $\nu_e$ received by an infinitely distant observer can be expressed as
\begin{eqnarray}
I_o(r,v_o)=g^3I_e(r,\nu_e),
\end{eqnarray}
where $g=\nu_o/\nu_e=\sqrt{f(r)}$ is the redshift factor, $\nu_o$ and $I_e(r,\nu_e)$ are the observed frequency and the specific intensity of the accretion disk, respectively. The total observed intensity $I_{obs}(r)$ is obtained by integrating over all observed frequencies $I_o(r,\nu_o)$ and can be written as
\begin{eqnarray}
I_{obs}(r)=\int I_o(r,\nu_o)d\nu_o=\int g^4I_e(r,\nu_e)d\nu_e=f(r)^2I_{em}(r),
\end{eqnarray}
where we denote $I_{em}(r)=\int I_e(r,\nu_e)d\nu_e$ as the total emitted intensity. The light ray tracing back from the observer may intersect the accretion disk multiple times, depending on the type of emission, and extract energy from it. Thus, ignoring the absorption, the total observed intensity is the sum of the intensities from each intersection \cite{Gralla:2019xty}, yielding
\begin{eqnarray}
I_{obs}(b)=\sum_m f(r)^2I_{em}(r)\mid_{r=r_m(b)},
\label{eq:Iobs}
\end{eqnarray}
here, $r_m(b)$ is the transfer function, which describes the radial coordinate of the $m$-th intersection of the light ray with impact parameter $b$ with the accretion disk. The slope $dr/db$, being the scaling factor of the image, indicates that a larger value corresponds to strong demagnification and a smaller total luminosity. Subsequently, the transfer functions of the Bardeen black hole image for different dark matter parameters $\alpha$ are plotted in Fig.\ref{fig:transfer}. The black, gold, and red curves correspond to the transfer functions of direct emission, lensed ring emission, and photon ring emission, respectively. The slope of the transfer function of direct emission is almost $1$, which corresponds to the source profile after redshift and makes the main contribution to the total observed intensity. Compared with direct emission, the transfer function slope of the lensed ring emission is larger, which makes a bit of a contribution to the total photon flux. The slope of the transfer function of the photon ring emission is much larger than $1$, and its contribution can be ignored. In addition, the increase of dark matter parameters $\left|\alpha\right|$ causes the first three transfer functions to move toward the direction of increasing impact parameter $b$, and the widths of the lensed ring and photon ring also significantly widen. This result is consistent with the data in Table \ref{table01}.

To obtain the optical appearance of the Bardeen black hole surrounded by PFDM, we need to give the emission function of the accretion disk according to Eq.(\ref{eq:Iobs}). Here, we choose two toy models with different emission functions \cite{Wang:2022yvi,Yang:2022btw}. In Model I, consider that the emission profile of accretion disk is the power of the second-order decay function starting from the innermost stable circular orbit $r_{isco}$ of the particle
\begin{align}
	\begin{split}
		I_{em1}(r)=\left \{
		\begin{array}{ll}
			I_o\frac{1}{[r-(r_{isco}-1)]^2} ,                ~~~~~~~~~~~~& r>r_{isco}\\\\
			0,                                            ~~~~~~~~~~~~& r\leq r_{isco}
		\end{array},
		\right.
	\end{split}\label{obser1}
\end{align}
here, $I_o$ is the maximum intensity, and the innermost stable circular orbit $r_{isco}$ can be obtained by \cite{Wang:2023vcv}
\begin{equation}
	r_{isco}=\frac{3f(r_{isco})f'(r_{isco})}{2f'(r_{isco})^2-f(r_{isco})f''(r_{isco})},
\label{eq:isco}
\end{equation}
in which $f$ is the metric function of the Bardeen black hole surrounded by PFDM, and the prime represents the derivative with respect to the radial coordinate $r$. In Model II, we consider a more moderate decay emission function starting from the event horizon $r_h$
\begin{align}
	\begin{split}
		I_{em2}(r)=\left \{
		\begin{array}{ll}
			I_o\frac{\frac{\pi}{2}-\arctan(r-r_{risco}+1)}{\frac{\pi}{2}-\arctan(r_{h}-r_{isco}+1)},                 ~~~~~~~~~~~~& r>r_{h}\\\\
			0,                                                                                        ~~~~~~~~~~~~& r\leq r_{h}
		\end{array}.
		\right.
	\end{split}\label{obser3}
\end{align}

\begin{figure}[H]
    \centering
    \begin{subfigure}[t]{\textwidth}
        \centering
        \begin{minipage}[t]{0.32\textwidth}
            \centering
            \includegraphics[width=\linewidth]{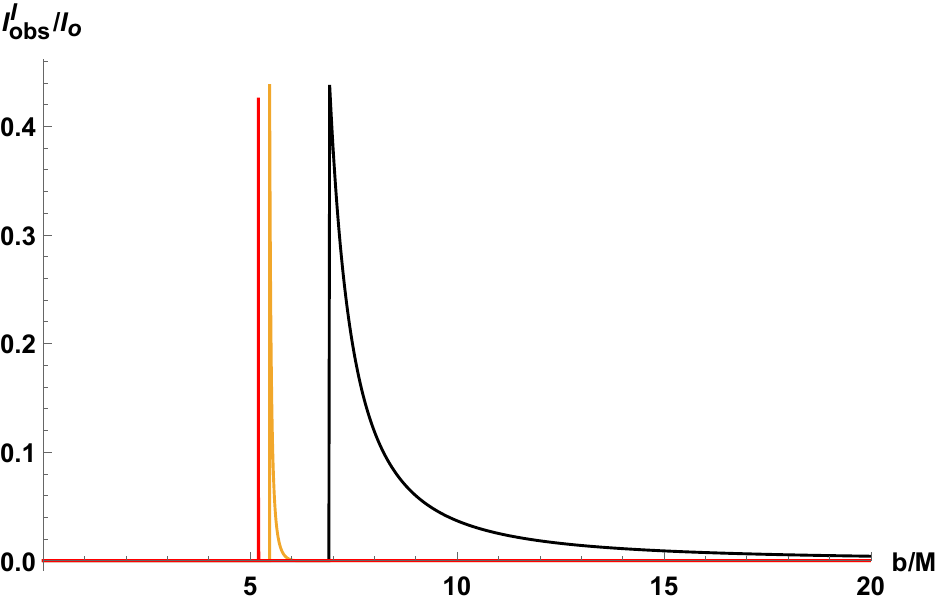}
        \end{minipage}
        \hfill            
\begin{minipage}[t]{0.32\textwidth}
            \centering
            \includegraphics[width=\linewidth]{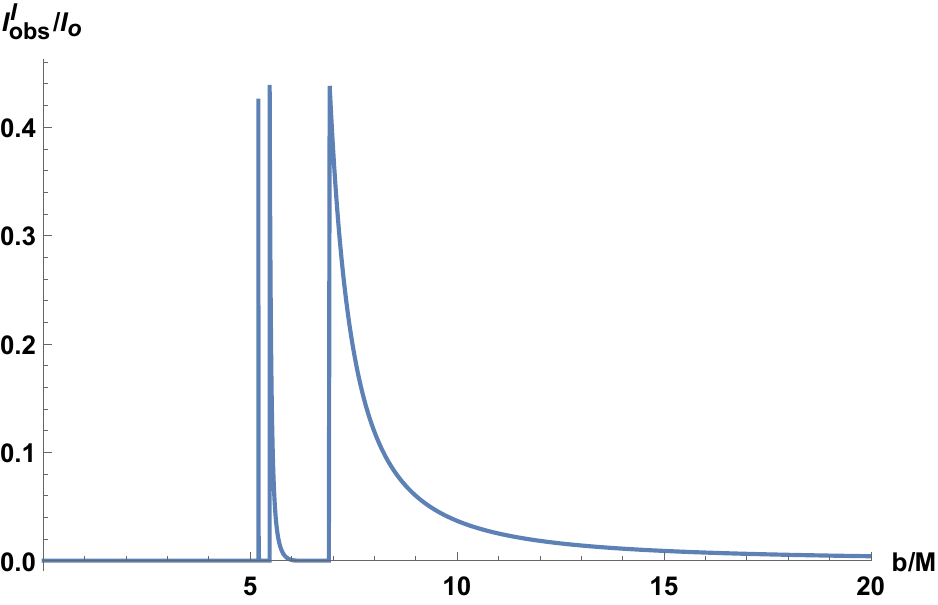}
        \end{minipage}
        \hfill   
\begin{minipage}[t]{0.32\textwidth}
            \centering
            \includegraphics[width=\linewidth]{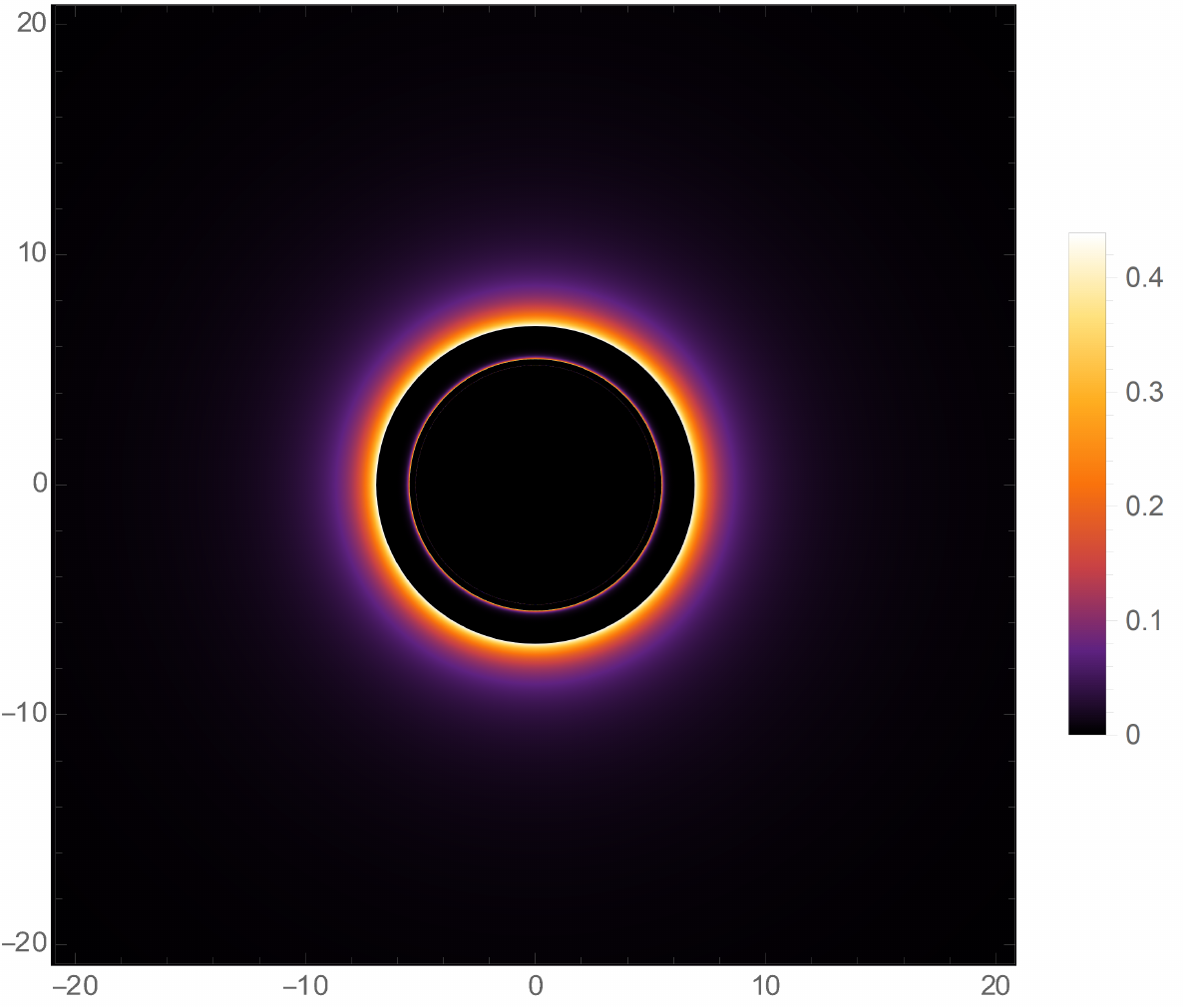}
        \end{minipage}
        \par\centering
        \makebox[\textwidth]{$g=0.1,\alpha=0$}
    \end{subfigure}
    \vspace{0.5em}    
\begin{subfigure}[t]{\textwidth}
        \centering
        \begin{minipage}[t]{0.32\textwidth}
            \centering
            \includegraphics[width=\linewidth]{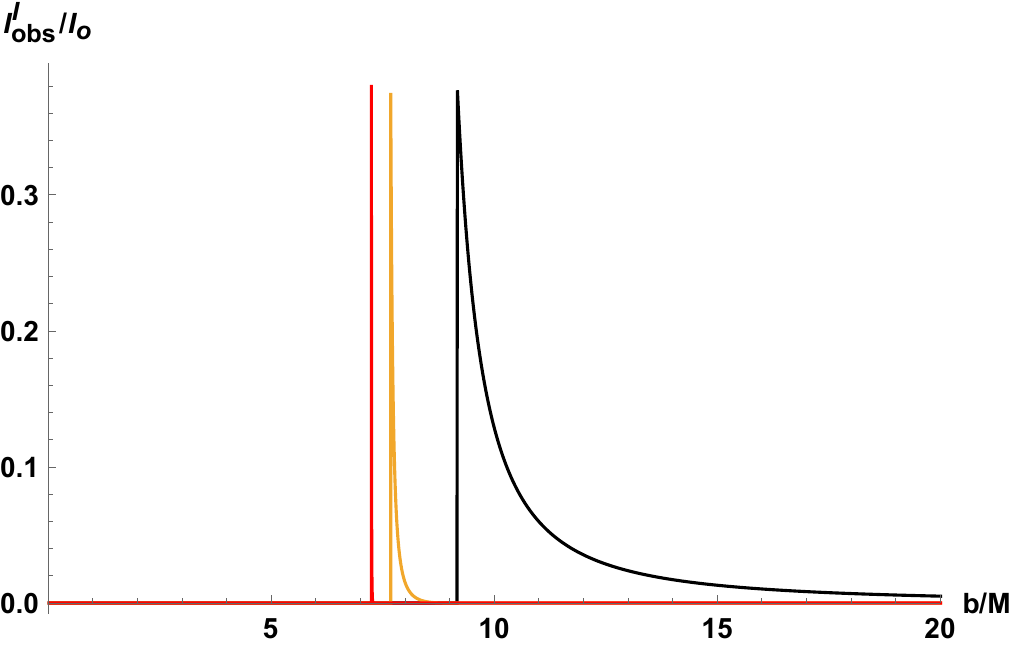}
        \end{minipage}
        \hfill      
\begin{minipage}[t]{0.32\textwidth}
            \centering
            \includegraphics[width=\linewidth]{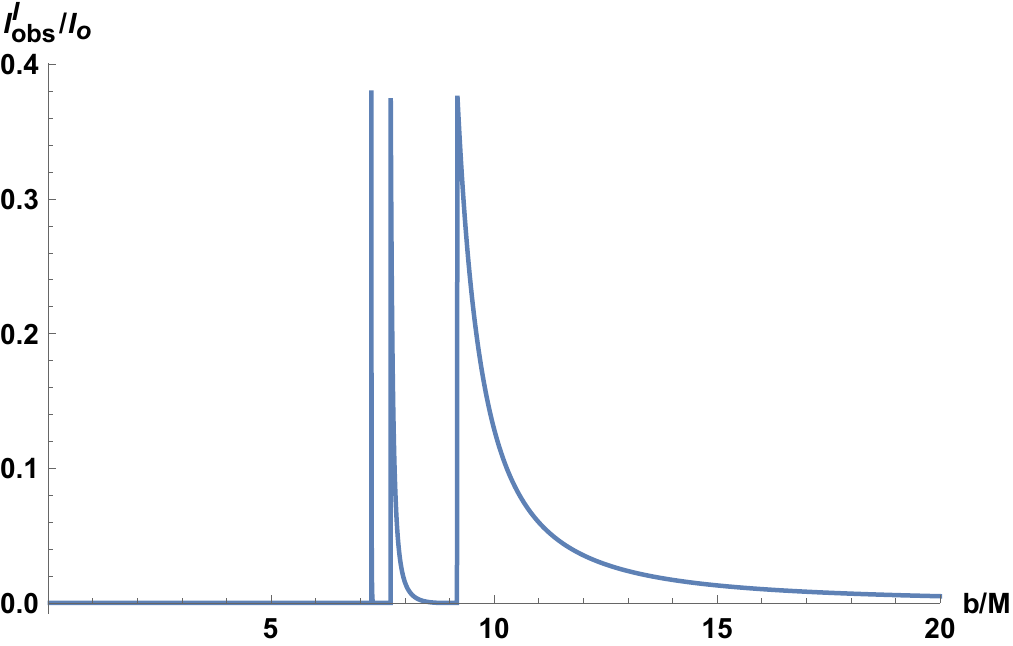}
        \end{minipage}
        \hfill       
\begin{minipage}[t]{0.32\textwidth}
            \centering
            \includegraphics[width=\linewidth]{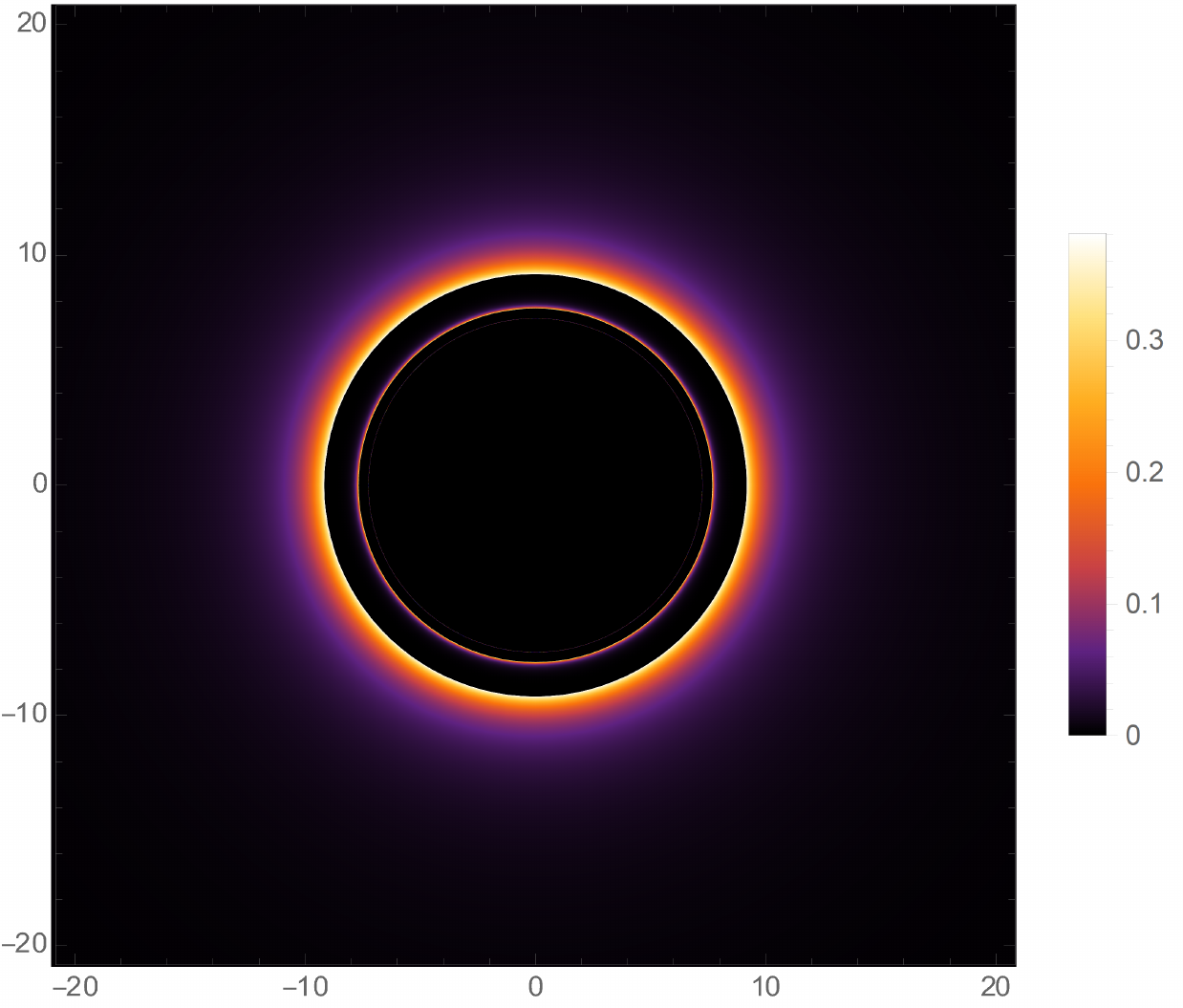}
        \end{minipage}
        \par\centering
        \makebox[\textwidth]{$g=0.1,\alpha=-0.3$}
    \end{subfigure}
    \vspace{0.5em}
\begin{subfigure}[t]{\textwidth}
        \centering
        \begin{minipage}[t]{0.32\textwidth}
            \centering
            \includegraphics[width=\linewidth]{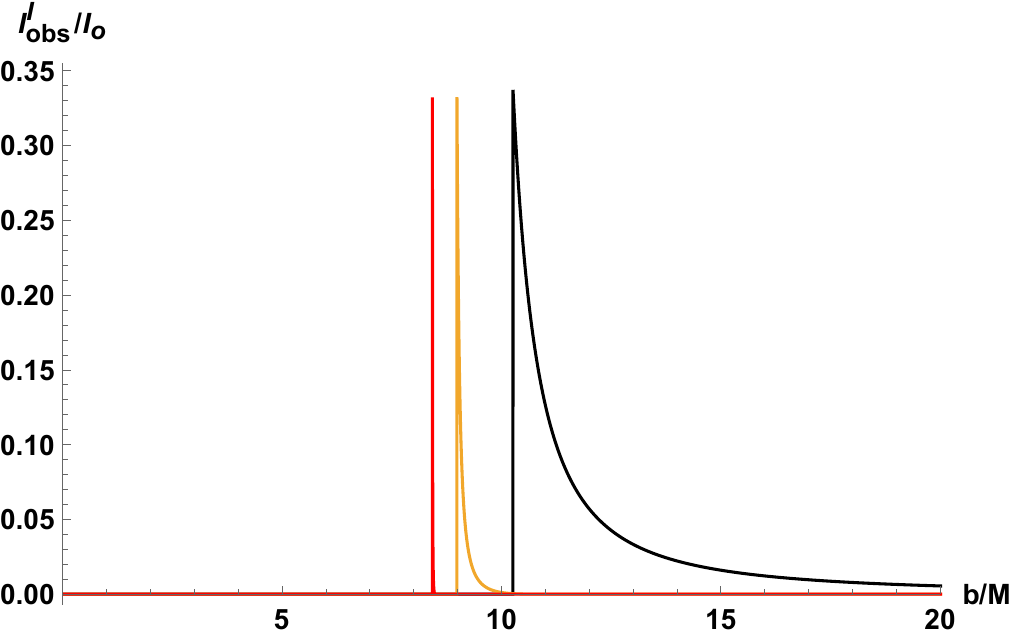}
        \end{minipage}
        \hfill   
\begin{minipage}[t]{0.32\textwidth}
            \centering
            \includegraphics[width=\linewidth]{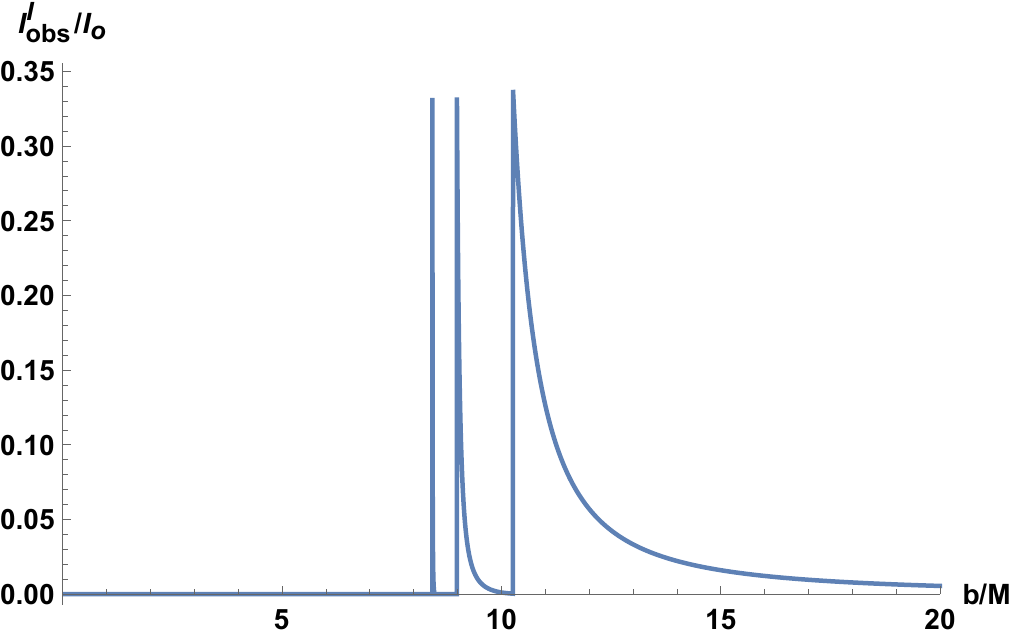}
        \end{minipage}
        \hfill      
\begin{minipage}[t]{0.32\textwidth}
            \centering
            \includegraphics[width=\linewidth]{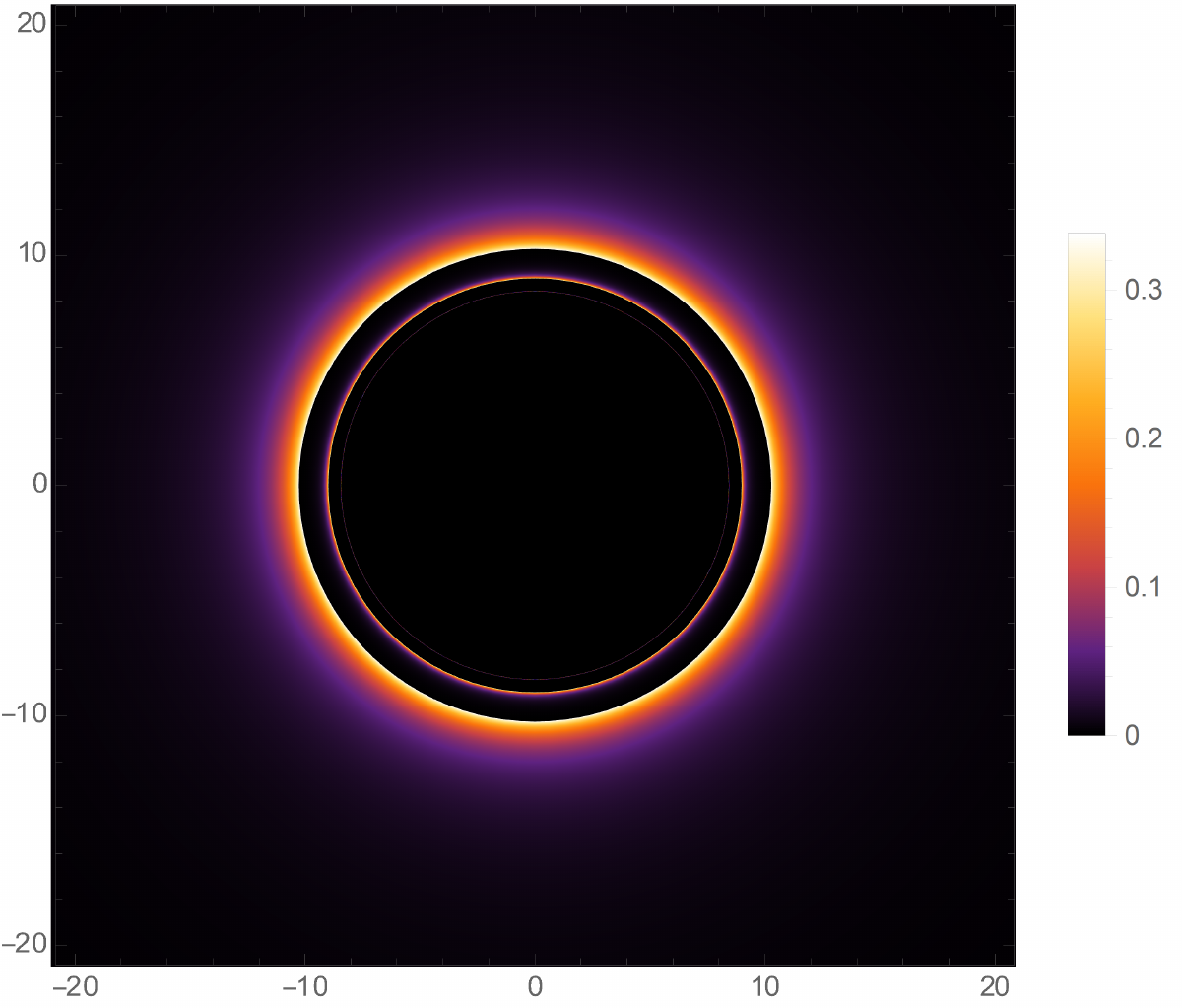}
        \end{minipage}
        \par\centering
        \makebox[\textwidth]{$g=0.1,\alpha=-0.6$}
    \end{subfigure}
    \vspace{0.5em} 
\begin{subfigure}[t]{\textwidth}
        \centering
        \begin{minipage}[t]{0.32\textwidth}
            \centering
            \includegraphics[width=\linewidth]{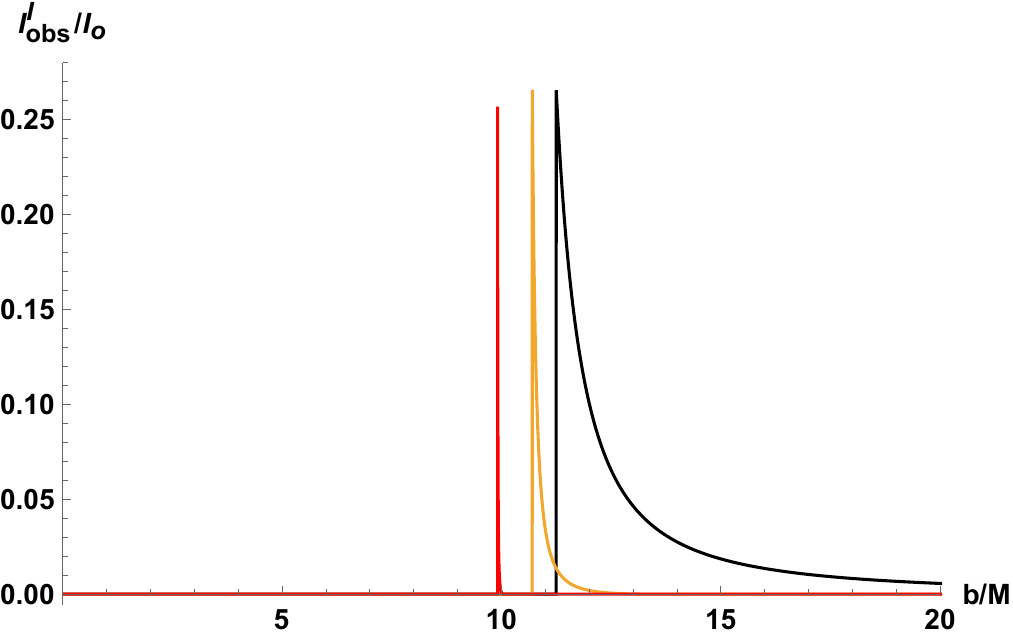}
        \end{minipage}
        \hfill
\begin{minipage}[t]{0.32\textwidth}
            \centering
            \includegraphics[width=\linewidth]{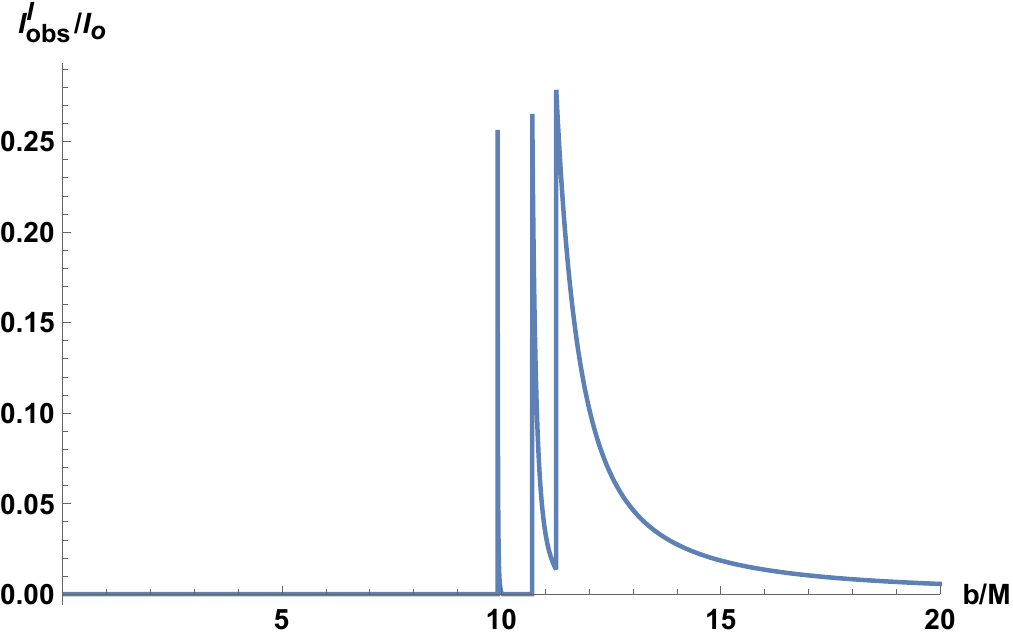}
        \end{minipage}
        \hfill     
\begin{minipage}[t]{0.32\textwidth}
            \centering
            \includegraphics[width=\linewidth]{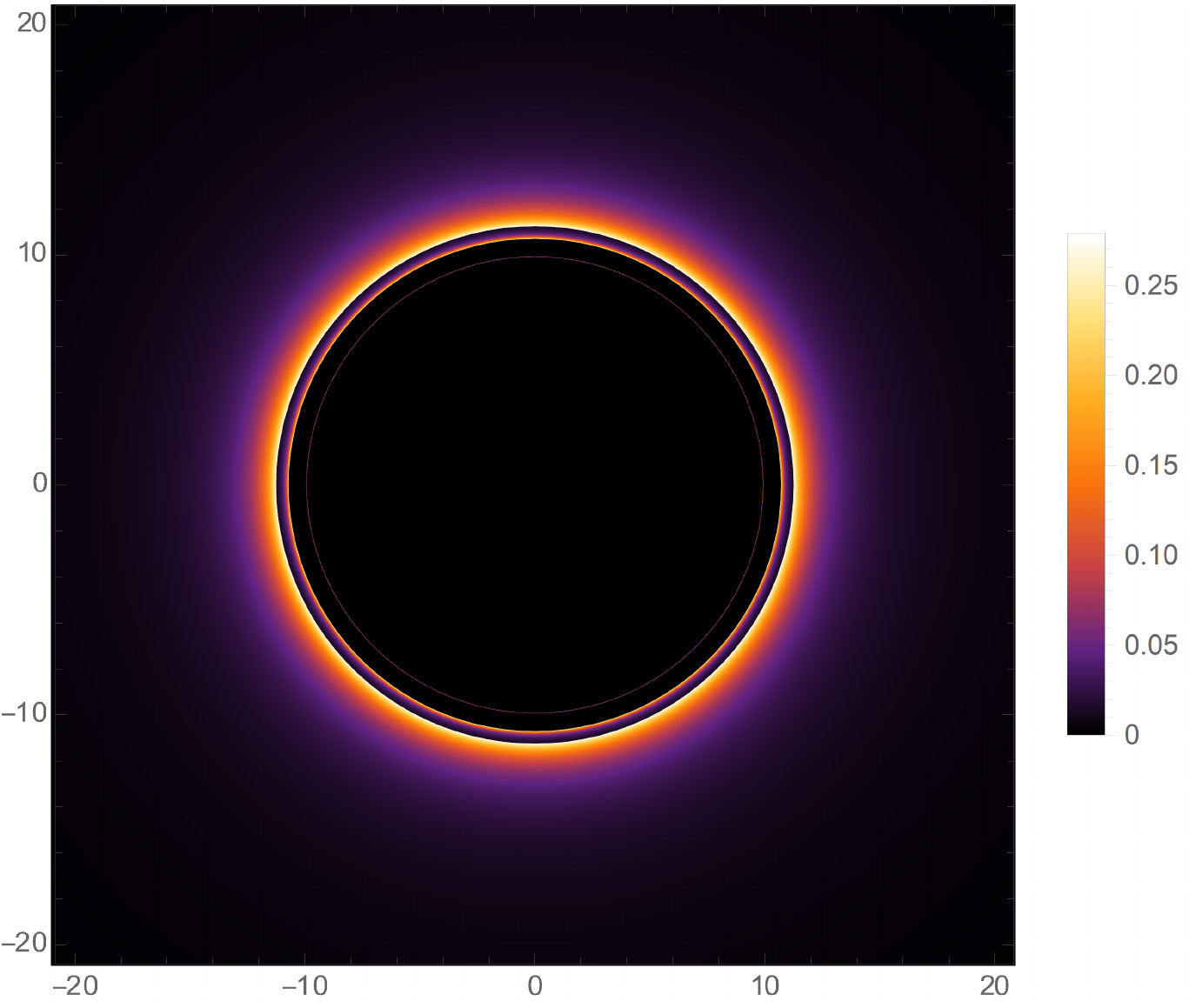}
        \end{minipage}
        \par\centering
        \makebox[\textwidth]{$g=0.1,\alpha=-1.2$}
    \end{subfigure}
    \caption{
        Observational appearances of a thin disk with the first emission profile $I_{em1}(r)$ for different $\alpha$ with $g=0.1$, $M=1$. First column: the different observed intensities originated from the first (black), second (gold), and third (red) transfer function in Eq.\eqref{eq:Iobs} respectively. Second column: the total observed intensities $I_{obs}/I_o$ as a function of impact parameter $b$. Third column: the optical appearance: the distribution of observed intensities into two-dimensional disks.
    }
    \label{fig:Iemit-observed1}
\end{figure}

\begin{figure}[H]
    \centering
    \begin{subfigure}[t]{\textwidth}
        \centering
        \begin{minipage}[t]{0.32\textwidth}
            \centering
            \includegraphics[width=\linewidth]{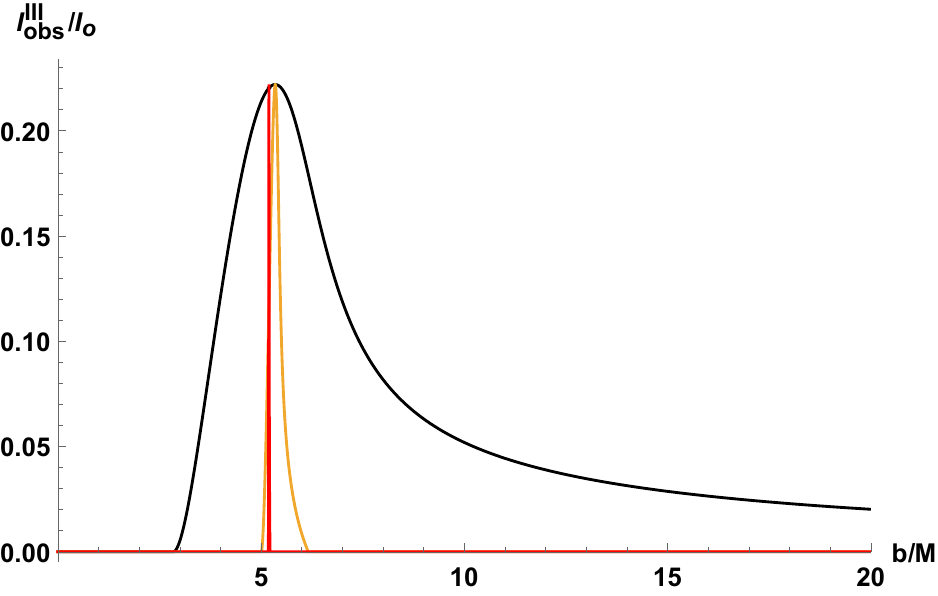}
        \end{minipage}
        \hfill       
\begin{minipage}[t]{0.32\textwidth}
            \centering
            \includegraphics[width=\linewidth]{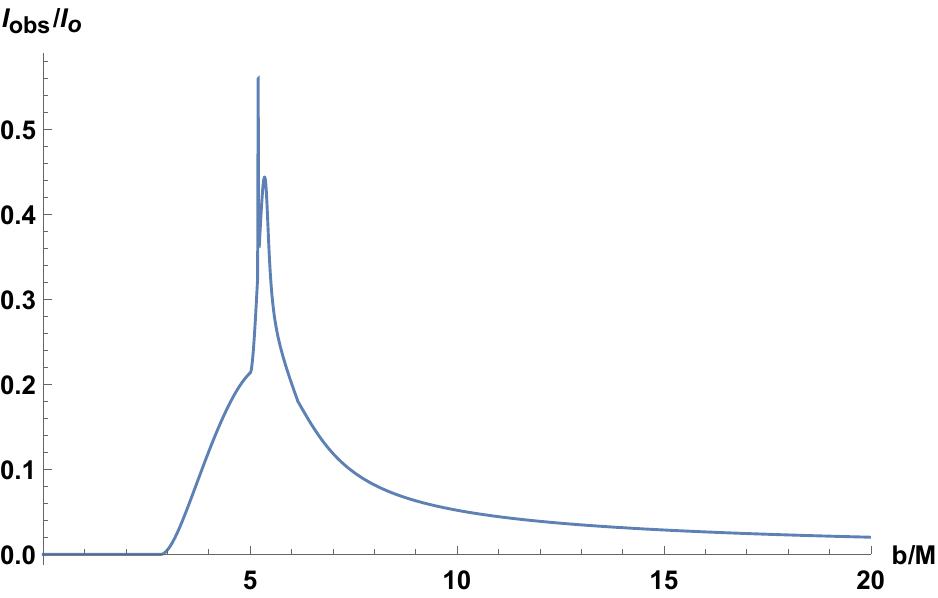}
        \end{minipage}
        \hfill      
\begin{minipage}[t]{0.32\textwidth}
            \centering
            \includegraphics[width=\linewidth]{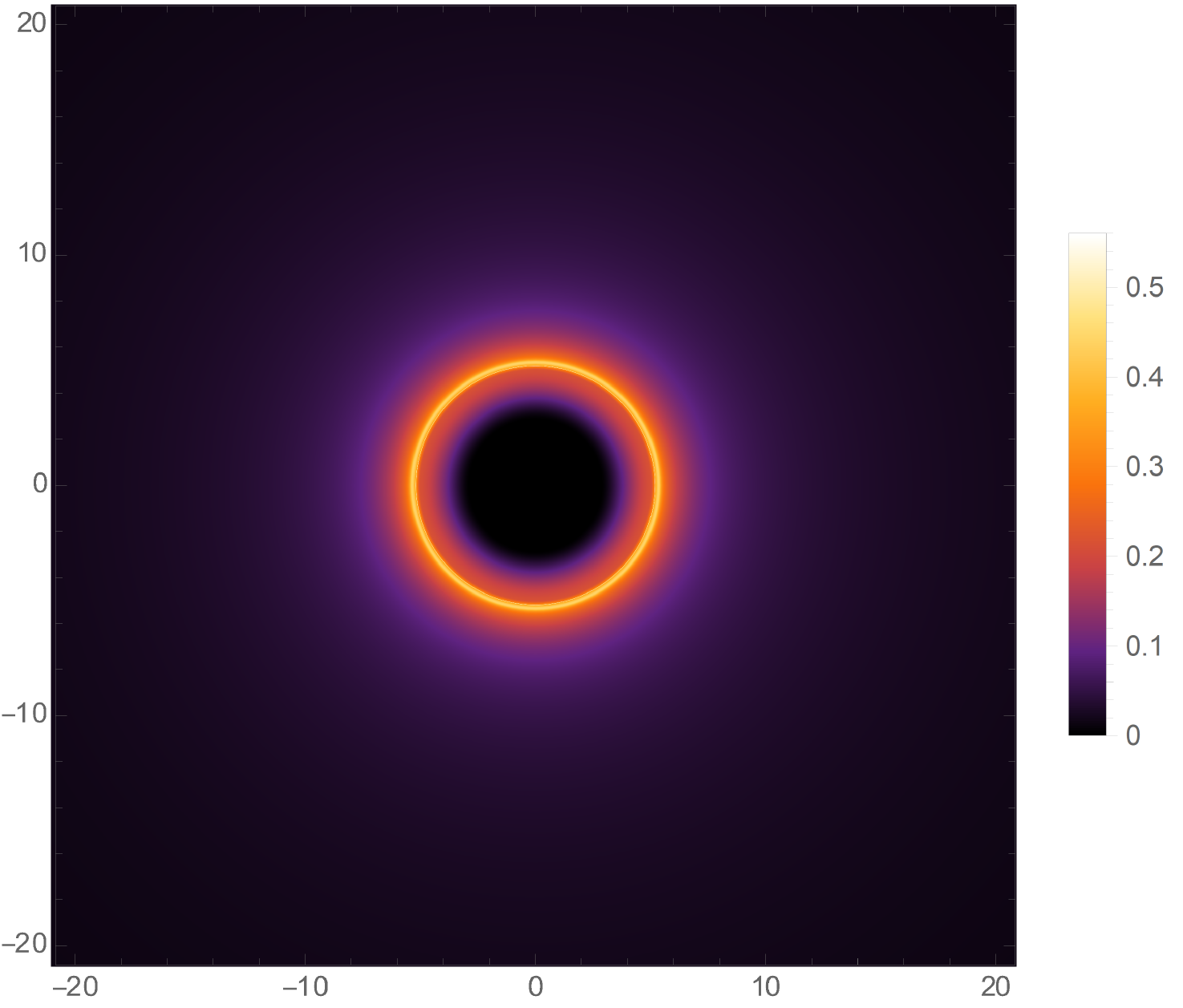}
        \end{minipage}
        \par\centering
        \makebox[\textwidth]{$g=0.1,\alpha=0$}
    \end{subfigure}
    \vspace{0.5em}
\begin{subfigure}[t]{\textwidth}
        \centering
        \begin{minipage}[t]{0.32\textwidth}
            \centering
            \includegraphics[width=\linewidth]{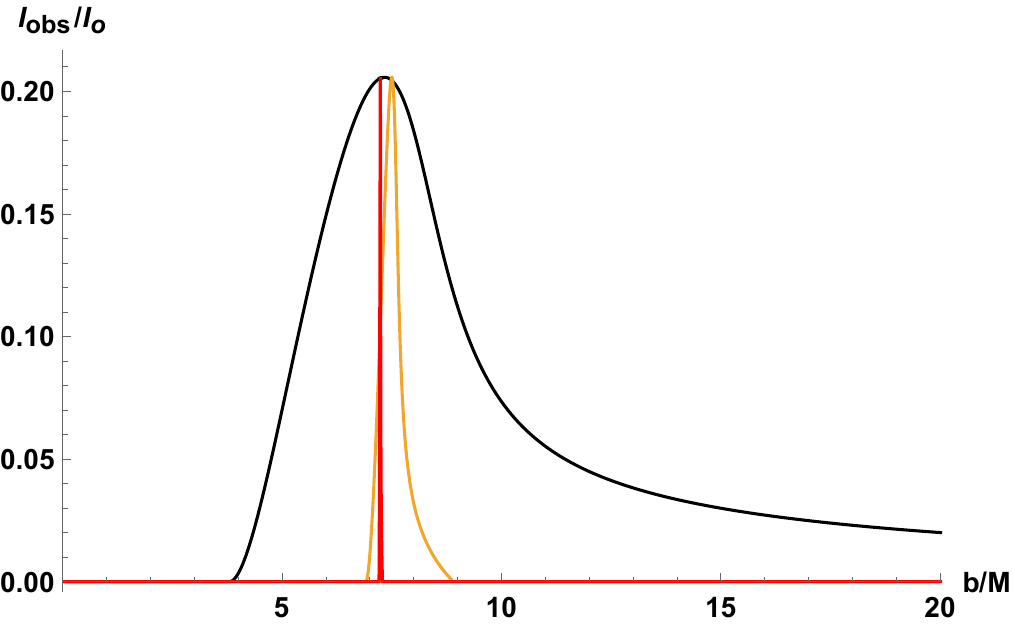}
        \end{minipage}
        \hfill        
\begin{minipage}[t]{0.32\textwidth}
            \centering
            \includegraphics[width=\linewidth]{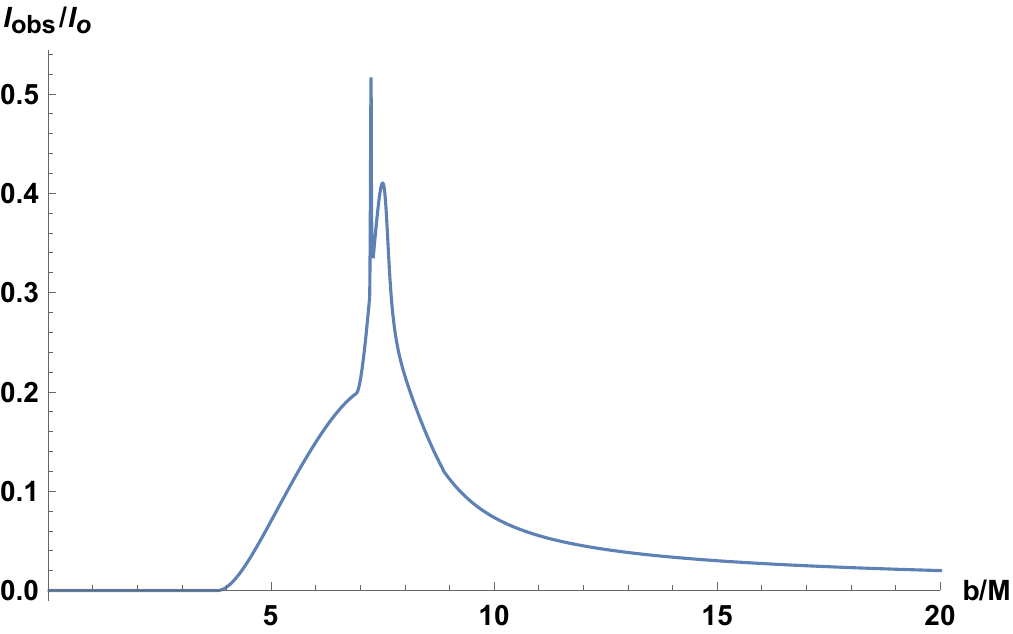}
        \end{minipage}
        \hfill         
\begin{minipage}[t]{0.32\textwidth}
            \centering
            \includegraphics[width=\linewidth]{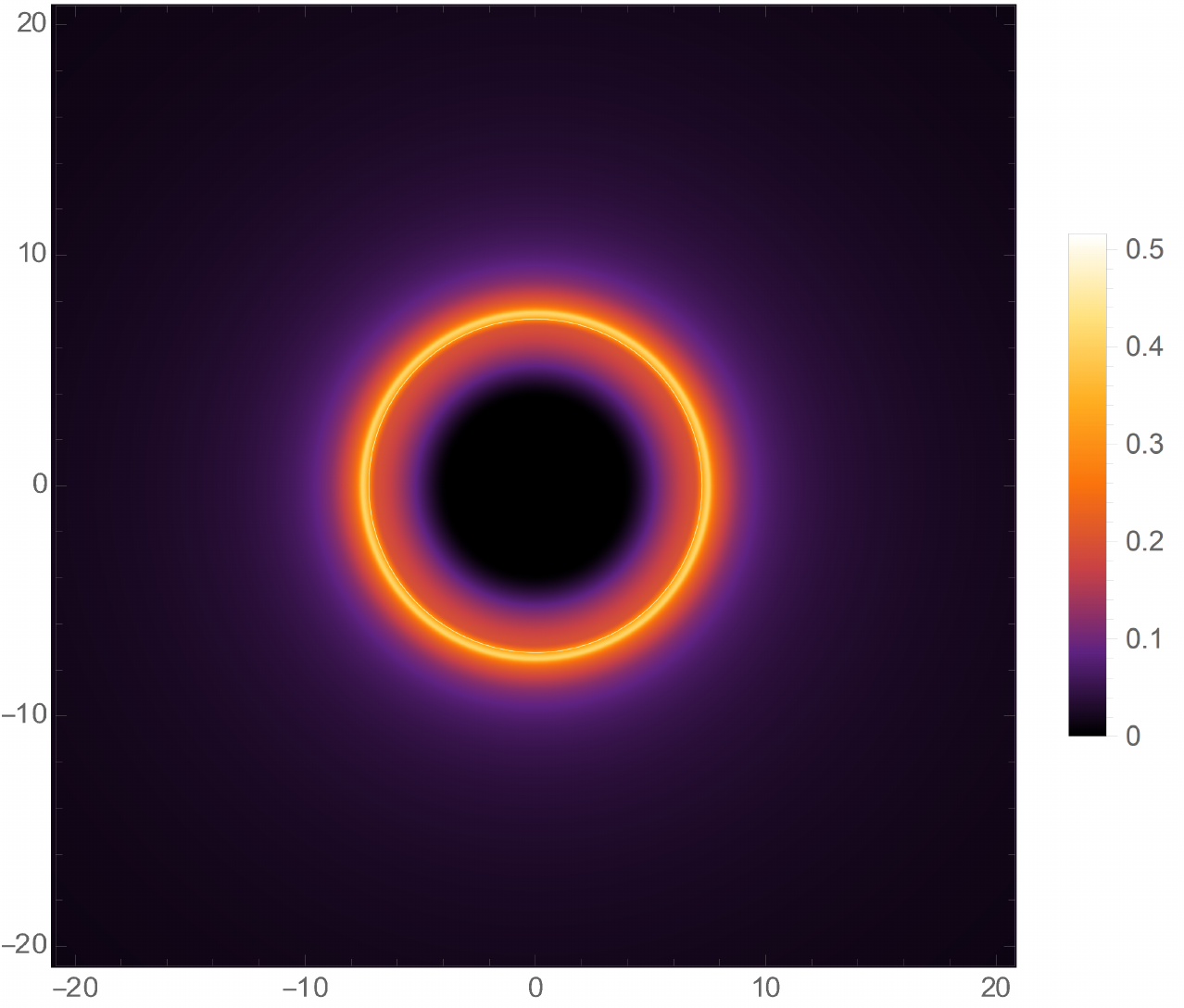}
        \end{minipage}
        \par\centering
        \makebox[\textwidth]{$g=0.1,\alpha=-0.3$}
    \end{subfigure}
    \vspace{0.5em}    
\begin{subfigure}[t]{\textwidth}
        \centering
        \begin{minipage}[t]{0.32\textwidth}
            \centering
            \includegraphics[width=\linewidth]{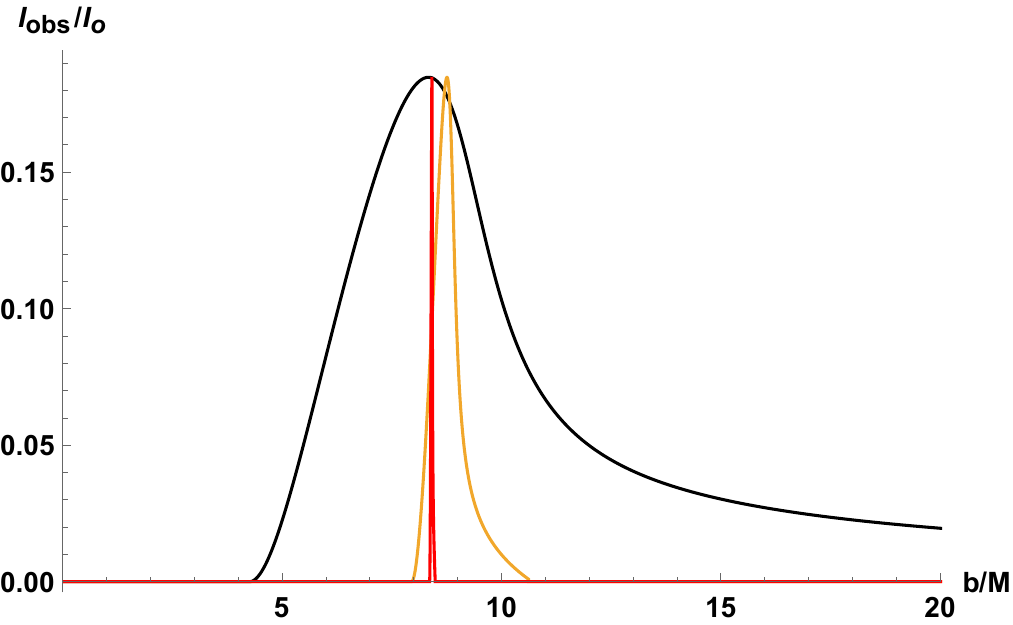}
        \end{minipage}
        \hfill       
\begin{minipage}[t]{0.32\textwidth}
            \centering
            \includegraphics[width=\linewidth]{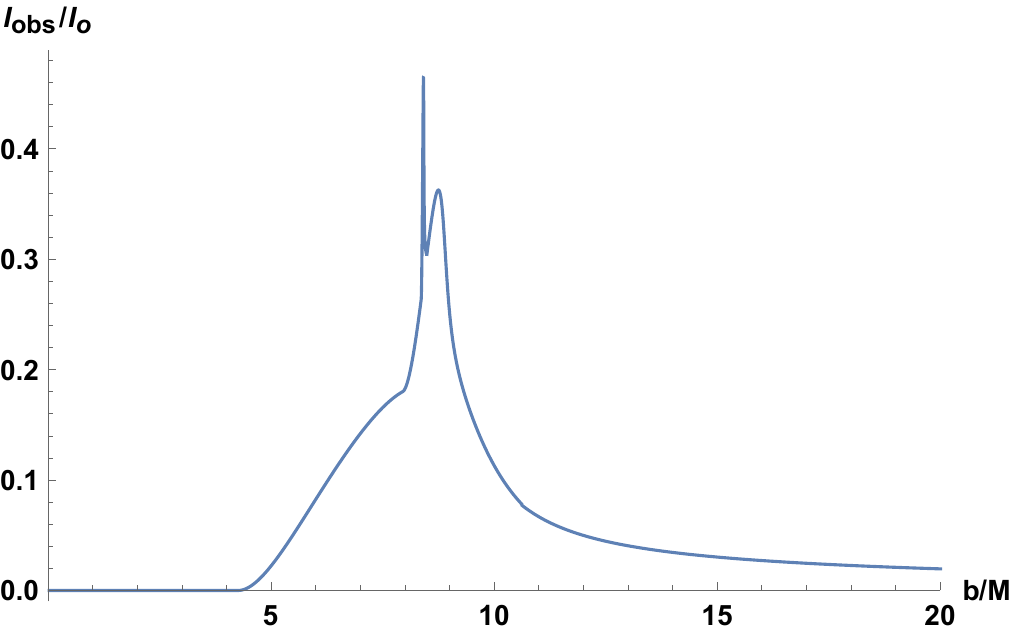}
        \end{minipage}
        \hfill         
\begin{minipage}[t]{0.32\textwidth}
            \centering
            \includegraphics[width=\linewidth]{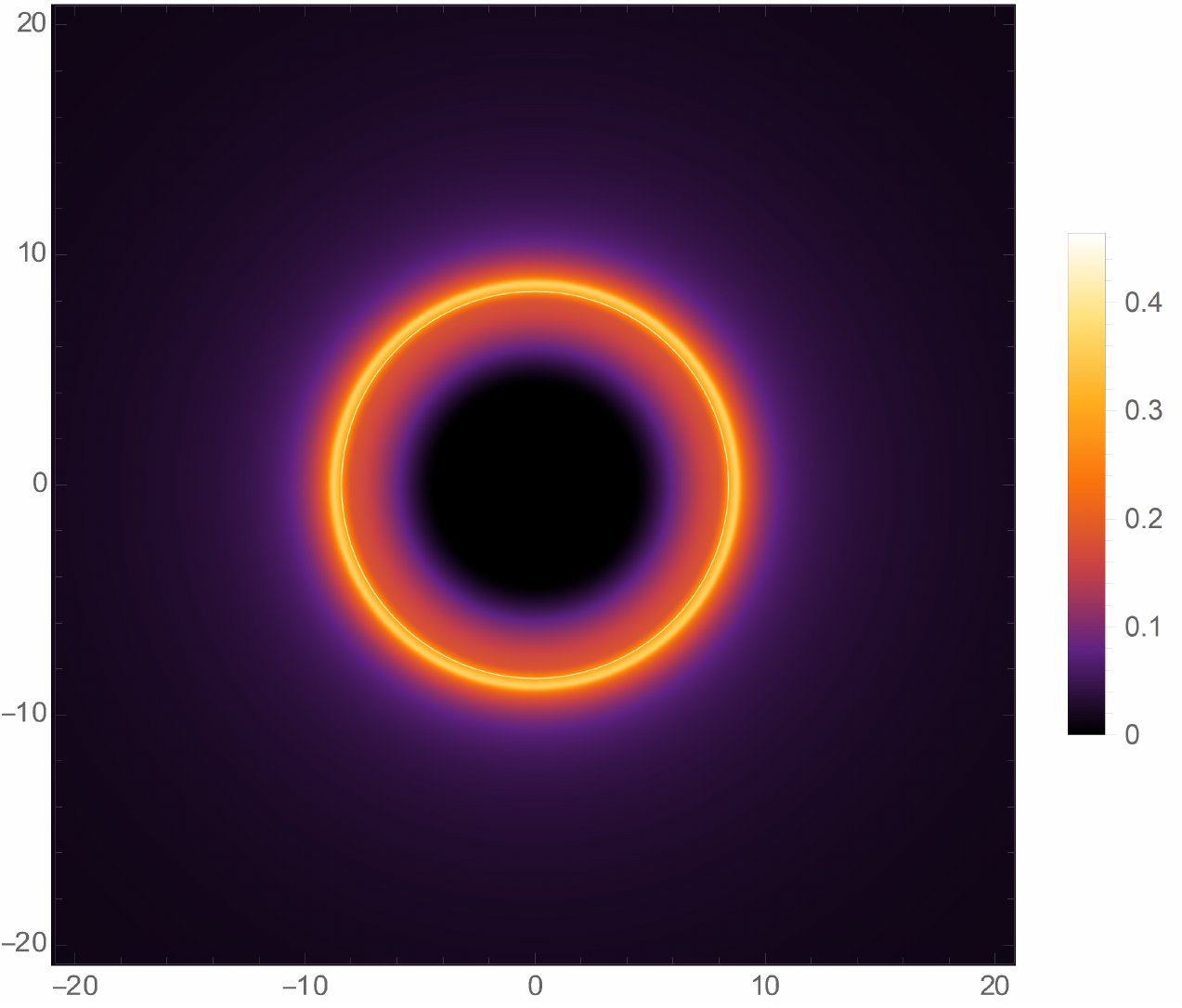}
        \end{minipage}
        \par\centering
        \makebox[\textwidth]{$g=0.1,\alpha=-0.6$}
    \end{subfigure}
    \vspace{0.5em}
\begin{subfigure}[t]{\textwidth}
        \centering
        \begin{minipage}[t]{0.32\textwidth}
            \centering
            \includegraphics[width=\linewidth]{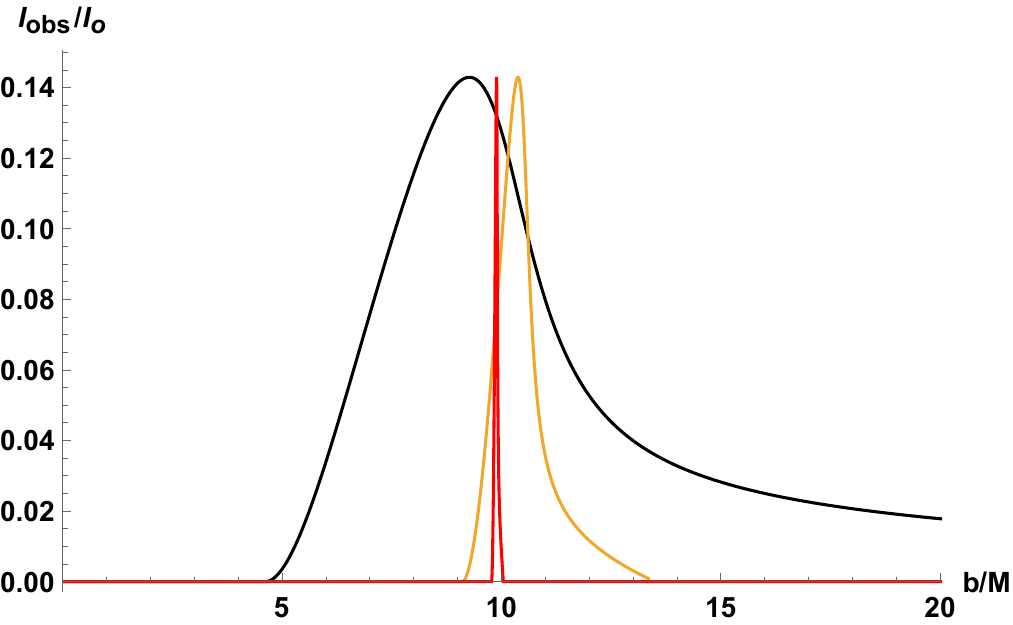}
        \end{minipage}
        \hfill        
\begin{minipage}[t]{0.32\textwidth}
            \centering
            \includegraphics[width=\linewidth]{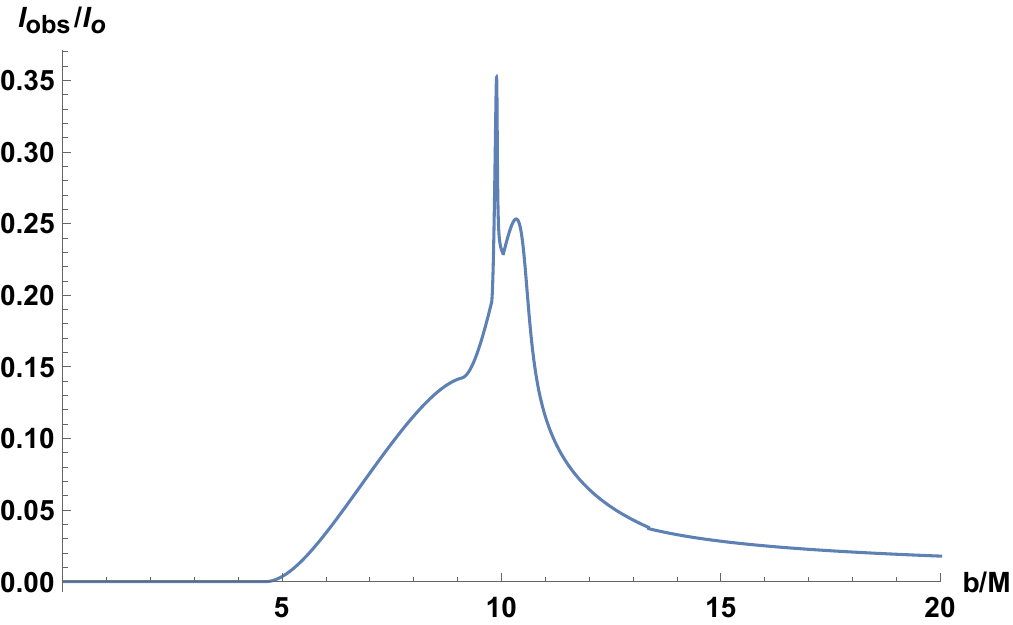}
        \end{minipage}
        \hfill     
\begin{minipage}[t]{0.32\textwidth}
            \centering
            \includegraphics[width=\linewidth]{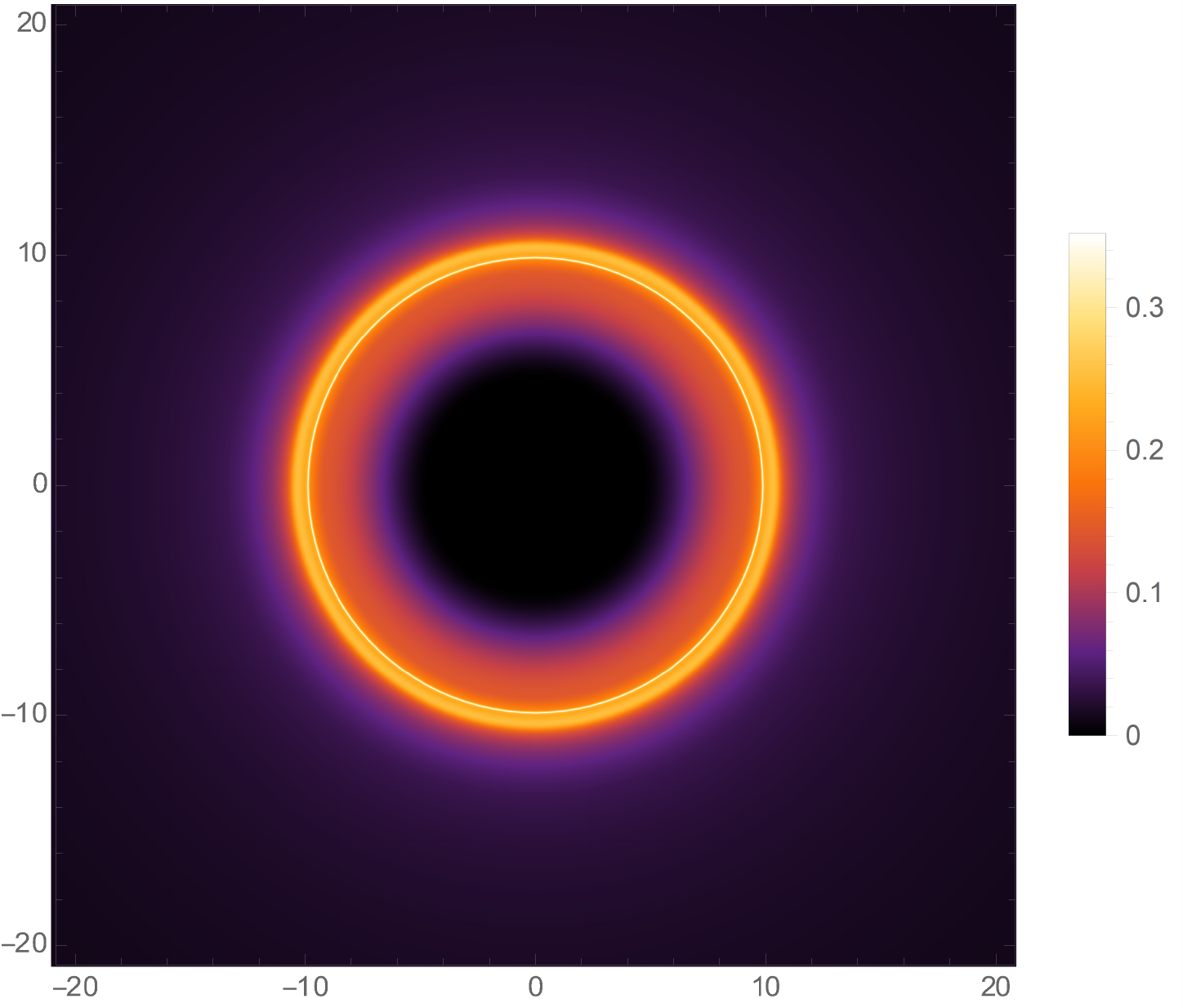}
        \end{minipage}
        \par\centering
        \makebox[\textwidth]{$g=0.1,\alpha=-1.2$}
    \end{subfigure}
    \caption{
        Observational appearances of a thin disk with the third emission profile $I_{em2}(r)$ for different $\alpha$ with $g=0.1$, $M=1$.First column: the different observed intensities originated from the first (black), second (gold), and third (red) transfer function in Eq.\eqref{eq:Iobs} respectively. Second column: the total observed intensities $I_{obs}/I_o$ as a function of impact parameter $b$.Third column: the optical appearance: the distribution of observed intensities into two-dimensional disks.
    }
    \label{fig:Iemit-observed3}
\end{figure}

Next, we will investigate the images of Bardeen black hole surrounded by PFDM illuminated by two different accretion models. For a given emission function model, we can obtain the observed intensity of black hole via Eq.(\ref{eq:Iobs}) and figure out the black hole image. The results for the selected parameters are shown in Fig.\ref{fig:Iemit-observed1} and Fig.\ref{fig:Iemit-observed3}, where each figure can be divided into three columns. The black, gold, and red curves in the first column correspond to the observed intensities of direct, lensed ring, and photon ring emissions, respectively. The second column is the total observed intensity as a function of the impact parameter, and then project them into a two dimensional plane to explicitly present the optical appearance of the accretion disk in the third column. It is obvious from Fig.\ref{fig:Iemit-observed1} and Fig.\ref{fig:Iemit-observed3} that the two different emission models of accretion disk lead to different black hole optical appearances. In the first emission model, the contributions of direct, lensed ring, and photon ring emission are separated, which allows us to clearly observe a thin bright ring contributed by the photon ring emission. In contrast, in model II, the bright ring in the image is the result of the combined effects of these three types of emission, making it difficult to directly distinguish their individual contributions from the black hole's optical appearance. This suggests that it is possible to infer the emission position and physical model of the accretion disk by analyzing the composition of the bright rings in astronomical observations. Of course, this requires more precise observation results in the future. In addition, as the dark matter parameter $\left|\alpha\right|$ increases, the diameter of the inner shadow and bright ring of the black hole image significantly enlarges, while the total observed intensity decreases.

\section{Images of Bardeen black hole surrounded by PFDM illuminated by static spherical accretions}\label{sec:thin spherical accretion}

In this section, we investigate the optical appearance of Bardeen black hole surrounded by PFDM by a static spherical accretion. When matter is trapped by a black hole, the accreting matter usually forms a disk-shaped accretion flow and rotates with a large angular momentum. However, when the angular momentum is extremely small or even negligible, the matter will flow radially to the black hole, forming spherically symmetric accretion \cite{yuan2014hot,jaroszynski1997optics}. For an optically and geometrically thin static accretion with spherical symmetry, the observed intensity $I(\nu_o)$ detected by an observer at infinity $r=\infty$ (measured in $\mathrm{erg~s^{-1}~cm^{-2}~str^{-1}~Hz^{-1}}$ )
$I(\nu_o)$ can be evaluated by integrating the specific emissivity along the photon path $\gamma$\cite{bambi2013can}
\begin{equation}I(\nu_o)=\int_\gamma g^3j_e(\nu_e)dl_\mathrm{prop},
\label{eq:int-Ivo}
\end{equation}
here, $\nu_o$ and $\nu_e$ denote the observed and emitted photon frequencies, respectively. The redshift factor is $g=\nu_o/\nu_e=f(r)^{1/2}$. $j_e(\nu_e)$ is the emissivity per unit volume in the rest frame and usually taken the form $j_e(\nu_e)\propto \delta(\nu_r-\nu_e)/r^2$, where $\nu_r$ is the emitter’s rest-frame frequency. Additionally, $dl_{\text{prop}}$ is the infinitesimal proper length given by
\begin{equation}
	dl_{prop}=\sqrt{\frac{1}{f(r)}dr^2+r^2d\phi^2}=\sqrt{\frac{1}{f(r)}+r^2\left(\frac{d\phi}{dr}\right)^2}dr.
	\label{eq:dlprop}
\end{equation}
By integrating Eq.(\ref{eq:int-Ivo}) over all observed frequencies, the total observed intensity can be obtained
\begin{equation}I_{\mathrm{obs}}=\int_{\nu_{o}} I\left(\nu_{o}\right) d \nu_{o}=\int_{\nu_{e}} \int_{\gamma} g^{4} j_{e}\left(\nu_{e}\right) d l_{\mathrm{prop}} d \nu_{e} =\int_\gamma\frac{f(r)^2}{r^2}\sqrt{\frac{1}{f(r)}+r^2\left(\frac{d\phi}{dr}\right)^2}dr.
\label{eq:int1-Ivo}
\end{equation}

\begin{figure}[H]
    \centering
    \begin{subfigure}[t]{\textwidth}
        \centering
        \begin{minipage}[t]{0.5\textwidth}
            \centering
            \includegraphics[width=\linewidth]{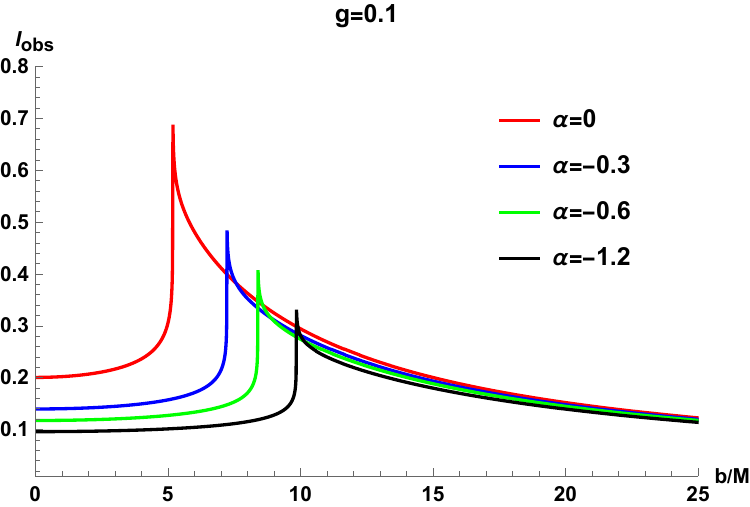}
        \end{minipage}
        \par\centering
        \makebox[\textwidth]{\textbf{(a)} Fixing $g=0.1$ and changing $\alpha$}
        \label{subfig:params}
    \end{subfigure}
    \vspace{0.5em}  
\begin{subfigure}[t]{\textwidth}
        \centering
        \begin{minipage}[t]{0.24\textwidth}
            \centering
            \includegraphics[width=\linewidth]{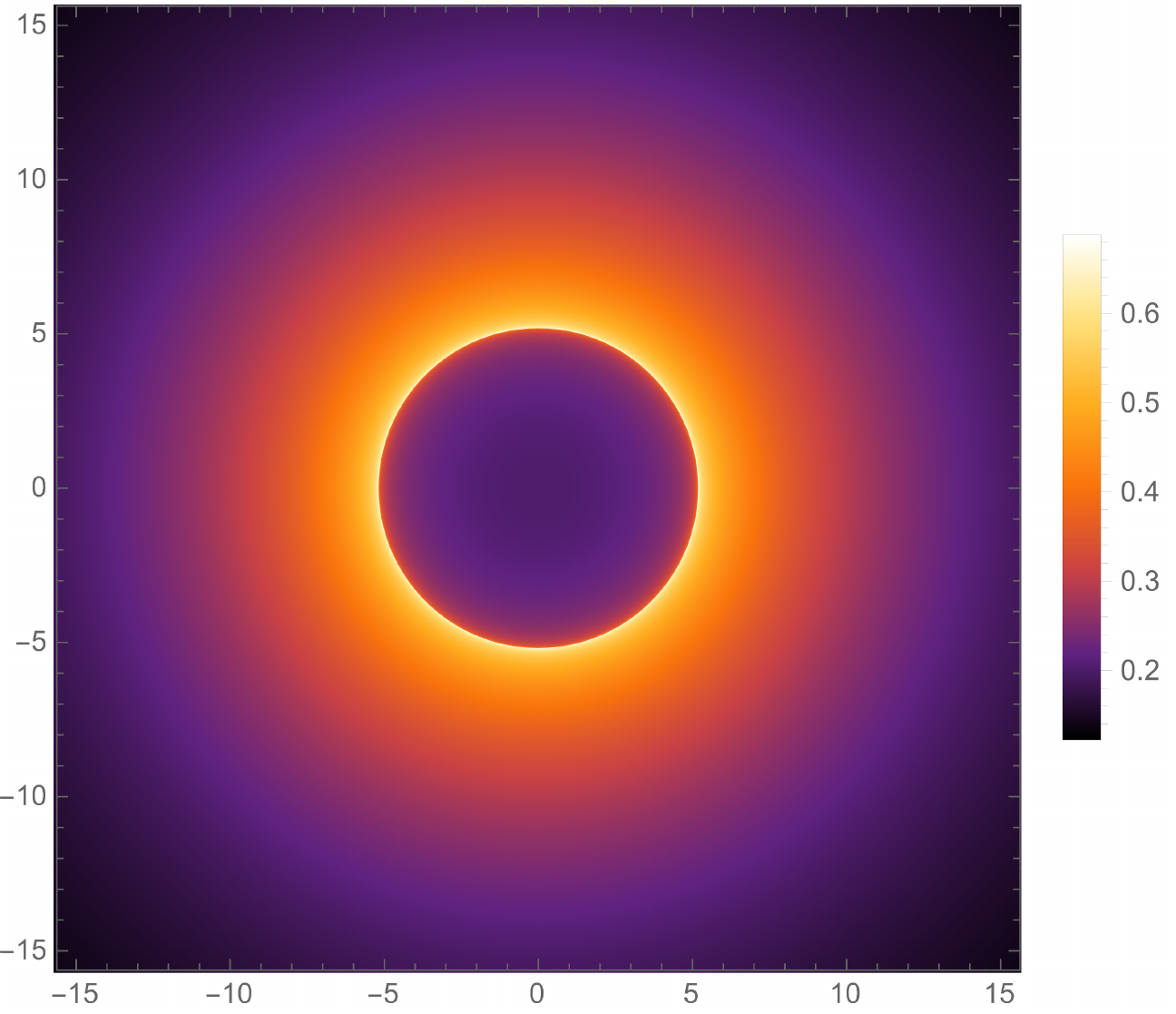}
        \end{minipage}
        \hfill               
\begin{minipage}[t]{0.24\textwidth}
            \centering
            \includegraphics[width=\linewidth]{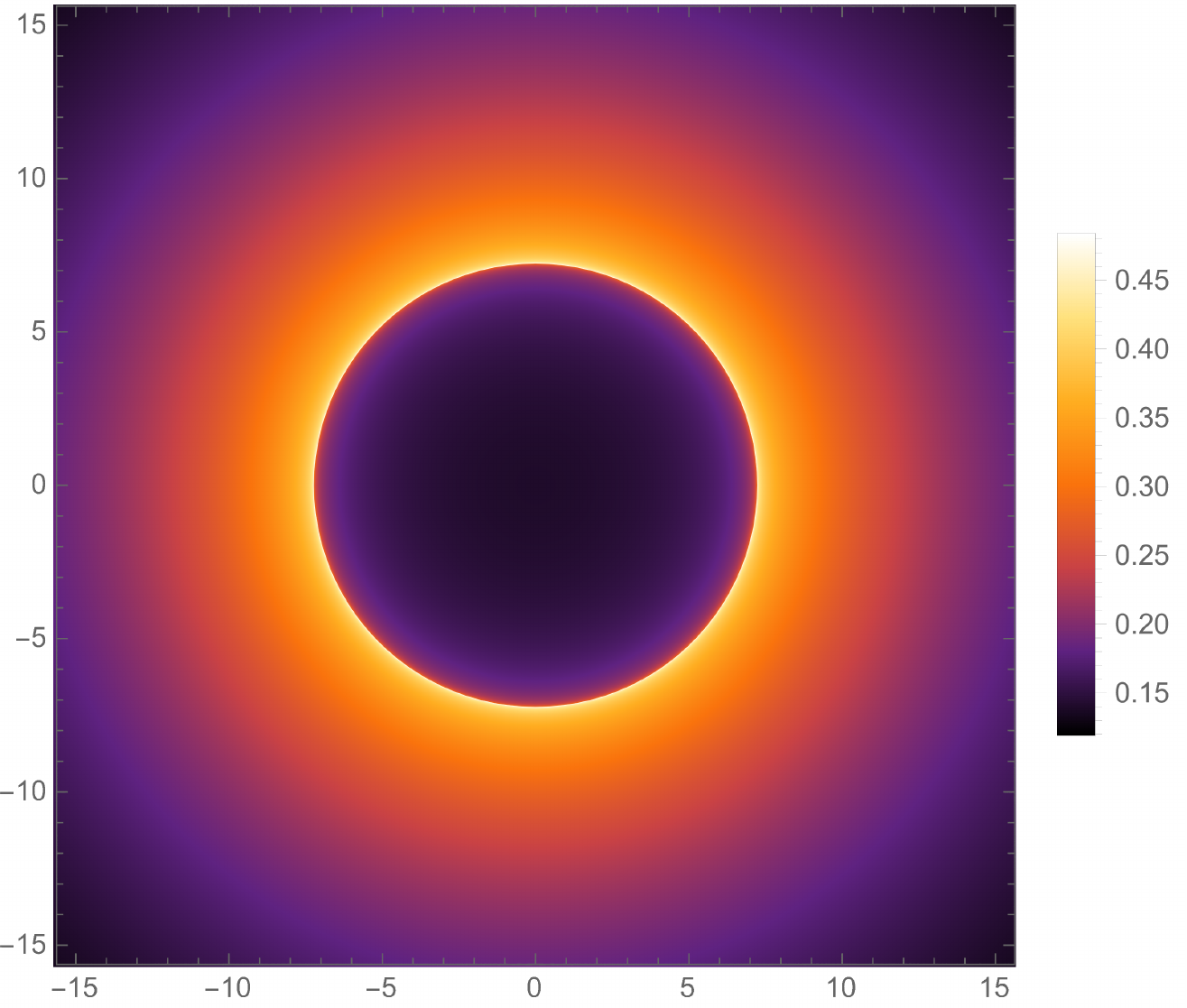}
        \end{minipage}
        \hfill            
\begin{minipage}[t]{0.24\textwidth}
            \centering
            \includegraphics[width=\linewidth]{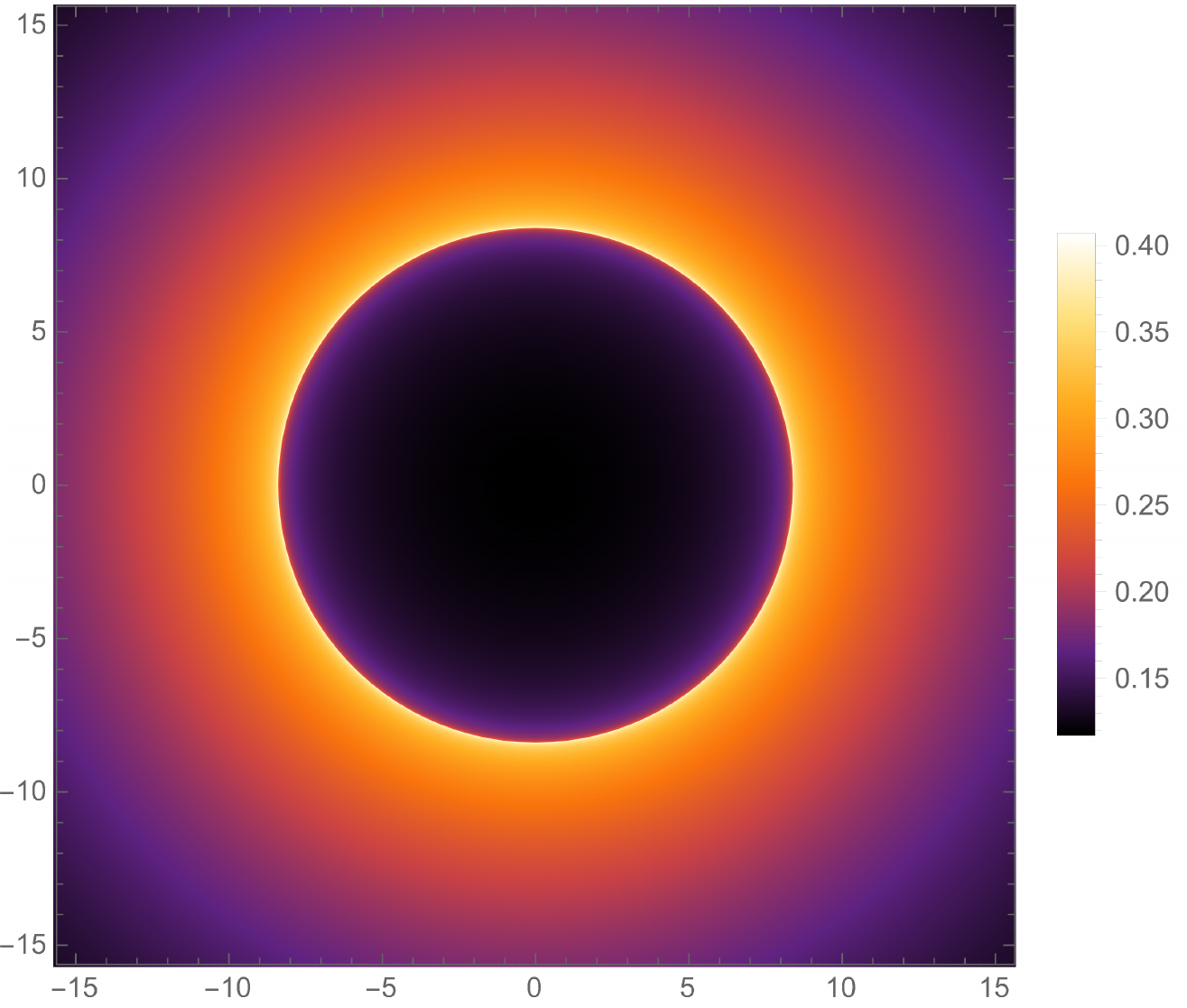}
        \end{minipage}
        \hfill           
\begin{minipage}[t]{0.24\textwidth}
            \centering
            \includegraphics[width=\linewidth]{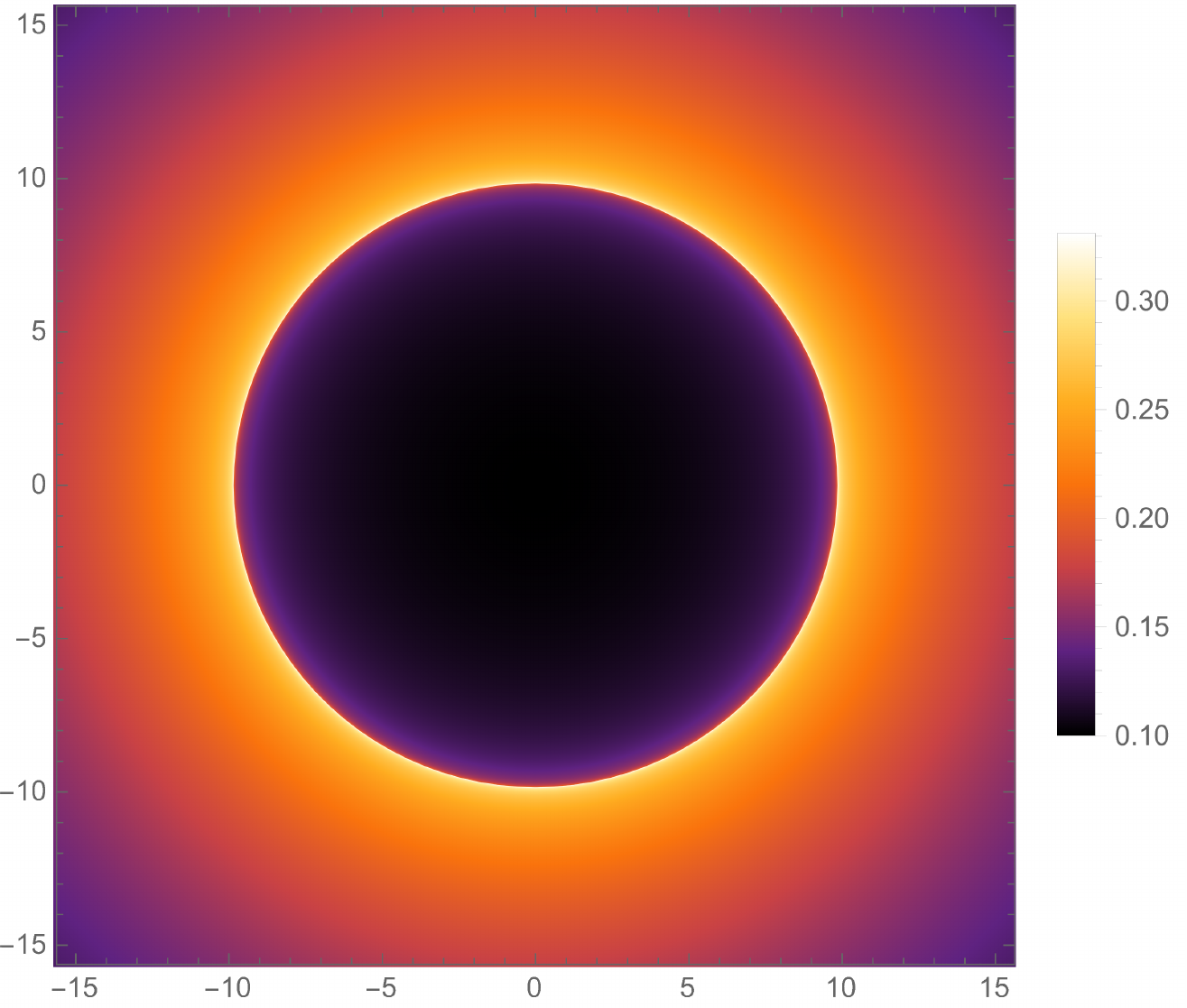}
        \end{minipage}
        \par\centering
        \makebox[\textwidth]{\textbf{(b)} Fixing $g=0.1$, from left to right: $\alpha=0$, $-0.3$, $-0.6$, $-1.2$}
        \label{subfig:alphas}
    \end{subfigure}
    \caption{
        The total observed intensity Eq.(\ref{eq:int1-Ivo}) radiated from the static spherical accretion as a function of impact parameter, and the optical appearances of the Bardeen black hole surrounded by PFDM for selected parameters.
    }
    \label{fig:staticsph}
\end{figure}

Furthermore, we investigated the observed intensity of Bardeen black holes surrounded by PFDM with different dark matter parameters $\alpha$ and depicted them into a two dimensional plane, as shown in Fig.\ref{fig:staticsph}. In Fig.\ref{fig:staticsph}a, the peak in the total observed intensity at $b=b_{ph}$ manifests as the bright ring seen in the black hole image of Fig.\ref{fig:staticsph}b. For $b<b_{ph}$, the total observed intensity increases with the increase of the impact parameter $b$. On the contrary, when $b>b_{ph}$, a larger impact parameter $b$ corresponds to a smaller observed intensity. In addition, for a larger dark matter parameter $\left|\alpha\right|$, the Bardeen black hole surrounded by PFDM exhibits a fainter bright ring and a larger central faint illuminating region in its image. Compared with the Bardeen black hole, the bright ring in the image of the Bardeen black hole surrounded by PFDM is fainter, and the faint illuminating region in the center is larger, which is consistent with our previous conclusions.

\section{Conclusion and discussion}\label{sec:conclusion}

The imaging of black holes by the EHT collaboration provides a new approach to studying physical processes in strong fields regime. Its observational data also offer an opportunity to constrain various theoretical models and theories of gravity. This paper investigates the shadow and accretion images of Bardeen black hole surrounded by PFDM and constrains the dark matter parameters $\alpha$.

As preparation, we first investigate the influence of the dark matter parameter $\alpha$ and the magnetic charge parameter $g$ on the fundamental characteristic quantities of the Bardeen black hole surrounded by PFDM, including the event horizon $r_h$, photon sphere radius $r_{ph}$, critical impact parameter $b_{ph}$, and effective potential $V_{\text{eff}}$. The results show that as the dark matter parameter $\left|\alpha\right|$ increases, $r_h$, $r_{ph}$, and $b_{ph}$ all increase accordingly, while the increase of the magnetic charge parameter $g$ leads to these characteristic quantities to decrease. Furthermore, when $\left|\alpha\right|$ decreases, the peak of the effective potential rises and shifts in the direction of decreasing $r$. Conversely, a decrease in $g$ causes the peak of the effective potential to fall and shift in the direction of increasing $r$. Therefore, for the characteristic quantities of black holes, $r_h$, $r_{ph}$, $b_{ph}$, and $V_{\text{eff}}$, there is a clear competing effect between the magnetic charge parameter $g$ and the dark matter parameter $\left|\alpha\right|$. In addition, we use the observational data of the supermassive black holes M87* and SgrA* to constrain the magnetic charge parameter $g$ and dark matter parameter $\alpha$ of the Bardeen black hole surrounded by PFDM. The magnetic charge parameter $g$ is consistent with observations throughout the entire parameter space ($g,\alpha$), while the dark matter parameter $\alpha$ is consistent with observations only when its value is above the red curve (see Fig. \ref{fig:M87-SgrA}). Further analysis shows that the black hole SgrA* imposes stricter constraints on the dark matter parameter $\alpha$ than black hole M87*.

Next, we analyze the images of the Bardeen black hole surrounded by PFDM illuminated by a geometrically and optically thin accretion disk. The optical appearance of the black hole shows significant differences for the two different emission profile models. In model I, the emissions of direct, lensed ring, and photon ring are separated from each other, and the thin ring produced by the photon ring emission can be clearly observed in the image. In model II, the emissions of the three are superimposed on each other to form the observed bright ring structure, and their respective contributions cannot be distinguished in the image. We look forward to discovering these differences through future higher-precision observational results, and also expect to use these differences to invert the emission profile model and location of the accretion disk. Furthermore, our results show that larger dark matter parameters $\left|\alpha\right|$ lead to an expansion of the inner shadow region of the black hole while reducing the observed intensity of the bright ring. Finally, we analyze the image of the Bardeen black hole surrounded by the PFDM by the illumination of a spherically symmetric accretion. Similar to the case of illumination from the thin accretion disk, larger dark matter parameters $\left|\alpha\right|$ increase the central faint illuminating region and reduce the brightness of the bright ring. Therefore, compared with the classical Bardeen black hole, the image of Bardeen black hole surrounded by PFDM exhibits a larger inner shadow or faint illuminating region. It is worth noting that the images of other black holes surrounded by dark matter also exhibit the same feature: as the density of dark matter increases, the inner shadow in the black hole images enlarges \cite{Zeng:2025kqw,Luo:2025xjb,Li:2025ver}.

In conclusion, we constrain the dark matter parameters $\alpha$ using EHT observational data. Based on the investigation of different emission profile models, it is found that different models will significantly affect the image of the black hole. Furthermore, our research demonstrates that dark matter parameters also significantly influence the black hole image, which provides a potential method to distinguish Bardeen black holes surrounded by PFDM from classical Bardeen black holes. Therefore, we expect that future higher-precision observation data will be able to observe these differences.

In fact, in the field of black hole observation, a series of achievements have been made, especially in accretion images and magnetic field deflection characteristics \cite{EventHorizonTelescope:2019dse,EventHorizonTelescope:2022wkp,EventHorizonTelescope:2025vum}, but the understanding of the structure of black hole photon rings and the internal properties of accretion disks is still inadequate \cite{EventHorizonTelescope:2024whi,Johnson:2019ljv}. We expect these preliminary results to provide some insights into using black hole shadows and images to study dark matter in the future.

\begin{acknowledgments}
This work is partly supported by the Natural Science Foundation of China under Grant No.12375054. The authors thank Professor Xiao-Mei Kuang and Dr. Xi-Jing Wang for their valuable help and guidance.

\textbf{Note added:} When we finished this article, we noticed a related independent work by arXiv:2512.00824 \cite{Feng:2025ljz}. Although both studies investigate the images of Bardeen black holes surrounded by PFDM, our work distinguishes itself in the following key aspects: 1. By investigating the fundamental characteristic quantities of black holes, we discovered a competing effect between dark matter parameter $\left|\alpha\right|$ and magnetic charge parameter $g$; 2. More importantly, we additionally discussed images of black holes illuminated by a spherical accretion model and investigated the relationship between the images and the accretion model. In addition, our article and reference \cite{Feng:2025ljz} set different dark matter parameters, but both found that dark matter increases the inner shadow of the black hole and reduces the observation intensity, and the conclusions remained consistent.

\end{acknowledgments}

\bibliography{ref}
\bibliographystyle{apsrev}

\end{document}